\documentclass[12pt,prd,aps,amssymb,amsmath,tightenlines,showpacs]{article}
\usepackage[utf8]{inputenc}
\usepackage[a4paper,top=3cm,bottom=3cm,left=2.5cm,right=2.5cm]{geometry}
\usepackage{amssymb}
\usepackage{amsmath}
\usepackage{amsfonts}
\usepackage{mathtools}
\usepackage{amsthm}
\usepackage{tipa}
\usepackage{yhmath} 
\usepackage{latexsym} 
\usepackage{graphicx}
\usepackage{comment}
\usepackage{color}
\usepackage{tikz}
\usepackage[compat=1.1.0]{tikz-feynman}

\newcommand{\be}{\begin{equation}}
\newcommand{\ee}{\end{equation}}
\newcommand{\bea}{\begin{eqnarray}}
\newcommand{\eea}{\end{eqnarray}}

\begin{document}
	\graphicspath{{FIGURE/}}
	\topmargin=0.0cm
	
\begin{center} {\Large{\bf Non-linear and weak-coupling expansion in \\Quantum Field Theory}}\\	
	\vspace*{0.8 cm}

Vincenzo Branchina$^{a,b,}$\,\footnote{branchina@ct.infn.it}, 
Alberto Chiavetta$^{a,c}$\,\footnote{albertochiavetta@gmail.com}, 
Filippo Contino$^{a,b}$\,\footnote{filippo.contino@ct.infn.it}
	\vspace*{0.4cm}
	
	{\it ${}^a$Department of Physics and Astronomy, University of Catania,\\
Via Santa Sofia 64, 95123 Catania, Italy,} \\
	\vspace*{0.02cm}
{\it ${}^b$
INFN, Sezione di Catania, Via Santa Sofia 64, 95123 Catania, Italy,} \\
	\vspace*{0.02cm}
{\it ${}^c$Scuola Superiore di Catania, Via Valdisavoia 9, 95123 Catania, Italy} \\
	
	\vspace*{1 cm}
	
	{\LARGE Abstract}\\
	
\end{center}
	
A formal expansion for the Green's functions of an interacting quantum field theory in a parameter that somehow encodes its ``distance" from the corresponding non-interacting one was introduced more than thirty years ago, and has been recently reconsidered in connection with its possible application to the renormalization of non-hermitian theories. Besides this new and interesting application, this expansion has special properties
already when applied to ordinary (i.e. hermitian) theories, and in order to disentangle the peculiarities of the expansion itself from those of non-hermitian theories, it is worth to push further the investigation limiting first the analysis to ordinary theories. In the present work we study some aspects related to the renormalization of a scalar theory within the framework of such an expansion. Due to its peculiar properties, it turns out that at any finite order in the expansion parameter the theory looks as non-interacting. We show that when diagrams of appropriate classes are resummed, this apparent drawback disappears and the theory recovers its interacting character. In particular we have seen that with a certain class of diagrams, the weak-coupling expansion results are recovered, thus establishing a bridge between the two expansions.

\section{Introduction}
Even though almost hundred years passed since the pioneeristic work of Dirac on the quantization of the electromagnetic field\,\cite{Dirac:1927dy}, it can hardly be said that the subject of Quantum Field Theory (QFT) is a closed chapter of theoretical physics. This is strikingly  evident with the appearance of divergent contributions to physical amplitudes, whose  presence prompted the introduction of renormalization\,\cite{Tomonaga:1946zz,Koba:1947rzy,Bethe:1947id,Feynman:1948km,Feynman:1948fi,Schwinger:1948yk,Schwinger:1948iu,Dyson:1949bp}, that independently of the divergences themselves has proved to be a fruitful and insightful approach to QFTs\,\cite{Petermann:1953wpa,GellMann:1954fq,Wilson:1971bg,Wilson:1971dh,Wegner:1972ih,Nicoll:1974zz,Wilson:1973jj,Wilson:1974mb,Hasenfratz:1985dm}. 
However, in spite of the efforts made along several decades for implementing QFTs and the related renormalization program in perturbative\,\cite{tHooft:1972tcz,tHooft:1972ikm,Gross:1973id,Politzer:1973fx} and/or non-perturbative\,\cite{Creutz:1980zw,Polchinski:1983gv,Luscher:1987ay,Luscher:1987ek,Hasenfratz:1987mv,Wetterich:1989xg,Liao:1992fm,Morris:1995af} frameworks, these still remain evergreen subjects for new discoveries,
sometimes even for frustration. 

The Green's functions, the main tools for calculating physical quantities in QFT, have been approximated according to different expansions and/or resummations: semiclassical or $\hbar$ expansion, weak-coupling expansion, truncations of Dyson-Schwinger-type, Borel-resummations, Padé approximants, resummation of infrared and ultraviolet renormalons\,\cite{Coleman,Rivers,Peskin,Zinn}, just to mention few well known approaches.

By considering the case of a scalar theory with a typical polynomial interaction, almost thirty years ago an expansion in terms of a non-linearity parameter that somehow encodes the ``distance" between the linear (non-interacting) theory and the interacting one was proposed in\, \cite{Bender:1987dn,Bender:1988rq}. Such a formal expansion was reconsidered in a recent paper\, \cite{Bender:2018pbv}, where it was suggested that it could be the key for implementing a systematic renormalization program for non-hermitian (in particular PT-symmetric) quantum field theories. 

In these latter years great interest has grown around PT-symmetric theories\,\cite{Bender:1998ke,Bender:1998gh,Bender:2007nj}, and impressive successful applications to many different areas of quantum mechanics have been realized\,\cite{Guo:2009yqd,Longhi,Ruter,Lin,Ayache,Schindler,Liertzer,Feng,Peng,Brandstetter,Hodaei,Wong,Konotop,Ge,El-Ganainy,Kawabata,Miri,Wu,Li,Huanan}. The extension to QFTs has also been considered\,\cite{Shalaby:2009re,Shalaby:2009xda,Bender:2012ea,Bender:2013qp,Bender:2014bna,Alexandre:2018uol,Alexandre:2019jdb,Fring:2019hue}.
However, while in the case of PT-symmetric quantum mechanics the boundary conditions on the Schrodinger equation are imposed in complex Stokes sectors, the search for the Stokes wedges for the integration over infinitely many complex field variables is an insurmountable task. 
The great advancement that the  expansion in the parameter $\epsilon$ seems to bring is that it allows to perform the functional integration along the real-$\phi$ axis (rather than in the complex-$\phi$ domain), as the functional integral that defines the theory (see Eqs.\,(\ref{lagrnew}), (\ref{Ln}) and (\ref{Zpartition}) below) converges term-by-term in powers of $\epsilon$\,\cite{Bender:2018pbv}. 

PT-symmetric theories, however, have their own peculiar features, so that, in order to make progress in understanding the expansion itself, it is useful to disentangle its intrinsic properties from those that are specific to this kind of theories\footnote{The extension of this analysis to PT-symmetric theories is left for future work\,\cite{future}.}. To this end, it is useful to deepen the analysis by studying first the application of this expansion to ordinary theories.
Earlier attempts in this direction  where made in \cite{Bender:1988ux}, however a more complete analysis, and in particular the implementation of a systematic renormalization program within the framework of such an expansion is still missing.
An ambitious program would be to study the renormalization of ordinary theories with the same degree of systematicity realized in the context of the weak-coupling expansion. Clearly this is an enormous task, and the aim of the present work is more modest.

Our goal is to investigate some aspects related to the renormalization of theories for which the Green's functions are calculated within the framework of such an expansion. In order to better appreciate the peculiar properties of this approach, in this paper we consider the simplest possible framework, by studying the case of an ordinary scalar theory with interaction term 
$\phi^{2(1+\epsilon)}$. For $\epsilon=0$ the non-interacting linear theory is obtained, and this is why we said that the expansion is organized in terms of a parameter (the parameter $\epsilon$) that  measures the distance between the interacting and the corresponding non-interacting theory. Note that for $\epsilon=1$ the physically relevant $\phi^4$ theory is recovered.

The starting point of the procedure introduced in\,\cite{Bender:1987dn,Bender:1988rq}, and later reconsidered in\,\cite{Bender:1988ux,Bender:1988ig}, consists in expanding the interaction term $\phi^{2(1+\epsilon)}$ in powers of $\epsilon$,
\begin{equation} \label{interacting}
\phi^{2(1+\epsilon)} \,=\, \phi^2 \sum_{k=0}^{\infty} \frac{\epsilon^k}{k!}\left[ \log\left(\phi^2\right) \right]^k\,.
\end{equation}
For our scopes it is important to note that each order in $\epsilon$ contains a power of the logarithm of the scalar field, and that such logs are treated by considering an adaptation of the replica-trick \cite{Parisi} that is suitable to the expansion of the theory's Green's functions. This is obtained with the help of the formal identity:
\begin{equation}                                        
\phi^2 \log^k (\phi^2) = \lim_{N \rightarrow 1} \frac{d^k}{dN^k} \phi^{2N} \,.
\end{equation}
The whole procedure is built around this identity\,\cite{Bender:1987dn}, and aims at finding approximations to the Green's functions $\mathcal{G}_n$ in terms of an expansion in powers of $\epsilon$. 
The actual implementation of this program, however, presents several delicate aspects that will be discussed in detail in the paper. 

In previous works the calculation of the Green's functions
$\mathcal{G}_n$ for any $n$ was performed at first order in $\epsilon$, while at order $\epsilon^2$ only  $\mathcal{G}_2$ and  $\mathcal{G}_4$ were calculated \cite{Bender:1988ig}.
One of our goals is to push the calculation of the generic $\mathcal{G}_n$  beyond the leading order. The general framework and the notation will be established by repeating first the calculation of the $\mathcal{O}(\epsilon)$, for which we reobtain the results of \cite{Bender:1987dn}, and later the full $\mathcal{O}(\epsilon^2)$ contribution will be obtained.

To this end we introduce a systematic diagrammatic expansion for the $\mathcal{G}_n$'s, and successively define ``effective vertices" that allow to express the final results in an effective and simple way, even at higher orders in $\epsilon$.
However we will show that more and more delicate issues appear when higher and higher order contributions are considered, and we will show that this is due to a peculiar ultraviolet behavior of the $\mathcal{G}_n$'s that is related to the properties of this expansion.
We will be confronted with the appearance of series for which the convergence is not always guaranteed, and we will need appropriate procedures to regularize these series.

The ultraviolet behaviour of the Green's functions will be studied  with the introduction of a physical momentum cut-off $\Lambda$, and the different contributions to the $\mathcal{G}_n$ will be classified with respect to their dependence on $\Lambda$. 
As an outcome of this analysis, we will find an unexpected and in a sense astonishing behaviour with $\Lambda$. More specifically, we will see that (barring unconventional ways to renormalize the theory that at present are not supported and/or justified and on which we will comment in due time) as long as we confine ourselves to a finite order in $\epsilon$, for space-time dimensions $d>2$, the $\mathcal{G}_n$ with $n>2$ vanish with $\Lambda\to\infty$, thus suggesting that at any finite order the theory turns out to be non-interacting.

Let us apply these results to a $\phi^4$ theory. Although for $d=4$ a final word on the interacting character of the theory is still missing, we certainly know that in $d=3$ dimensions the theory is interacting. In principle we cannot exclude that the interacting character of the theory could be recovered with the help of a non-conventional renormalization, however it seems to us that these results signal a weakness of the expansion when we limit ourselves to consider finite order approximations to the Green's functions.

Motivated by these observations, we looked for better approximations to the $\mathcal{G}_n$'s exploring the possibility of resumming classes of diagrams from all orders in $\epsilon$. In particular, resorting to the classification in terms of the physical cut-off $\Lambda$ referred above, we started by picking up the leading contributions (in terms of their dependence on $\Lambda$) from each order in $\epsilon$.
Interestingly we found that this resummation provides a bridge between the expansion in $\epsilon$ and the leading order of the usual weak-coupling expansion in the coupling constant $g$. 
Even more interestingly we found that, extending this criterion also to the next-to-leading terms, it is possible to recover the results of the weak-coupling expansion also at order $g^2$. Along the same lines, we have also indicated the resummations that should lead to higher order results in $g$.

Besides being of great importance on their own, these results shade some light on the ``strange" ultraviolet behaviour (no interaction) of the Green's functions in the framework of this expansion: finite orders in $\epsilon$ seem to give a ``too poor" truncation for the interacting term (\ref{interacting}) in the lagrangian.

The rest of the paper is organized as follows. 
Section 2 is partially devoted to reviewing some of the results presented in\,\cite{Bender:1987dn,Bender:1988rq}. 
After introducing the strategy to realize the expansion in powers of $\epsilon$, we will define a set of Feynman rules useful to give a diagrammatic representation for the Green's functions. We will then consider the $\mathcal{O}(\epsilon)$ contributions to the $\mathcal{G}_n$'s with the help of the corresponding diagrammatic representation, and analyse their behaviour in terms of the physical cut-off $\Lambda$.
In Section 3 we will move to the $\mathcal{O}(\epsilon^2)$, and classify the different contributions with respect to their $\Lambda$-dependence.
In Section 4 we will consider higher order contributions to the Green's functions and again classify them according to their ultraviolet behaviour.
Section 5 is devoted to the resummation of the leading diagrams, while in Section 6 we will push forward the resummation to the order $g^2$.
Section 7 is for the conclusions.

\section{Expansion in $\epsilon$ and leading order results}
\subsection{Definition of the theory and Feynman rules}
\label{feynrules}
The original idea presented in\,\cite{Bender:1987dn,Bender:1988rq} for solving a self-interacting scalar theory with interaction of the kind $\phi^{2p}$, with integer $p$, consists in considering the (Euclidean) Lagrangian density (in  \textit{d}-dimensions with $\epsilon > 0$)
\begin{equation} \label{eq:L}
\mathcal{L} = \frac{1}{2}(\nabla\phi)^2+\frac{1}{2}m^2\phi^2+\frac{1}{2}g\mu^2\phi^2\left(\mu^{2-d}\phi^2\right)^{\epsilon} \,,
\end{equation}
and putting forward a scheme to calculate the Green's functions based on an expansion in the parameter $\epsilon$, considered as small. In this sense, the parameter $\epsilon$ measures the departure from the linear theory. Since the interaction term is defined as $\phi^2\,(\phi^2)^{\epsilon}$, varying $\epsilon$ from zero to $p-1$ the Lagrangian (\ref{eq:L}) ``interpolates" between a free Lagrangian and a Lagrangian with even interaction $\phi^{2p}$.
Note that this analytic continuation follows a path that preserves the parity symmetry, so that if for instance we consider  $\epsilon=\frac{1}{2}$, the interaction $\phi^2\,(\phi^2)^{\epsilon}$ gives rise to $\phi^2 \lvert\phi\rvert$ rather than $\phi^3$ (that would have been obtained from $\phi^2\,(\phi)^{2\epsilon}$). So doing, the different theories for different values of $\epsilon$ are uniquely and unambiguously defined.

Clearly, once the Green's functions are calculated to a given order in $\epsilon$, the final goal is to send $\epsilon\to p-1$ and we have to assume that the limits $\epsilon \to p-1$ provides sensible approximations to the desired Green's functions. Some arguments with respect to the convergence of these series are given in Ref.\,\cite{Bender:1987dn,Bender:1988rq}.

The first step to set up the expansion in the parameter $\epsilon$ consists in writing the term $(\mu^{2-d}\phi^2)^\epsilon$ as $e^{ \epsilon\,\log\left(\mu^{2-d}\phi^2\right) }$ and expanding the exponential in powers of $\epsilon$, so that (\ref{eq:L}) is replaced by the highly non-polynomial lagrangian:
\begin{equation} \label{eq:Lexp}
\mathcal{L} = \frac{1}{2}(\nabla\phi)^2 + \frac{1}{2}m^2\phi^2 + \frac{1}{2}g\mu^2\phi^2 \sum_{k=0}^{\infty} \frac{\epsilon^k}{k!}\left[ \log\left(\mu^{2-d}\phi^2\right) \right]^k \,.
\end{equation}
Successively, including the $\epsilon^0$ term in the ``free lagrangian" $\mathcal L_0$
\begin{equation}
\label{L0} \mathcal{L}_0 \equiv \frac{1}{2}(\nabla\phi)^2 + \frac{1}{2}\left(m^2+g\mu^2\right)\phi^2 \,,
\end{equation}
the lagrangian $\mathcal L$ is written as
\begin{equation}\label{lagrnew}
\mathcal{L}= \mathcal{L}_0+ \mathcal{L}_{int}\equiv \mathcal L_0+\sum_{k=1}^\infty \mathcal L_k\,,
\end{equation}
where 
\begin{equation}\label{Ln}
\mathcal{L}_k=  \frac{1}{2}g\mu^2\phi^2  \frac{\epsilon^k}{k!}\left[ \log\left(\mu^{2-d}\phi^2\right) \right]^k \,.
\end{equation}

It is worth to stress that from now we will consider the theory as defined by the lagrangian (\ref{lagrnew}) rather than by (\ref{eq:L}). This is however not the last step of the strategy. As we will see other steps are needed to give a precise meaning to the expansion of the Green's functions in powers of $\epsilon$.

The partition function $Z$ and the $n$-points Green's functions are given as usual\,\cite{Peskin}:
\begin{equation} \label{Zpartition}
Z = \int \mathcal D\phi \, e^{-\int d^du \, \mathcal{L}}\,,
\end{equation}
\begin{equation} \label{Gn}
\mathcal{G}_n(x_1,\ldots,x_n) = \frac{1}{Z}\int \mathcal D\phi \, \phi(x_1)\ldots\phi(x_n)\, e^{-\int d^du \, \mathcal{L}}\,.
\end{equation}
where from now on we consider even values of $n$ for parity reasons.

As mentioned above, the aim of the present section is to provide the building blocks for the calculation of the connected Green's functions to any given order in $\epsilon$. 
To this end, we consider $e^{-S_{int}}=e^{-\int d^du \mathcal{L}_{int}}$ as a perturbation term, then expand it in Eqs.\,(\ref{Zpartition}) and (\ref{Gn}) and collect different powers of $\epsilon$, thus getting:
\begin{align}\label{interactioneps}
&e^{-S_{int}} = \quad 1 \, - {\int d^du\,\mathcal{L}_1} \,\, + \,\, \Bigg[-\int d^du\,\mathcal{L}_2 \,{+\,\frac{1}{2}\int d^du\,\mathcal{L}_1 \int d^dw\,\mathcal{L}_1}\Bigg] \nonumber \\
& + \Bigg[-\int d^du\,\mathcal{L}_3 \,+\,\int d^du\,\mathcal{L}_1 \int d^dw\,\mathcal{L}_2 \,{-\,\frac{1}{6}\int d^du\,\mathcal{L}_1\int d^dw\,\mathcal{L}_1\int d^dz\,\mathcal{L}_1}\Bigg] + \, \ldots 
\end{align}
Following then the usual procedure, for the calculation of the Green's functions (\ref{Gn}) we invert the order of the space and functional integrals.

At this point, however, we note that all the ``interaction terms" $\mathcal{L}_k$ are of the kind $\phi^2\log^k (\mu^{2-d}\phi^2)$. In\,\cite{Bender:2018pbv}, and previously in\,\cite{Bender:1987dn,Bender:1988rq} in an implicit manner, it was suggested that each of these interactions can be rewritten in terms of powers of the field with the help of the formal identity
\begin{equation} \label{replica}                                       
\mu^{2-d}\phi^2 \log^k (\mu^{2-d}\phi^2) = \lim_{N \rightarrow 1} \frac{d^k}{dN^k} \mu^{(2-d)N}\phi^{2N} \,,
\end{equation}
that is an adaptation of the replica trick introduced in\,\cite{Parisi}. The actual calculation of the Green's functions follows a procedure based on this formal replacement, that consists of several steps.

First of all the operators $\lim_{N_1\to 1}\frac{d^{k_1}}{dN_1^{k_1}} \,,\lim_{N_2\to 1}\frac{d^{k_2}}{dN_2^{k_2}}\,,\dots $ have to be taken on the left of the functional integrals (actually, we have one of these operators for each of the $\mathcal{L}_k$ appearing in the products in (\ref{interactioneps})). Then, at a first stage the variables $N_i$ have to be considered as integers numbers, so that the calculation of the functional integrals is traced back to an application of the Wick's theorem. 
Subsequently, the resulting functions have to be extended to real values of the $N_i$'s, and then the derivatives with respect to $N_i$'s and the limits $N_i\to1$ can be taken.

As we are interested in the calculation of the connected Green's functions, we can replace in (\ref{Gn}) the partition function $Z$ with its zeroth-order (in $\epsilon$) expression $Z_0$, and calculate the functional integrals involved in the described procedure considering only connected diagrams. From now on, we will refer to $\mathcal{G}_n$ as to the connected Green's functions.

This whole procedure can be resumed by introducing a set of Feynman rules from which we obtain the diagrams corresponding to the desired Green's functions. To this end we first introduce the ``auxiliary" interaction lagrangian:
\begin{equation}\label{Laux}
\mathcal L^{aux}_{int}= \sum_{k=1}^\infty \lambda_{k}(N)\, \phi^{2N} \,,
\end{equation}
with
\begin{equation} \label{lambdak}
\lambda_{k}(N)\equiv\frac 12 g \mu^{2+(2-d)(N-1)} \frac{\epsilon^k}{k!} \,,
\end{equation}
that gives rise to infinitely many $2N$-legs vertices, each related to the $\mathcal{O}(\epsilon^k)$ ``coupling" $\lambda_k(N)$. This brings us to the definition of the vertices
\begin{equation} \label{auxvertex}
{\small
	\begin{tikzpicture}[baseline=-3]
	\begin{feynman}
	\vertex[blob, label={center:$\lambda_{k}$},pattern color=white] (m) at (0, 0) {};
	\vertex (a) at (-0.9,0.9) {\small $2$};
	\vertex (b) at (-0.9,-0.9) {$1$};
	\vertex (c) at (1.24421,0) {$2N$};
	\vertex (l1) at (0.5,0.35) {};
	\vertex (l2) at (-0.2,0.55) {};
	\vertex (d1) at (0.495294,-0.308824) {};
	\vertex (d2) at (0.824706,0.308824) {};
	\vertex (d3) at (-0.700391,0.226862) {};
	\vertex (d4) at (-0.370979,0.844509) {};
	\vertex (d5) at (-0.700391,-0.226862) {};
	\vertex (d6) at (-0.370979,-0.844509) {};
	\diagram* {
		(a) -- (m) -- (b);
		(m) -- (c);
		(l1) -- [thick,dotted,out=-280, in=-305](l2);
	};
	\draw (d1) to (d2);
	\draw (d3) to (d4);
	\draw (d5) to (d6);
	\end{feynman}
	\end{tikzpicture}
}
\equiv-\lambda_{k}(N)\,.
\end{equation}

With the definitions given above the $\mathcal{O}(\epsilon^m)$ contribution to the $n$-points connected Green's functions is obtained by considering first the connected diagrams that can be drawn to that given order in $\epsilon$ with the vertices (\ref{auxvertex}) (that amounts to perform the usual Wick's contractions with the appropriate combinatorial factors), and then, before integrating in the coordinates of the vertices, applying to each of these diagrams the operator  
\begin{equation}\label{limit}
\lim_{N \to 1} \frac{d^k}{dN^k}
\end{equation}
for each $\lambda_{k}$-vertex that appears.

Naturally we should ask ourselves if with these prescriptions we recover the correct terms of the expansion in $\epsilon$ of the Green's functions. Arguments in this sense are given in \cite{Bender:1987dn,Bender:1988rq}, and for those cases for which the check could be performed the coincidence was explicitly shown.

In the next subsection we consider the connected Green's functions to $\mathcal{O}(\epsilon)$ and draw the corresponding Feynman diagrams according to the Feynman rules given above.

\subsection{Green's functions and Feynman diagrams up to $\mathcal O (\epsilon)$}
The systematic expansion of the Green's functions in the non-linearity parameter $\epsilon$ starts with the contributions of order $\epsilon^0$ and $\epsilon^1$, that were already calculated in\,\cite{Bender:1987dn,Bender:1988rq}. The main goal of the present subsection is to derive these results in terms of Feynman diagrams obtained with the rules introduced above. We will also express the $\mathcal{O}(\epsilon)$ contribution to the Green's functions in terms of $\mathcal{O}(\epsilon)$ ``effective vertices", to be later defined. The usefulness of these definitions will be more apparent when moving to higher order diagrams.

Let us begin by noting that, since the interaction Lagrangian contains terms that are at least $\mathcal{O} (\epsilon)$, the two-points function $\mathcal{G}_2(x_1,x_2)$ is the only connected Green's function that has an order $\epsilon^0$ contribution, that we denote with $\mathcal{G}^{(0)}_2(x_1,x_2)$. For the free lagrangian given in Eq.\,(\ref{L0}), this is nothing but the Feynman propagator (from now on $M^2=m^2+g \mu^2$)
\begin{equation}\label{Delta}
\mathcal{G}^{(0)}_2(x_1,x_2)=\Delta(x_1-x_2)=\int \frac{d^dp}{(2\pi)^d} \frac{1}{p^2+M^2} e^{-ip(x_1-x_2)}\,.
\end{equation}

Moving to $\mathcal O (\epsilon)$, we see that at this order we need to consider only the vertex $\lambda_1$, so that the $\mathcal O (\epsilon)$ contribution to the generic $n$-points Green's function is:
\begin{equation} \label{Gn0}
	\mathcal{G}^{(1)}_n(x_1,\dots,x_n) =\int d^du
	\lim_{N \to 1} \frac{d}{dN}\, \left\{ -\lambda_{1}(N) \, \frac{1}{Z_0} \left( \int \mathcal D\phi \,e^{-S_0} \phi(x_1)\ldots\phi(x_n) \phi(u)^{2N} \right)_C \right\}
\end{equation}
where $C$ in the path integral is for connected and where, as explained before, the derivative and the limit with respect to $N$ will be properly defined when the analytic extension to real values of $N$ near $N=1$ will be considered.
Focusing now on the path integral in the curly brackets, we note that for integers $N<\frac{n}{2}$ a vanishing contribution to $\mathcal{G}^{(1)}_n$ is obtained (there are no connected diagrams in this case), while for $N\geq \frac{n}{2}$ we get
\begin{align}\label{Gn1}
&- \frac{\lambda_{1}(N)}{Z_0} \left( \int \mathcal D\phi \,e^{-S_0} \phi(x_1)\ldots\phi(x_n) \phi(u)^{2N} \right)_{C} ={
	\begin{tikzpicture}[baseline=(b)]
	\begin{feynman}[inline=(b)]
	\vertex (b)[blob,label={center:{\small $\lambda_1$}},pattern color=white] at (0, 0) {};
	\vertex (a) at (-1.2, -0.64) {$x_1$};
	\vertex (d) at ( 1.2, -0.64) {$x_n$};
	\vertex (c) at (-0.357948,-1.07532) {$x_2$};
	\vertex (e) at ( 0.357948,-1.07532) {};
	\vertex (l1) at (0.4,0.3) {};
	\vertex (l2) at (-0.1,0.45) {};
	\vertex (p1) at (-0.16,-0.7) {};
	\vertex (p2) at (0.35,-0.5) {};
	\diagram* {
		(a) -- (b) -- [out=35, in=0, loop, min distance=1.2cm, edge label={\footnotesize $N-\frac{n}{2}$} ] b -- [out=135, in=100, loop, min distance=1.2cm, edge label={\footnotesize $2$}] b -- [out=185, in=150, loop, min distance=1.2cm,edge label={\footnotesize $1$}] b -- (d);
		(c) -- (b);
		(l1) -- [thick,dotted,out=-280, in=-315](l2);
		(p1) -- [thick,dotted,out=0, in=20](p2);
	};
	\end{feynman}
	\end{tikzpicture}
}
 \nonumber \\
&= -\lambda_{1}(N)\left[\Delta(0)\right]^{N-\frac{n}{2}} C_n(N) \prod_{i=1}^{n}\Delta(x_i-u)\,,
\end{align}
where $C_n(N)$ is the combinatorial factor coming from the contractions:
\begin{equation} \label{CnN0}
C_n(N)= \left[2N(2N-1)\dots(2N-n+1)\right]\,(2N-n-1)!! 
\end{equation}
with the first $n$ factors related to the contractions of $n$ of the $2N$ fields of the vertex with $n$ fields in different spacetime points, while the double factorial comes from the $N-\frac{n}{2}$ self-loops obtained from the leftover fields. Putting together all the odd factors, we can rewrite (\ref{CnN0}) as:
\begin{equation}\label{CnN1}
C_n(N)=2^{\frac{n}{2}}\,\left[N(N-1)\dots(N-\frac{n}{2}+1)\right](2N-1)!!=2^{\frac{n}{2}}\,(N)_{\frac{n}{2}}\,(2N-1)!!
\end{equation}
where we used the symbol $(x)_m$ to denote the falling factorial, defined as
\begin{equation}\label{fallfact}
(x)_m=x(x-1)\dots(x-m+1) \qquad \forall x \in \mathbb{R}, m \in \mathbb{N^+}
\end{equation}  
for positive integer values of $m$ and extended to $(x)_0=1$ for the case $m=0$\,.
From (\ref{fallfact}) we see that $C_n(N)$ vanishes for integers $N<\frac{n}{2}$. For this reason, the expression in the last member of (\ref{Gn1}) gives the correct result for the path integral also when $N<\frac{n}{2}$.

This observation is crucial to the procedure. In a sense, we aim at a ``definition" of the path integral in (\ref{Gn1}) even for real values of $N$, and in particular around $N=1$, where the derivative with respect to $N$ has to be calculated.
This is achieved by considering the analytic extension of the last member of (\ref{Gn1}), more precisely of the factor $-\lambda_{1}(N)\left[\Delta(0)\right]^{N-\frac{n}{2}} C_n(N)$\,. It is worth to stress that, while this procedure provides the analytically extended function of real variable $N$ even for $N<\frac{n}{2}$, the latter keeps memory of the fact that the original path integral in (\ref{Gn1}) is actually performed for integer values of $N$ with $N\geq\frac{n}{2}$, and this is why the factor $\left[\Delta(0)\right]^{N-\frac{n}{2}}$ appears.

Once this analytic extension is performed, the $\mathcal{O}(\epsilon)$ contribution to the $n$-point Green's function can be expressed in terms of an $n$-legs ``effective vertex" defined as 
\begin{equation} \label{Pi_n^1def}
\Pi^{(1)}_{n}\,=\,
\begin{tikzpicture}[baseline=-3]
\begin{feynman}
\vertex[draw, circle, minimum size=0.85cm,very thick] (m) at (0, 0) {1};
\vertex (a) at (-0.9,0.9) {$2$};
\vertex (b) at (-0.9,-0.9) {$1$};
\vertex (c) at (1.24421,0) {$n$};
\vertex (l1) at (0.5,0.35) {};
\vertex (l2) at (-0.2,0.55) {};
\vertex (d1) at (0.595294,-0.308824) {};
\vertex (d2) at (0.924706,0.308824) {};
\vertex (d3) at (-0.700391,0.226862) {};
\vertex (d4) at (-0.370979,0.844509) {};
\vertex (d5) at (-0.700391,-0.226862) {};
\vertex (d6) at (-0.370979,-0.844509) {};
\diagram* {
	(a) -- (m) -- (b);
	(m) -- (c);
	(l1) -- [thick,dotted,out=-280, in=-305](l2);
};
\draw (d1) to (d2);
\draw (d3) to (d4);
\draw (d5) to (d6);
\end{feynman}
\end{tikzpicture} \,\equiv\,
\lim_{N \rightarrow 1} \frac{d}{dN} \left\{-\lambda_{1}(N)\left[\Delta(0)\right]^{N-\frac{n}{2}} C_n(N)\right\}
\end{equation}
where the superscript $(1)$ indicates that $\Pi^{(1)}_{n}$ is $\mathcal{O}(\epsilon)$. 
With this definition:
\begin{equation} \label{Gn1Pi}
\mathcal{G}^{(1)}_n(x_1,\dots,x_n) = \Pi^{(1)}_{n} \int d^du \prod_{i=1}^{n}\Delta(x_i-u) \,.
\end{equation}

We are then left with the calculation of the $\Pi^{(1)}_{n}$'s. Replacing in (\ref{Pi_n^1def}) the combinatorial factor $C_n(N)$ in (\ref{CnN1}) and the coupling $\lambda_{1}(N)$ in (\ref{lambdak}), we obtain:
\begin{align} \label{Pi_n^1def2}
\Pi^{(1)}_{n} &= \lim_{N \to 1} \frac{d}{dN} \left[-\frac{\epsilon}{2}\,g\,\mu^{2+(2-d)(N-1)}\,\left[\Delta(0)\right]^{N-\frac{n}{2}}\, 2^{\frac{n}{2}} (N)_{\frac{n}{2}} (2N-1)!! \right] \nonumber\\
&=-\epsilon\,g\,\mu^2\left[\frac{2}{\Delta(0)}\right]^{\frac{n}{2}-1} \lim_{N \rightarrow 1} \frac{d}{dN}\, f_n(N)
\end{align}
where
\begin{equation} \label{fnNint}
f_n(N) = (N)_{\frac{n}{2}} (2N-1)!!	\left[\mu^{2-d}\Delta(0)\right]^{N-1} \,,
\end{equation}
and is the function that need to be analytically extended.
The falling factorial $(N)_{\frac{n}{2}}$ is just a polynomial in $N$ and as such it is an entire function of $N$. On the contrary, the semifactorial $(2N-1)!!$ is defined in terms of an integer $N$. With the help of the relation $(2N-1)!!=2^{N-1}\frac{\Gamma(N+\frac{1}{2})}{\Gamma(\frac{3}{2})}\,$\,\cite{Abra}
 valid for integers $N$, we can finally define the analytic extension of $f_n(N)$ to real values of its variable as 
\begin{equation} \label{fnNcont}
f_n(x)=(x)_{\frac{n}{2}} \frac{\Gamma(x+\frac{1}{2})}{\Gamma(\frac{3}{2})} \left[2 \mu^{2-d}\Delta(0)\right]^{x-1}\,,
\end{equation}  
that is analytic around $x=1$. For simplicity in the following we continue to indicate the real variable as $N$. We are now in the position to take the derivative and the limit. 

Due to the presence of the factor $(N-1)$ in the falling factorial, the derivative of $f_n(N)$ with respect to $N$ has a simple behaviour in the limit $N\to1$ for the cases $n=4,6,\dots$, while the case $n=2$ has to be treated separately. Let us begin with the latter.\\

\textit{\underline{Case $n=2$\,.}}\quad We have 
\begin{equation} \label{f2N}
f_2(N)= N \frac{\Gamma\left(N+\frac{1}{2}\right)}{\Gamma\left(\frac{3}{2}\right)}	\left[2\mu^{2-d}\Delta(0)\right]^{N-1}\,,
\end{equation}
so that
\begin{equation} \label{K}
\lim_{N \rightarrow 1} \frac{d}{dN} f_2(N) = \log\left[2\mu^{2-d}\Delta(0)\right]+1+\frac{\Gamma'(\frac{3}{2})}{\Gamma(\frac{3}{2})} \equiv K\,,
\end{equation}
and the $2$-points effective vertex becomes
\begin{equation} \label{Pi_2^1}
\Pi^{(1)}_{2}\,=\,
\begin{tikzpicture}[baseline=-3]
\begin{feynman}
\vertex[draw, circle, minimum size=0.85cm,very thick] (m) at (0, 0) {1};
\vertex (a) at (-1.24421,0) {};
\vertex (b) at (1.24421,0) {};
\vertex (d1) at (0.595294,-0.308824) {};
\vertex (d2) at (0.924706,0.308824) {};
\vertex (d3) at (-0.924706,-0.308824) {};
\vertex (d4) at (-0.595294,0.308824) {};
\diagram* {
	(a) -- (m) -- (b);
};
\draw (d1) to (d2);
\draw (d3) to (d4);
\end{feynman}
\end{tikzpicture} = -\,\epsilon\,g\,\mu^2\, K \,.
\end{equation}

\textit{\underline{Case $n>2$\,.}}\quad As stressed above, in these cases the falling factorial contains the factor ${(N-1)}$, so that in the limit $N\to1$ only the term that comes from the derivative of $(N-1)$ is non-vanishing. Using the symbolic expression $x^{(m)}$ for the rising factorial $x^{(m)}=x(x+1)\dots(x+m-1)$, and noting that $(x)_m=(-1)^m\,(-x)^{(m)}$, we can rewrite the function $f_n$ as
\begin{equation} \label{fnN}
f_n(N)= N (N-1) (-1)^{\frac{n}{2}-2} (2-N)^{(\frac{n}{2}-2)}  \frac{\Gamma\left(N+\frac{1}{2}\right)}{\Gamma\left(\frac{3}{2}\right)}	\left[2\mu^{2-d}\Delta(0)\right]^{N-1} \,.
\end{equation}
Being $1^{(m)}=m!$, we easily obtain:
\begin{equation}
\lim_{N \rightarrow 1} \frac{d}{dN} f_n(N) = (-1)^{\frac{n}{2}-2} \left(\frac{n}{2}-2\right)!
\end{equation}
that in turn gives for the $n$-legs effective vertex $(n>2)$:
\begin{equation} \label{Pi_n^1}
\Pi^{(1)}_{n}\,=\,
\begin{tikzpicture}[baseline=(m)]
\begin{feynman}[inline=(m)]
\vertex[draw, circle, minimum size=0.85cm,very thick] (m) at (0, 0) {1};
\vertex (a) at (-0.9,0.9) {$2$};
\vertex (b) at (-0.9,-0.9) {$1$};
\vertex (c) at (1.24421,0) {$n$};
\vertex (l1) at (0.5,0.35) {};
\vertex (l2) at (-0.2,0.55) {};
\vertex (d1) at (0.595294,-0.308824) {};
\vertex (d2) at (0.924706,0.308824) {};
\vertex (d3) at (-0.700391,0.226862) {};
\vertex (d4) at (-0.370979,0.844509) {};
\vertex (d5) at (-0.700391,-0.226862) {};
\vertex (d6) at (-0.370979,-0.844509) {};
\diagram* {
	(a) -- (m) -- (b);
	(m) -- (c);
	(l1) -- [thick,dotted,out=-280, in=-305](l2);
};
\draw (d1) to (d2);
\draw (d3) to (d4);
\draw (d5) to (d6);
\end{feynman}
\end{tikzpicture} 
= (-1)^{\frac{n}{2}-1} \left(\frac{n}{2}-2\right)!\,\, \epsilon\,g\,\mu^2\left[\frac{2}{\Delta(0)}\right]^{\frac{n}{2}-1} \,.
\end{equation}
\vskip10pt
We can now write the Green's functions up to $\mathcal{O}(\epsilon)$ in a convenient form.
\paragraph{Two-points Green's function $\mathcal{G}_2$.}
Up to order $\epsilon$ we have
\begin{equation} \label{G2x}
\mathcal{G}_2(x_1,x_2)= \Delta(x_1-x_2) + \Pi^{(1)}_{2} \int d^du \, \Delta(x_1-u)\Delta(x_2-u) \,.
\end{equation}
Going to Fourier space, and using (\ref{Pi_2^1}) for $\Pi^{(1)}_{2}$, for $\widetilde{{G}}_2(p)$ we obtain
\begin{align}\label{G2fourier}
\widetilde{{G}}_2(p) \equiv&
\begin{tikzpicture}[baseline=-3]
\begin{feynman}
\vertex[large,blob] (m) at (0, 0) {};
\vertex (a) at (-1.5,0) {};
\vertex (b) at ( 1.5,0) {};
\diagram* {
	(a) -- [fermion](m) -- [fermion](b),
};
\end{feynman}
\end{tikzpicture}
= 
\begin{tikzpicture}[baseline=-3]
\begin{feynman}
\vertex (b) at (0, 0) {};
\vertex (a) at (-1.13333, 0) {};
\vertex (d) at ( 1.13333, 0) {};
\diagram* {
	(a) -- [fermion](d);
};
\end{feynman} 
\end{tikzpicture}
+
\begin{tikzpicture}[baseline=-3]
\begin{feynman}
\vertex[draw, circle, minimum size=0.85cm,very thick] (m) at (0, 0) {1};
\vertex (a) at (-1.5,0) {};
\vertex (b) at ( 1.5,0) {};
\diagram* {
	(a) -- [fermion](m) -- [fermion](b),
};
\end{feynman}
\end{tikzpicture}
\nonumber \\
= &\frac{1}{p^2+M^2} - \frac{1}{p^2+M^2}\, \epsilon\,g\, \mu^2 \,K\,\frac{1}{p^2+M^2}\,.
\end{align}
Eq.\,(\ref{G2fourier}) has a resemblance with the corresponding $2$-point Green's function of the ordinary weak-coupling expansion  at $\mathcal{O}(g)$. The $\mathcal{O}(\epsilon^0)$ term is the free propagator (as the corresponding $\mathcal{O}(g^0)$ term in the weak-coupling expansion), while the $\mathcal{O}(\epsilon)$ term contains the radiative corrections, here carried by the residual loop factor $K$ of the effective vertex $\Pi^{(1)}_{2}$.

Noting that the bubble diagram in Eq.\,(\ref{G2fourier}) is the lowest 1PI self-energy diagram within the $\epsilon$-expansion, we can get the first approximation to the self-energy by resumming the following geometric series:
\begin{equation}
\begin{tikzpicture}[baseline=-3]
\begin{feynman}
\vertex[large,blob] (m) at (0, 0) {};
\vertex (a) at (-1.5,0) {};
\vertex (b) at ( 1.5,0) {};
\diagram* {
	(a) -- [fermion](m) -- [fermion](b),
};
\end{feynman}
\end{tikzpicture}
= 
\begin{tikzpicture}[baseline=-3]
\begin{feynman}
\vertex (b) at (0, 0) {};
\vertex (a) at (-1.1, 0) {};
\vertex (d) at ( 1.1, 0) {};
\diagram* {
	(a) -- [fermion](d);
};
\end{feynman} 
\end{tikzpicture}
+
\begin{tikzpicture}[baseline=-3]
\begin{feynman}
\vertex[draw, circle, minimum size=0.85cm,very thick] (m) at (0, 0) {1};
\vertex (a) at (-1.5,0) {};
\vertex (b) at ( 1.5,0) {};
\diagram* {
	(a) -- [fermion](m) -- [fermion](b),
};
\end{feynman}
\end{tikzpicture}
+
\begin{tikzpicture}[baseline=-3]
\begin{feynman}
\vertex[draw, circle, minimum size=0.85cm,very thick] (m) at (-0.75, 0) {1};
\vertex[draw, circle, minimum size=0.85cm,very thick] (n) at (0.75, 0) {1};
\vertex (a) at (-2,0) {};
\vertex (b) at ( 2,0) {};
\diagram* {
	(a) -- [fermion](m) -- [fermion](n) -- [fermion](b),
};
\end{feynman}
\end{tikzpicture}
+ \ldots
\end{equation}
getting
\begin{equation}\label{G2p}
\widetilde{G}_2(p) = \frac{1}{p^2+M^2 + \epsilon g\mu^2\,K}
\end{equation}
from which the radiatively corrected mass turns out to be:
\begin{equation} \label{mReps}
m_{R}^2=M^2 + \epsilon g\mu^2 \left\{\log\left[2\mu^{2-d}\Delta(0)\right]+1+\frac{\Gamma'(\frac{3}{2})}{\Gamma(\frac{3}{2})}\right\}\,.
\end{equation}
Few comments are in order. We know that the loop integral $\Delta(0)$, which is nothing but
\begin{equation}\label{delta0}
\Delta(0)=\int \frac{d^dp}{(2\pi)^d} \frac{1}{p^2+M^2} \,,
\end{equation}
is a divergent quantity when $d \geq 2$. In these cases, we need to regularize this divergence. In this respect we note that, with the exception of the Theory of Everything, any quantum field theory is an effective theory. From the physical point of view this means that we can encode our ignorance on physics above a given scale $\Lambda$ by introducing this scale in the theory as a physical momentum cut-off when summing the quantum fluctuations. Having this in mind, we regularize $\Delta(0)$ with the help of a momentum cut-off $\Lambda$. 

To be more specific, it is worth at this point to focus on the $d=4$ case, where
\begin{equation} \label{delta0reg}
\Delta(0)=\frac{1}{16 \pi^2}\left(\Lambda^2-M^2\log \frac{\Lambda^2}{M^2}\right)+\mathcal{O}(\Lambda^{-2}) \,.
\end{equation}
Replacing this result in (\ref{mReps}), we obtain
\begin{equation}\label{mR1epsreg}
m_{R}^2=M^2 + \epsilon g\mu^2 \left\{\log{\frac{\Lambda^2}{\mu^2}}+ 1+\frac{\Gamma'(\frac{3}{2})}{\Gamma(\frac{3}{2})} - \log(8\pi^2) \right\} + \mathcal{O}(\Lambda^{-2})\,,
\end{equation}
from which we see that the radiative correction to the mass diverges as $\log \Lambda$, irrespectively of the value of $\epsilon$, i.e. irrespectively of the degree of the self-interaction.

Considering the $\frac{\lambda}{4!}\phi^4$ theory, that corresponds to the case $\epsilon=1$ (with the replacement $g=\frac{\lambda}{12}$), it is worth to remind that within the framework of the weak coupling expansion for the radiatively corrected mass at $\mathcal{O}(g)$ we have
\begin{equation}\label{mR1ord}
m_R^2=m^2 + \frac{\lambda}{32\pi^2}\left(\Lambda^2-m^2\log{\frac{\Lambda^2}{m^2}}\right)+\mathcal{O}(\Lambda^{-2})\,.
\end{equation}
Eq.\,(\ref{mR1epsreg}) is greatly intriguing. As compared with Eq.\,(\ref{mR1ord}), it seems to suggest that within the $\epsilon$-expansion the \textit{unnatural} quadratically divergent correction to the scalar mass is healed, as the correction diverges only logarithmically. If confirmed at higher orders, such a result could be of great interest for the naturalness/hierarchy problem.
In the following we will further investigate this question. 

\paragraph{Green's functions $\mathcal{G}_n$ for $n>2$.} 
As mentioned above, in these cases there is no order $\epsilon^0$ term. As for the $\mathcal{O}(\epsilon)$ contribution to $\mathcal G_n$, replacing in Eq.\,(\ref{Gn1Pi}) the vertex factor (\ref{Pi_n^1}) we obtain:
\begin{equation} \label{Gngeneric}
\mathcal{G}^{(1)}_n(x_1,\ldots,x_n)=(-1)^{\frac n2-1} \epsilon g\mu^2\left[\frac{2}{\Delta(0)}\right]^{\frac{n}{2}-1}\left(\frac{n}{2}-2\right)! \int d^du \prod_{i=1}^{n}\Delta(x_i-u) \,,
\end{equation}
that in momentum space is:
\begin{equation}\label{Gngenericp}
\widetilde{G}_n(p_1,\ldots,p_n)=(-1)^{\frac n2-1}\left(\frac{n}{2}-2\right)!\,\, \epsilon g\mu^2\left[\frac{2}{\Delta(0)}\right]^{\frac{n}{2}-1} \prod_{i=1}^{n}\frac{1}{p_i^2+M^2}\,,
\end{equation}
with the external momenta conservation to be imposed. 

It is worth to compare the result (\ref{Gngenericp}) with the corresponding one in weak coupling expansion. While in the latter case for an interaction of the kind $\phi^{2p}$ there are non-vanishing Green's functions at first order only for $n\leq2p$, Eq.\,(\ref{Gngenericp}) apparently shows a non-zero result at $\mathcal{O}(\epsilon)$ for any value of $n$. 

However, sticking again to the $d=4$ case, from (\ref{Gngenericp}) we see that all the amputated Green's functions seem to vanish in the limit $\Lambda\to\infty$ as inverse powers of the cut-off, again irrespectively of the value of $\epsilon$:
\begin{equation}\label{Gnzero}
\widetilde{G}_n(p_1,\ldots,p_n) \sim \frac{1}{\Lambda^{n-2}}\,.
\end{equation}
This would mean that for $d=4$ all the Green's functions starting from $n=4$ (and then all the scattering amplitudes) vanish in the limit $\Lambda\to\infty$, thus suggesting (at least at this order in $\epsilon$) that the theory is non-interacting. Similar results (with different degrees of vanishing with $\Lambda$) hold for any $d\geq2$. If confirmed at higher orders, this result would have serious consequences for the theory.
One of the scopes of the present paper is to further investigate this issue.

Moreover Eq.\,(\ref{Gnzero}) shows that the $\widetilde{G}_n$'s are suppressed by higher inverse powers of the cut-off as $n$ increases.
The reason is intrinsic to the procedure outlined above. In fact, before proceeding to the analytic extension, the path integral in (\ref{Gn1}) was calculated by considering sufficiently large values of the integer $N$ ($N>\frac{n}{2}$), and the limit $N\to1$ was taken only at the end. As an outcome, Eq.\,(\ref{Gngenericp}) contains the factor $\Delta(0)^{1-\frac{n}{2}}$ that gives the suppressed behaviour shown in (\ref{Gnzero}).

In passing we note that a possible way out from this apparent non-interacting character of the theory would be to consider unconventional renormalization patterns that could make the scattering amplitudes finite and non-vanishing. If we again consider the $d=4$ case and focus our attention on the four-points Green's function, from Eq.\,(\ref{Gngenericp}) we see that the only way to keep $\mathcal{G}_4$ finite (non-vanishing) is to require for the coupling constant $g$ a behaviour with $\Lambda$ such that $g\sim\Lambda^2$.
From Eq.\,(\ref{mR1epsreg}) we then see that the radiative correction $\delta m^2$ to the mass is no longer proportional to $\log\Lambda$, but rather goes as $\delta m^2 \sim \Lambda^2 \log\Lambda$, so that the original hope of having a less severe naturalness problem is immediately disattended.
Moreover we note that, even imposing such a behaviour with $\Lambda$ to the coupling constant $g$, from (\ref{Gnzero}) we see that the Green's functions starting from $n=6$ would still vanish.

In order to address the two questions posed in this section, namely the degree of divergence of the radiative correction to the mass (hierarchy problem) and the vanishing of the Green's functions with $n>2$, we have to move to higher orders in $\epsilon$. In the following section we begin with the $\mathcal O(\epsilon^2)$.

\section{Green's functions and Feynman diagrams at $\mathcal O(\epsilon^2)$}
In this section we push forward the analysis to second order in $\epsilon$. Even though the two- and four-points Green's functions have already been considered in \cite{Bender:1988ig}, here we perform the calculations for all the $\mathcal{G}_n$, showing that in the general case there are delicate problems to be taken into account.

Going back to Eq.\,(\ref{interactioneps}) we see that the $\mathcal O(\epsilon^2)$ contributions to the Green's functions, that we denote with $\mathcal G^{(2)}_n(x_1,\dots,x_n)$, come from the third and fourth terms of this equation. This can be read in terms of the Feynman rules for the auxiliary vertices defined in Eq.\,(\ref{lambdak}): for each of the Green's function we have two $\mathcal{O}(\epsilon^2)$ classes of diagrams, one with a single vertex $\lambda_2$ and one with two vertices $\lambda_1$. Every vertex carries a power of the coupling constant $g$. Therefore, disregarding the trivial dependence on $g$ carried by the mass $M^2$ in the factored out propagators, the first class of diagrams is $\mathcal{O}(\epsilon^2 g)$ while the second one is $\mathcal{O}(\epsilon^2 g^2)$. Let us analyse separately these two contributions, that we indicate respectively with $\mathcal{G}^{(2,g)}_n$ and $\mathcal{G}^{(2,g^2)}_n$.

\subsection{The order $\epsilon^2 g$ and $\epsilon^2 g^2$ contributions}

\paragraph{The order\, $\boldsymbol{\epsilon^2 g}$.}
The contribution at this order comes from the vertex $\lambda_{2}(N)$ and is:
\begin{align}
&\mathcal{G}^{(2,g)}_n(x_1,\dots,x_n) = \int d^du
\lim_{N \to 1} \frac{d^2}{dN^2}\, \left\{ -\lambda_{2}(N) \, \frac{1}{Z_0} \left(\int \mathcal D\phi \,e^{-S_0} \phi(x_1)\ldots\phi(x_n) \phi(u)^{2N} \right)_C \right\} \nonumber
\end{align}
\begin{align}\label{Gn2A}
&=\lim_{N \to 1} \frac{d^2}{dN^2}\,\left\{
	\begin{tikzpicture}[baseline=(b)]
	\begin{feynman}[inline=(b)]
	\vertex (b)[blob,label={center:{\small $\lambda_2$}},pattern color=white] at (0, 0) {};
	\vertex (a) at (-1.2, -0.64) {$x_1$};
	\vertex (d) at ( 1.2, -0.64) {$x_n$};
	\vertex (c) at (-0.357948,-1.07532) {$x_2$};
	\vertex (e) at ( 0.357948,-1.07532) {};
	\vertex (l1) at (0.4,0.3) {};
	\vertex (l2) at (-0.1,0.45) {};
	\vertex (p1) at (-0.16,-0.7) {};
	\vertex (p2) at (0.35,-0.5) {};
	\diagram* {
		(a) -- (b) -- [out=35, in=0, loop, min distance=1.2cm,edge label={\footnotesize $N-\frac{n}{2}$}] b -- [out=135, in=100, loop, min distance=1.2cm,edge label={\footnotesize $2$}] b -- [out=185, in=150, loop, min distance=1.2cm,edge label={\footnotesize $1$}] b -- (d);
		(c) -- (b);
		(l1) -- [thick,dotted,out=-280, in=-315](l2);
		(p1) -- [thick,dotted,out=0, in=20](p2);
	};
	\end{feynman}
	\end{tikzpicture}
\right\} 
= \lim_{N \rightarrow 1} \frac{d^2}{dN^2} \left\{-\lambda_{2}(N)\left[\Delta(0)\right]^{N-\frac{n}{2}} C_n(N)\right\}\int d^du \prod_{i=1}^{n}\Delta(x_i-u)\,.
\end{align}
As can be easily verified, the diagram in (\ref{Gn2A}) is practically the same as the one in (\ref{Gn1}), with the only difference that the $\mathcal{O}(\epsilon^2)$ vertex  $\lambda_{2}$ (and correspondingly the second derivative operator) appears rather than the $\mathcal{O}(\epsilon)$ vertex $\lambda_{1}$.
As for the self-loops and the combinatorial factor, they are exactly the same. 
Similarly to what we have done at $\mathcal{O}(\epsilon)$, we now define the $n$-legs $\mathcal{O}(\epsilon^2)$ effective vertex $\Pi^{(2)}_{n}$ as:
\begin{equation} \label{Pi_n^2def}
\Pi^{(2)}_{n}\,=\,
\begin{tikzpicture}[baseline=-3]
\begin{feynman}
\vertex[draw, circle, minimum size=0.85cm,very thick] (m) at (0, 0) {2};
\vertex (a) at (-0.9,0.9) {\small$2$};
\vertex (b) at (-0.9,-0.9) {\small$1$};
\vertex (c) at (1.24421,0) {\small$n$};
\vertex (l1) at (0.5,0.35) {};
\vertex (l2) at (-0.2,0.55) {};
\vertex (d1) at (0.595294,-0.308824) {};
\vertex (d2) at (0.924706,0.308824) {};
\vertex (d3) at (-0.700391,0.226862) {};
\vertex (d4) at (-0.370979,0.844509) {};
\vertex (d5) at (-0.700391,-0.226862) {};
\vertex (d6) at (-0.370979,-0.844509) {};
\diagram* {
	(a) -- (m) -- (b);
	(m) -- (c);
	(l1) -- [thick,dotted,out=-280, in=-305](l2);
};
\draw (d1) to (d2);
\draw (d3) to (d4);
\draw (d5) to (d6);
\end{feynman}
\end{tikzpicture} \,\equiv\,
\lim_{N \rightarrow 1} \frac{d^2}{dN^2} \left\{-\lambda_{2}(N)\left[\Delta(0)\right]^{N-\frac{n}{2}} C_n(N)\right\}\,.
\end{equation}
Replacing in (\ref{Pi_n^2def}) the expressions for $\lambda_{2}(N)$ and $C_n(N)$ in Eqs.\,(\ref{lambdak}) and (\ref{CnN1}), we have:
\begin{equation} 
\Pi^{(2)}_n = -\frac{\epsilon^2}{2}\,g\,\mu^2\left[\frac{2}{\Delta(0)}\right]^{\frac{n}{2}-1} \lim_{N \rightarrow 1} \frac{d^2}{dN^2}\, f_n(N)\,,
\end{equation}
where the function $f_n(N)$ is defined in Eq.\,(\ref{fnNint}) for integers $N$, and its analytic extension is given in (\ref{fnNcont}). Due to the presence of the factor $(N-1)$, once again we have to treat separately the cases $n=2$ and $n>2$. Performing the calculations, we get:
\begin{align}
\underline{n=2} \hspace*{30pt} \Pi^{(2)}_2 & = -\frac{\epsilon^2}{2}\,g\,\mu^2 \left[K^2 - 1 + \psi'\left(\frac{3}{2}\right)\right] \label{Pi_2^2}\\
\underline{n>2} \hspace*{30pt} \Pi^{(2)}_n & = (-1)^{\frac{n}{2} -1}\left(\frac n2-2\right)!\,\, \epsilon^2 g \mu^2 \left[\frac{2}{\Delta(0)}\right]^{\frac{n}{2}-1}  \left(K- H_{\frac n2 -2}\right) \label{Pi_n^2}
\end{align}
where $K$ is defined in Eq.\,(\ref{K}), while $H_n$ stands for the $n$-th Harmonic number. 

\vskip 10pt

\paragraph{The order\, $\boldsymbol{\epsilon^2 g^2}$.}
The calculation of this term is somehow more complicated than the previous one. Due to the presence of two  vertices $\lambda_{1}$, the number of different possible diagrams grows enormously, and there are subtleties that need to be carefully treated. From (\ref{interactioneps}), the contribution to $\mathcal{G}_n$ to this order is:
\begin{align}\label{Gn2B}
\mathcal{G}^{(2,g^2)}_n(x_1,\dots,x_n) &=  \frac{1}{2} \int d^du\, d^dw\,
\lim_{N \to 1} \, \lim_{M \to 1} \, \frac{d}{dN} \, \frac{d}{dM}\, \Bigg\{ \left(-\lambda_{1}(N)\right) \left(-\lambda_{1}(M)\right)\,\nonumber \\ & \times \frac{1}{Z_0} \left(\int \mathcal D\phi \,e^{-S_0} \phi(x_1)\ldots\phi(x_n) \phi(u)^{2N} \phi(w)^{2M}\right)_C \Bigg\}\,.
\end{align}

In order to list the different diagrams that can be built, we can begin by considering the contractions of the $n$ ``external" fields $\phi(x_1),\dots,\phi(x_n)$ with the fields $\phi(u)$ in the point $u$. We can connect $r$, with $0 \leq r \leq n$, of them with $\phi(u)$, leaving the remaining $n-r$ fields for contractions with $\phi(w)$. For each of these choices, we still have the freedom to use the leftover fields $\phi(u)$ and $\phi(w)$ to link the two vertices $u$ and $w$ with as many links as we like, with the upper limit posed by the number of available fields, that in turn depends on the values of $N$ and $M$. Moreover, the number of $u-w$ links must have the same parity of $r$, so that all the remaining fields at each vertex are contracted in pairs, giving rise to self-loops. For this reason, the even ($r=2j$) and odd ($r=2j+1$) cases have to be treated separately. 
\vskip10pt
$\underline{\textit{``Even" contribution.}}$\quad As for the $\mathcal{O}(\epsilon)$ case, the path integral in (\ref{Gn2B}) gives non vanishing results only for sufficiently large integers $N$ and $M$, namely $N>j$ and $M>\frac{n}{2}-j$, so that connected diagrams can be drawn. In this case, indicating with $\mathcal{G}^{(2,g^2)}_{n,\,E}$ the sum of all the diagrams of this kind\footnote{As explained above, the two vertices $\lambda_{1}$ are connected with an even number $2l$ of lines. As we start with $l=1$, the minimal number of internal lines is $2$, while the maximal number is $2$\,Min. In (\ref{Gn2Be}) this is indicated by connecting these two vertices with two continuous lines and two dashed ones that represent the sum over $l$.}, we have:
\begin{align} \label{Gn2Be}
&\mathcal{G}^{(2,g^2)}_{n,\,E} = \frac{1}{2} \lim_{\substack{N \to 1 \\ M \to 1}} \frac{d}{dN} \frac{d}{dM} \sum_{j=0}^{n/2} \sum_{l=1}^{\text{Min}} \left\{ 
\begin{tikzpicture}[baseline=(m)]
\begin{feynman}[inline=(m)]
\vertex[blob,label={center:{\small $\lambda_1$}},pattern color=white] (m) at (-1, 0) {};
\vertex[label={center:{\footnotesize $u$}}] (m0) at (-1,-0.5) {};
\vertex[blob,label={center:{\small $\lambda_1$}},pattern color=white] (n) at (1, 0) {};
\vertex[label={center:{\footnotesize $w$}}] (n0) at (1,-0.5) {};
\vertex (a) at (-2.3,1.2) {$x_{1}$};
\vertex (b) at (-2.3,-1.2) {$x_{2j}$};
\vertex (c) at (2.3,1.2) {$x_{2j+1}$};
\vertex (d) at (2.3,-1.2) {$x_{n}$};
\vertex (l1) at (1.7,0.6) {};
\vertex (l2) at (1.7,-0.6) {};
\vertex (p1) at (-1.7,0.6) {};
\vertex (p2) at (-1.7,-0.6) {};
\vertex (q1) at (-1.3,0.65) {};
\vertex (q2) at (-0.7,0.65) {};
\vertex (t1) at (1.3,0.65) {};
\vertex (t2) at (0.7,0.65) {};
\diagram* {
	(a) -- (m) -- [out=70, in=50, loop, min distance=0.9cm] m -- [out=130, in=110, loop, min distance=0.9cm] m -- (b);
	(m) -- [half left, out=17, in=163](n);
	(m) -- [half left, out=-17, in=-163](n);
	(m) -- [half left, out=36, in=144,dashed](n);
	(m) -- [half left, out=-36, in=-144,edge label'={\small $2l$},dashed](n);
	(c) -- (n) -- [out=70, in=50, loop, min distance=0.9cm] n -- [out=130, in=110, loop, min distance=0.9cm] n -- (d);	
	(l1) -- [thick,dotted,out=-60, in=60](l2);
	(p2) -- [thick,dotted,out=120, in=-120](p1);
	(q2) -- [thick,dotted,out=160, in=20, edge label'={\tiny $N\!-j\!-l$}](q1);
	(t2) -- [thick,dotted,out=20, in=160, edge label={\tiny $M\!-\frac{n}{2}\!+j\!-l$}](t1);
};
\end{feynman}
\end{tikzpicture}
+ \binom{n}{2j} -1 \,\, \text{perm.} \right\}  \nonumber \\
&= \frac{1}{2} \int d^du \, d^dw\, \lim_{\substack{N \to 1 \\ M \to 1}} \frac{d}{dN} \, \frac{d}{dM} \sum_{j=0}^{n/2} \sum_{l=1}^{\text{Min}} \left[-\lambda_1(N) \Delta(0)^{N-j-l}\right]\left[-\lambda_1(M) \Delta(0)^{M-\frac{n}{2}+j-l}\right] \nonumber \\
&\times  C_{n,j,l}(M,N) \left[ \prod_{i=1}^{2j}\Delta(x_i-u) \prod_{h=2j+1}^n \!\Delta(x_h-w) \,\, \Delta(u-w)^{2l}\,\, +\, \binom{n}{2j} - 1\, {\rm perm.}\right]
\end{align}
where the coefficient $C_{n,j,l}(M,N)$ contains all the combinatorial factors, the $\binom{n}{2j}$ permutations corresponds to permutations of external points in the diagram, and the upper limit of the $l$-sum is $\text{Min}\equiv min(N-j,M-\frac{n}{2}+j)$\,. 
Rigorously speaking, Eq.\,(\ref{Gn2Be}) takes a precise mathematical meaning only for real values of the variables $N$ and $M$, with the the functions of $N$ and $M$ in (\ref{Gn2Be}) being analytic around $(N=1,M=1)$. This means that we have to search for the analytic extensions of these functions.
This is what we are going to do, but before reaching that point let us continue to consider $N$ and $M$ as integers, and stick on the above mentioned conditions
$N>j$ and $M>\frac{n}{2}-j$\,.

We now take out from the limits and the derivatives in (\ref{Gn2Be}) those factors that do not depend on $N$ and $M$, thus getting:
\begin{align} \label{Gn2Be3}
\mathcal{G}^{(2,g^2)}_{n,\,E}& = \frac{1}{2} \int d^du \, d^dw\, \sum_{j=0}^{n/2} \Bigg\{ \prod_{i=1}^{2j}\Delta(x_i-u) \prod_{h=2j+1}^n \!\Delta(x_h-w) \,\,\lim_{\substack{N \to 1 \\ M \to 1}} \frac{d}{dN} \, \frac{d}{dM} \sum_{l=1}^{\text{Min}} C_{n,j,l}(M,N) \nonumber \\ &\times \left[-\lambda_1(N) \Delta(0)^{N-j-l}\right]\left[-\lambda_1(M) \Delta(0)^{M-\frac{n}{2}+j-l}\right] \, \Delta(u-w)^{2l}\,\, +\, \binom{n}{2j}-1\,{\rm perm.} \Bigg\}\,,
\end{align}
and focus our attention on the function of which we have to take the derivatives and the limits with to respect to $M$ and $N$.

Let us consider first the combinatorial factor $C_{n,j,l}(M,N)$. It is:
\begin{align}\label{CnjlMN}
C_{n,j,l}(M,N)&=2N(2N-1)\dots(2N-2j-2l+1)\,(2N-2j-2l-1)!! \nonumber \\
&\times 2M(2M-1)\dots(2M-n+2j-2l+1)\,(2M-n+2j-2l-1)!!\, \frac{1}{(2l)!} \nonumber\\
&=\frac{1}{(2l)!}\,C_{2j+2l}(N)\,C_{n-2j+2l}(M)\,.
\end{align}
The last line of (\ref{CnjlMN}) is obtained by noting that each of the two factors in the second member is of the same form of the combinatorial coefficients $C_n$ in Eq.\,(\ref{CnN0}). Such a factorization is quite remarkable, and can be understood by looking at the diagram in\,(\ref{Gn2Be}). Actually, the combinatorial factor $C_{n,j,l}(M,N)$ takes into account:
\begin{enumerate}
	\item the contractions of $2j+2l$ (out of $2N$) $\phi(u)$ fields with fields at other space-time points: this is the definition of $C_{2j+2l}(N)$;
	\item the contractions of $n-2j+2l$ (out of $2M$) $\phi(w)$ fields with fields at other space-time points: this is the definition of $C_{n-2j+2l}(M)$;
	\item the fact that the $2l$ contractions made between fields $\phi(u)$ and $\phi(w)$ are indistinguishable: the factor $\frac{1}{(2l)!}$ cures the over-counting of these permutations.
\end{enumerate}
Thanks to this factorization, the sum over $l$ can then be written as:
\begin{equation} \label{function}
\sum_{l=1}^{\text{Min}} \left[-\lambda_1(N) C_{2j+2l}(N) \Delta(0)^{N-j-l}\right]\left[-\lambda_1(M) C_{n-2j+2l}(M) \Delta(0)^{M-\frac{n}{2}+j-l}\right] \frac{\Delta(u-w)^{2l}}{(2l)!}\,.
\end{equation}
Replacing in (\ref{function}) the expression for $\lambda_{1}$ given in (\ref{lambdak}), together with the combinatorial factors in (\ref{CnN1}), we have:
\begin{equation} \label{function2}
\sum_{l=1}^{\text{Min}} \left[-\epsilon g\mu^2\left[\frac{2}{\Delta(0)}\right]^{j+l-1} f_{2j+2l}(N)\right]\left[-\epsilon g\mu^2\left[\frac{2}{\Delta(0)}\right]^{\frac{n}{2}-j+l-1} f_{n-2j+2l}(M)\right] \frac{\Delta(u-w)^{2l}}{(2l)!}\,,
\end{equation}
where we used the functions $f_m$ defined in (\ref{fnNint}) as:
\begin{equation} \label{fmNint}
f_m(N)=(N)_{\frac{m}{2}} (2N-1)!! \left[2 \mu^{2-d}\Delta(0)\right]^{N-1}\,.
\end{equation} 
From (\ref{fmNint}) we see that, as the only dependence on the parameter $m$ is contained in the falling factorial $(N)_{\frac{m}{2}}$, using the property
\begin{equation}
(x)_{a+b}=(x)_a\,(x-a)_b\qquad \forall a,b\in\mathbb{Z} \,\, \text{with} \,\, a,b\geq0\,, \,\,\forall x\in \mathbb{R}\,,
\end{equation}
we can split the two falling factorials in $f_{2j+2l}(N)$ and $f_{n-2j+2l}(M)$ as 
\begin{align*}
(N)_{j+l}=(N)_{j}(N-j)_{l} \qquad ; \qquad
(M)_{\frac{n}{2}-j+l}=(M)_{\frac{n}{2}-j}\left(M-\frac{n}{2}+j\right)_{l}
\end{align*} 
so that:
\begin{align} \label{ffact}
f_{2j+2l}(N)=f_{2j}(N) \,(N-j)_l \qquad ; \qquad 
f_{n-2j+2l}(M)&=f_{n-2j}(M)\,\left(M-\frac{n}{2}+j\right)_{l}\,.
\end{align} 
Replacing Eqs.\,(\ref{ffact}) in (\ref{function2}), we can take out from the sum the $l$-independent terms, thus getting:
\begin{align} \label{function3}
&\left[-\epsilon g\mu^2\left[\frac{2}{\Delta(0)}\right]^{j-1} f_{2j}(N)\right]\left[-\epsilon g\mu^2\left[\frac{2}{\Delta(0)}\right]^{\frac{n}{2}-j-1} f_{n-2j}(M)\right] \nonumber \\
&\times\sum_{l=1}^{\text{Min}} \left[\frac{2}{\Delta(0)}\right]^{2l} \frac{\Delta(u-w)^{2l}}{(2l)!}\, (N-j)_l \left(M-\frac{n}{2}+j\right)_l\,.
\end{align}
Using the identity $(2l)!=(2l)!!(2l-1)!!=2^l \,l!\, 2^l\, \left(\frac{1}{2}\right)^{(l)}$, and turning the falling factorials to rising factorials through the relation $(x)_m=(-1)^m\,(-x)^{(m)}$, Eq.\,(\ref{function3}) becomes:
\begin{align} \label{before}
&\left[-\epsilon g\mu^2\left[\frac{2}{\Delta(0)}\right]^{j-1} f_{2j}(N)\right]\left[-\epsilon g\mu^2\left[\frac{2}{\Delta(0)}\right]^{\frac{n}{2}-j-1} f_{n-2j}(M)\right]\ \nonumber \\
& \times\sum_{l=1}^{\text{Min}} \frac{(j-N)^{(l)}\,\left(\frac{n}{2}-j-M\right)^{(l)}}{\left(\frac{1}{2}\right)^{(l)}} \frac{1}{l!} \left[\frac{\Delta(u-w)}{\Delta(0)}\right]^{2l} \,.
\end{align}

As Min$=min\left(N-j,M-\frac{n}{2}+j\right)$, and we are still considering $N$ and $M$ as integers, the sum over $l$ can be written as
\begin{equation} \label{hyper}
\sum_{l=1}^{\text{Min}} \frac{(j-N)^{(l)}\,\left(\frac{n}{2}-j-M\right)^{(l)}}{\left(\frac{1}{2}\right)^{(l)}} \frac{1}{l!} \left[\frac{\Delta(u-w)}{\Delta(0)}\right]^{2l} = {}_{2}{F}_{1}\left(j-N,\, \frac{n}{2}-j-M;\, \frac{1}{2}; \left[\frac{\Delta(u-w)}{\Delta(0)}\right]^{2} \right) \,-1\,,
\end{equation}
where the function ${}_2{F}_{1}(a,b;c;z)$ is the Gaussian hypergeometric function defined by
\begin{equation} \label{hyperdef}
{}_2{F}_{1}(a,b;c;z)=\sum_{l=0}^{\infty} \frac{(a)^{(l)}\,(b)^{(l)}}{\left(c\right)^{(l)}} \frac{z^l}{l!} \qquad \forall z\in\mathbb{C}\,\,\text{with}\,\,|z|<1 \,,
\end{equation}
which in (\ref{hyper}) is truncated to a polynomial in $z=\left[\frac{\Delta(u-w)}{\Delta(0)}\right]^{2}$, as the sum over $l$ runs up to Min due to the vanishing of the rising factorials in (\ref{hyper}) for $l>Min$. In passing we note that in this case the condition $|z|<1$ is unnecessary, as for a polynomial the convergence radius is infinite.

Eq.(\ref{before}) then takes the form
\begin{align} \label{analyticAPP}
&\left[-\epsilon g\mu^2\left[\frac{2}{\Delta(0)}\right]^{j-1} f_{2j}(N)\right]\left[-\epsilon g\mu^2\left[\frac{2}{\Delta(0)}\right]^{\frac{n}{2}-j-1} f_{n-2j}(M)\right]\ \nonumber \\
&\times \left[{}_{2}{F}_{1}\left(j-N,\, \frac{n}{2}-j-M;\, \frac{1}{2}; \left[\frac{\Delta(u-w)}{\Delta(0)}\right]^{2} \right) - 1\right] \,.
\end{align}

We got the expression we were looking for. Although Eq.\,(\ref{analyticAPP}) was obtained for $N$ and $M$ integers such that $N>j$ and $M>\frac{n}{2}-j$, we would like to use this expression to get an analytic extension of the ``even contribution" to the path integral in (\ref{Gn2B}) to real values of $N$ and $M$ close to $N=1$ and $M=1$.

To this end we begin by noting that, the functions $f_{2j}(N)$ and $f_{n-2j}(M)$ respectively vanish for $N<j$ and $M<\frac{n}{2}-j$ while for $N=j$ and/or $M=\frac{n}{2}-j$ the factor ${}_{2}{F}_{1} -1$ vanishes. These are also the conditions under which the ``even contribution" to the path integral for the connected Green's functions in (\ref{Gn2B}) vanishes. This is the key observation that allows to use Eq.(\ref{analyticAPP}) as the starting point for the desired analytic extension of the path integral.

First of all we note that the function $f_m(N)$ defined in (\ref{fmNint}) can be easily extended to real values of $N$, as already shown in Section 2 (see Eq.(\ref{fnNcont})). 

Let us consider next the extension of the function ${}_{2}{F}_{1}$ to real $N$ and $M$. 
This extension has to be done for any couple of points $(u,w)$ as these are the variables on which we have to integrate (see Eq.(\ref{Gn2Be3})). For a generic couple $(u,w)$ such that $u\neq w$ the condition $|\frac{\Delta(u-w)}{\Delta(0)}|<1$ holds, while for those points such that $u=w$ we have $|\frac{\Delta(u-w)}{\Delta(0)}|=1$. Let us begin by considering the case $u\neq w$, leaving aside for the moment the case $u=w$.

Under this condition ($u\neq w$), we can analytically extend ${}_{2}{F}_{1}$ in (\ref{analyticAPP}) to real values of $N$ and $M$ with the help of the definition (\ref{hyperdef}).
Moreover, as for any fixed couple of values $(z,c)$ the function ${}_2{F}_{1}(a,b;c;z)$ is an entire function of $a$ and $b$, the derivatives and the limits with respect to $N$ and $M$ can be performed term by term in the series. 

Inserting in (\ref{analyticAPP}) the left-hand side of (\ref{hyper}), with Min sent to infinity, appropriately recombining the rising factorials of (\ref{hyper}) with the functions $f_{2j}(N)$ and $f_{n-2j}(M)$ making use of Eqs.\,(\ref{ffact}), and then considering the derivatives and the limits of (\ref{analyticAPP}) we get:
\begin{align} \label{derseries}
\sum_{l=1}^{\infty} &\lim_{N \to 1} \frac{d}{dN}\left[-\epsilon g\mu^2\left[\frac{2}{\Delta(0)}\right]^{j+l-1} f_{2j+2l}(N)\right] \nonumber \\
\times &\lim_{M \to 1} \frac{d}{dM}   \left[-\epsilon g\mu^2\left[\frac{2}{\Delta(0)}\right]^{\frac{n}{2}-j+l-1} f_{n-2j+2l}(M)\right] \frac{\Delta(u-w)^{2l}}{(2l)!}\,.
\end{align}

Next we have to consider the case $u=w$, that corresponds to $z=|\frac{\Delta(u-w)}{\Delta(0)}|=1$. This case is more delicate as $z=1$ lies on the border of the  convergence circle of the series (\ref{hyperdef}). 
However the convergence of this power series is guaranteed even for $z=1$ under the further condition
\begin{equation}
	Re(c-a-b)>0\,,
\end{equation}
that in our case becomes
\begin{equation} \label{convcond}
N+M+\frac{1}{2}-\frac{n}{2}>0\,.
\end{equation}

Once again we note that, as long as $N>j$ and $M>\frac{n}{2}-j$ this condition is fulfilled.
However we need to consider the analytic extension around the point $(N=1,M=1)$, and in this region the condition (\ref{convcond}) is not fulfilled starting from $n=6$. 
As a consequence, for $n\geq6$ the series (\ref{derseries}) diverges when $u=w$.

However we have to stress that, even though for $n=2$ and $n=4$ the series (\ref{derseries}) converges in the whole domain of space-time integration, this does not mean that the final expression after integration is convergent. Actually we have checked (even numerically) that there are two distinct possible sources of divergences: (i) the series involved in the analytic extension could diverge in the ultraviolet regime $u \to w$ (that is the case we encountered for $n>4$); (ii) even when the series is convergent for $u\to w$, the whole expression could be ultraviolet divergent once the space-time integration is performed.
This is due to the fact that the resummed function, although everywhere finite, could be non-integrable or to the absence of dominated convergence that make it impossible to perform the integration by series.

These are genuinely novel features of the $\mathcal{O}(\epsilon^2)$ contributions to the Green's functions. At the order $\epsilon$ considered in the previous section, such a problem did not show up as, in that case, the functions to be analytically extended were factorized out of the space-time integral. Here, instead, we had to cope with analytic extensions of functions involved in space-time integrals that are related to the connection of two vertices at different points, and such analytic extension gave rise to hypergeometric infinite series that bring divergences in the ultraviolet regime.

These divergences pose delicate problems and deserve further investigation. One way to take care of them is through the introduction of a cut-off $L_{max}$ in the power series, and this will allow to treat together the cases $u\neq w$ and $u=w$. Once the infinite upper limit is replaced with $L_{max}$, Eq.(\ref{derseries}) can be extended to all the space-time domain of integration. We have not further studied this problem, but we are rather assuming that this is a suitable regularization and this is the implementation we will follow below. 

Under this assumption, we can finally write for $\mathcal{G}^{(2,g^2)}_{n,\,E}$
\begin{align} \label{Gn2Befact}
\mathcal{G}^{(2,g^2)}_{n,\,E} &= \frac{1}{2} \sum_{j=0}^{n/2} \sum_{l=1}^{L_{max}} \lim_{N \to 1} \frac{d}{dN} \left[-\epsilon g\mu^2\left[\frac{2}{\Delta(0)}\right]^{j+l-1} f_{2j+2l}(N) \right] \nonumber \\
&\times \lim_{M \to 1} \frac{d}{dM} \left[-\epsilon g\mu^2\left[\frac{2}{\Delta(0)}\right]^{\frac{n}{2}-j+l-1} f_{n-2j+2l}(M)\right] \frac{1}{(2l)!} \nonumber \\
&\times \left[\int d^du \, d^dw\, \prod_{i=1}^{2j}\Delta(x_i-u) \prod_{h=2j+1}^n \!\Delta(x_h-w) \,\, \Delta(u-w)^{2l}\,\, + \, \binom{n}{2j} -1 \, {\rm perm.}\right]\,.
\end{align}

Eq.(\ref{Gn2Befact}) is a very welcome result as each of the factorized limits in the sum over $l$ is nothing but an effective vertex as given in (\ref{Pi_n^1def2}).
For $\mathcal{G}^{(2,g^2)}_{n,\,E}$ we finally have the elegant and compact result:
\begin{align} \label{Gn2BePi}
\mathcal{G}^{(2,g^2)}_{n,\,E}&=\frac{1}{2} \sum_{j=0}^{n/2} \sum_{l=1}^{L_{max}} \,\,\,\Pi^{(1)}_{2j+2l} \,\,\, \Pi^{(1)}_{n-2j+2l}\,\,\frac{1}{(2l)!}\nonumber\\
&\times \left[\int d^du \, d^dw\, \prod_{i=1}^{2j}\Delta(x_i-u) \prod_{h=2j+1}^n \!\Delta(x_h-w) \,\, \Delta(u-w)^{2l}\,\, + \, \binom{n}{2j} -1\, {\rm perm.}\right]\,.
\end{align}

With the help of the diagrammatic representation of the effective vertices $\Pi^{(1)}_n$ given in (\ref{Pi_n^1}), we can write (\ref{Gn2BePi}) as:

\begin{equation} \label{Gn2Bediag}
\mathcal{G}^{(2,g^2)}_{n,\,E}= \frac{1}{2} \,\sum_{j=0}^{\frac{n}{2}}\,\sum_{l=1}^{L_{max}} \left\{\begin{tikzpicture}[baseline=(m)]
\begin{feynman}[inline=(m)]
\vertex[draw, circle, minimum size=0.85cm,very thick] (m) at (-1, 0) {1};
\vertex[label={center:{\footnotesize $u$}}] (m0) at (-1,-0.6) {};
\vertex[draw, circle, minimum size=0.85cm,very thick] (n) at (1, 0) {1};
\vertex[label={center:{\footnotesize $w$}}] (n0) at (1,-0.6) {};	
\vertex (a) at (-2.3,1.3) {$x_{1}$};
\vertex (b) at (-2.3,-1.3) {$x_{2j}$};
\vertex (c) at (2.3,1.3) {$x_{2j+1}$};
\vertex (d) at (2.3,-1.3) {$x_{n}$};
\vertex (l1) at (1.8,0.7) {};
\vertex (l2) at (1.8,-0.7) {};
\vertex (p1) at (-1.8,0.7) {};
\vertex (p2) at (-1.8,-0.7) {};
\diagram* {
	(a) -- (m) -- (b);
	(m) -- [half left, out=17, in=163](n);
	(m) -- [half left, out=-17, in=-163](n);
	(m) -- [half left, out=36, in=144,dashed](n);
	(m) -- [half left, out=-36, in=-144,edge label'={\small $2l$},dashed](n);
	(c) -- (n) -- (d);	
	(l1) -- [thick,dotted,out=-60, in=60](l2);
	(p2) -- [thick,dotted,out=120, in=-120](p1);
};
\end{feynman}
\end{tikzpicture}
+\,\,\binom{n}{2j}-1 \,\, \text{perm.}\right\}\,.
\end{equation}
Note that the factor $\frac{1}{(2l)!}$ of (\ref{Gn2BePi}) is recovered once the indistinguishability of the $2l$ lines connecting the effective vertices $\Pi^{(1)}$ in $u$ and $w$ is taken into account. It is worth to note that $\frac{1}{(2l)!}$ is the only genuine factorial term that is left in $\mathcal{G}^{(2,g^2)}_{n,\,E}$, the other factorial terms appearing in the intermediate steps of our derivation have been replaced by analytic continuations and then absorbed in the definition of the $\Pi^{(1)}$'s.

Finally we observe that, by keeping the points $u$ and $w$ attached to the vertices, we are considering them as two distinguishable points. Actually, as $u$ and $w$ are interchangeable, every pair of diagrams corresponding to $j=i$ and $j=n/2-i$  yields to the same contribution because they are equal under the exchange $u \iff w$. Therefore, if we only consider distinct diagrams under the exchange of vertices, we can cancel the prefactor $\frac{1}{2}$ in the right-hand side of (\ref{Gn2Bediag}). However, for generic values of $n$, it is not always true that $\frac{n}{4}$ is an integer number, and this is why we have to keep the factor $\frac{1}{2}$ in our expression. Finally when $j=n/2$ the number of permutations is halved.\\

$\underline{\textit{``Odd" contribution.}}$ By considering now the case where the two vertices are joined with an odd number of lines, we have\footnote{As compared to the diagrams appearing in the ``even" contribution (Eq.\,(\ref{Gn2Be})), in (\ref{Gn2Bo}) we have drawn only one continuous internal line, as in this case we have an odd number, namely $2l+1$, of internal lines, starting from $l=0$.}:
\begin{align} \label{Gn2Bo}
\mathcal{G}^{(2,g^2)}_{n,\,O} &=\frac{1}{2} \lim_{\substack{N \to 1 \\ M \to 1}} \frac{d}{dN} 		\frac{d}{dM} \sum_{j=0}^{n/2-1} \sum_{l=0}^{\rm Min} \left\{ 
\begin{tikzpicture}[baseline=(m)]
\begin{feynman}[inline=(m)]
\vertex[blob,label={center:{\small $\lambda_1$}},pattern color=white] (m) at (-1, 0) {};
\vertex[label={center:{\footnotesize $u$}}] (m0) at (-1,-0.5) {};
\vertex[blob,label={center:{\small $\lambda_1$}},pattern color=white] (n) at (0.8, 0) {};
\vertex[label={center:{\footnotesize $w$}}] (n0) at (0.8,-0.5) {};
\vertex (a) at (-2.3,1.3) {$x_{1}$};
\vertex (b) at (-2.3,-1.3) {$x_{2j+1}$};
\vertex (c) at (2.1,1.3) {$x_{2j+2}$};
\vertex (d) at (2.1,-1.3) {$x_{n}$};
\vertex (l1) at (1.5,0.6) {};
\vertex (l2) at (1.5,-0.6) {};
\vertex (p1) at (-1.7,0.6) {};
\vertex (p2) at (-1.7,-0.6) {};
\vertex (q1) at (-1.3,0.65) {};
\vertex (q2) at (-0.7,0.65) {};
\vertex (t1) at (1.1,0.65) {};
\vertex (t2) at (0.5,0.65) {};
\diagram* {
(a) -- (m) -- [out=70, in=50, loop, min distance=0.9cm] m -- [out=130, in=110, loop, min distance=0.9cm] m -- (b);
(m) -- [half left,out=30, in=150, dashed](n);
(m) -- [half left,out=-30, in=-150, edge label'={\small $2l+1$}, dashed](n);
(m) -- (n);
(c) -- (n) -- [out=70, in=50, loop, min distance=0.9cm] n -- [out=130, in=110, loop, min distance=0.9cm] n -- (d);	
(l1) -- [thick,dotted,out=-60, in=60](l2);
(p2) -- [thick,dotted,out=120, in=-120](p1);
(q2) -- [thick,dotted,out=160, in=20, edge label'={\tiny $N\!-j\!-l\!-1$}](q1);
(t2) -- [thick,dotted,out=20, in=160, edge label={\tiny $M\!-\frac{n}{2}\!+j\!-l$}](t1);
};
\end{feynman}
\end{tikzpicture} + \,\text{\small $\binom{n}{2j\!+\!1} -1$ \, perm.} \right\}\,.
\end{align}
Performing similar steps to those done for the even case, the contribution $\mathcal{G}^{(2,g^2)}_{n,\,O}$ to the Green's functions $\mathcal{G}^{(2,g^2)}_{n}$ turns out to be:

\begin{align} 
&\mathcal{G}^{(2,g^2)}_{n,\,O} =\frac{1}{2} \sum_{j=0}^{\frac{n}{2}-1} \sum_{l=0}^{L_{max}} \,\,\,\Pi^{(1)}_{2j+2l+2} \,\,\, \Pi^{(1)}_{n-2j+2l}\,\,\frac{1}{(2l+1)!}\times\nonumber\\
&\times \left[\int d^du \, d^dw\, \prod_{i=1}^{2j+1}\Delta(x_i-u) \prod_{h=2j+2}^n \!\Delta(x_h-w) \,\, \Delta(u-w)^{2l+1}\,\, + \,\, \binom{n}{2j\!+\!1} -1\,\, {\rm perm.}\right] \nonumber
\end{align}
\begin{align}\label{Gn2Bodiag}
&=\frac{1}{2} \,\sum_{j=0}^{\frac{n}{2}-1}\,\sum_{l=0}^{L_{max}} \left\{\begin{tikzpicture}[baseline=(m)]
\begin{feynman}[inline=(m)]
\vertex[draw, circle, minimum size=0.85cm,very thick] (m) at (-1, 0) {1};
\vertex[label={center:{\footnotesize $u$}}] (m0) at (-1,-0.6) {};
\vertex[draw, circle, minimum size=0.85cm,very thick] (n) at (1, 0) {1};
\vertex[label={center:{\footnotesize $w$}}] (n0) at (1,-0.6) {};	
\vertex (a) at (-2.3,1.3) {$x_{1}$};
\vertex (b) at (-2.3,-1.3) {$x_{2j+1}$};
\vertex (c) at (2.3,1.3) {$x_{2j+2}$};
\vertex (d) at (2.3,-1.3) {$x_{n}$};
\vertex (l1) at (1.8,0.7) {};
\vertex (l2) at (1.8,-0.7) {};
\vertex (p1) at (-1.8,0.7) {};
\vertex (p2) at (-1.8,-0.7) {};
\diagram* {
	(a) -- (m) -- (b);
	(m) -- [half left,out=30, in=150, dashed](n);
	(m) -- [half left,out=-30, in=-150, edge label'={\footnotesize $2l+1$}, dashed](n);
	(m) -- (n);
	(c) -- (n) -- (d);	
	(l1) -- [thick,dotted,out=-60, in=60](l2);
	(p2) -- [thick,dotted,out=120, in=-120](p1);
};
\end{feynman}
\end{tikzpicture}
+\,\,\binom{n}{2j\!+\!1} - 1\,\, \text{perm.}\right\}\,,
\end{align}
where $L_{max}$ is introduced for the same reasons encountered in the even case.

From Eqs.\,(\ref{Gn2Bediag}) and (\ref{Gn2Bodiag}) we see that the $\mathcal{O}(\epsilon^2)$ diagrams coming from two vertices $\lambda_{1}$ are easily expressed in terms of the effective vertices $\Pi^{(1)}_n$ of $\mathcal{O}(\epsilon)$\,. This allows to write all the $\mathcal{O}(\epsilon^2 g^2)$ contributions to the $n$-points Green's functions without referring to the original Feynman rules given in Section 2. What we need to do, instead, is to draw the diagrams with two effective vertices of the type $\Pi^{(1)}$, according to the following simple rules: 
\begin{enumerate}
	\item Draw all the different diagrams with $n$ external points and two ``effective vertices" $\Pi^{(1)}$ that fulfil the following conditions:
	\begin{enumerate}
		\item $r$ (with $0\leq r\leq \frac{n}{2}$) of the external points have to be attached at one of the two vertices, with the remaining points are attached to the other one;
		\item the two vertices have to be linked by a number of internal lines that is limited by the cut-off $L_{max}$, and this number must have the same parity of $r$ (we have at our disposal only vertices with an even number of legs);
	\end{enumerate}
	\item Attach to each vertex with $2i$ legs the effective vertex function $\Pi^{(1)}_{2i}$;
	\item Associate a propagator to each line, integrating over the internal spacetime points;
	\item Associate the combinatorial factor $\frac{1}{m!}$ if the two vertices are connected $m$ times\footnote{As we are considering only distinct diagrams, the factor $\frac{1}{2}$ in (\ref{Gn2Bediag}) and (\ref{Gn2Bodiag}) is cancelled by an exchange factor $2$ for the interchange of the vertices $u \iff w$, as explained in the text above.};
	\item For any $r$ with $0 \leq r \leq \frac{n}{2}-1$, consider the $\binom{n}{r}$ permutations of external points, each corresponding to a different channel. For $r=n/2$ the number of distinct permutations is $\frac{1}{2} \binom{n}{n/2}$.
\end{enumerate}

\subsection{Two-points Green's function}
Let us apply the above results to the two-points Green's function $\mathcal{G}_2$.

Collecting all the $\mathcal{O}(\epsilon^2)$ contributions we have:

\begin{align}\label{G_2^2diag}
\mathcal{G}^{(2)}_2(x_1,x_2) &=\,\,
\begin{tikzpicture}[baseline=-3]
\begin{feynman}
\vertex[draw, circle, minimum size=0.85cm,very thick] (m) at (0, 0) {2};
\vertex (a) at (-1.3,0) {$x_1$};
\vertex (b) at ( 1.3,0) {$x_2$};
\diagram* {
	(a) -- (m) -- (b),
};
\end{feynman}
\end{tikzpicture}
\,\,+\,\,
\sum_{l \geq 0}^{L_{max}}\,
\begin{tikzpicture}[baseline=-3]
\begin{feynman}
\vertex[draw, circle, minimum size=0.85cm,very thick] (m) at (-0.8, 0) {1};
\vertex[draw, circle, minimum size=0.85cm,very thick] (n) at (0.8, 0) {1};
\vertex (a) at (-2.2,0) {$x_1$};
\vertex (b) at ( 2.2,0) {$x_2$};
\diagram* {
	(a) -- (m) -- (n) -- (b);
	(m) -- [out=40, in=140,edge label={\footnotesize $2l+1$}, dashed](n);
	(m) -- [out=-40, in=-140, dashed](n);
};
\end{feynman}
\end{tikzpicture}
\,\,\nonumber\\
&+\,\,
\sum_{l\geq 1}^{L_{max}} \, 
\begin{tikzpicture}[baseline=-3]
\begin{feynman}
\vertex (m)[draw, circle, minimum size=0.85cm,very thick] at (0, 0) {1};
\vertex (n)[draw, circle, minimum size=0.85cm,very thick] at (0, 1.5) {1};
\vertex (a) at (-1.5, 0) {$x_1$};
\vertex (b) at ( 1.5, 0) {$x_2$};
\diagram* {
	(a) -- (m) -- (b);
	(m) -- [out=65, in=-65](n);
	(m) -- [out=115, in=-115](n);
	(m) -- [out=42, in=-42,edge label'={\footnotesize $2l$},dashed](n);
	(m) -- [out=138, in=-138,dashed](n);
};
\end{feynman}
\end{tikzpicture}
\end{align}
where
\begin{eqnarray} \label{G22A}
\begin{tikzpicture}[baseline=-3]
\begin{feynman}
\vertex[draw, circle, minimum size=0.85cm,very thick] (m) at (0, 0) {2};
\vertex (a) at (-1.5,0) {$x_1$};
\vertex (b) at ( 1.5,0) {$x_2$};
\diagram* {
	(a) -- (m) -- (b),
};
\end{feynman}
\end{tikzpicture}&=& \Pi^{(2)}_{2}\, \int d^du \, \Delta(x_1-u)\Delta(x_2-u)
\\ \label{G22C}
\begin{tikzpicture}[baseline=-3]
\begin{feynman}
\vertex[draw, circle, minimum size=0.85cm,very thick] (m) at (-0.8, 0) {1};
\vertex[draw, circle, minimum size=0.85cm,very thick] (n) at (0.8, 0) {1};
\vertex (a) at (-2.1,0) {$x_1$};
\vertex (b) at ( 2.1,0) {$x_2$};
\diagram* {
	(a) -- (m) -- (n) -- (b);
	(m) -- [out=40, in=140,edge label={\footnotesize $2l+1$}, dashed](n);
	(m) -- [out=-40, in=-140, dashed](n);
};
\end{feynman}
\end{tikzpicture} &=&\frac{\Pi^{(1)}_{2+2l}\,\Pi^{(1)}_{2+2l}}{(2l+1)!}\, \int d^du\, d^dw\, \Delta(x_1-u)  \Delta(u-w)^{2l+1} \Delta(x_2-w)\,\,\,
\nonumber\\ \\ \label{G22B}
\begin{tikzpicture}[baseline=-3]
\begin{feynman}[inline=(m)]
\vertex (m)[draw, circle, minimum size=0.85cm,very thick] at (0, 0) {1};
\vertex (n)[draw, circle, minimum size=0.85cm,very thick] at (0, 1.5) {1};
\vertex (a) at (-1.5, 0) {$x_1$};
\vertex (b) at ( 1.5, 0) {$x_2$};
\diagram* {
	(a) -- (m) -- (b);
	(m) -- [out=65, in=-65](n);
	(m) -- [out=115, in=-115](n);
	(m) -- [out=42, in=-42,edge label'={\footnotesize $2l$},dashed](n);
	(m) -- [out=138, in=-138,dashed](n);
};
\end{feynman}
\end{tikzpicture} &=&\frac{\Pi^{(1)}_{2+2l}\,\Pi^{(1)}_{2l}}{(2l)!}\, \int d^du \,d^dw\, \Delta(x_1-u) \Delta(x_2-u) \Delta(u-w)^{2l}
\end{eqnarray}
Going to momentum space, the $\mathcal{O}(\epsilon^2)$ contribution to $\widetilde{\mathcal{G}}_2(p)$ is then:
\begin{align}\label{G_2^2pdiag}
\widetilde{\mathcal{G}}^{(2)}_2(p) \,\,&=\,\,
\begin{tikzpicture}[baseline=-3]
\begin{feynman}
\vertex[draw, circle, minimum size=0.85cm,very thick] (m) at (0, 0) {2};
\vertex (a) at (-1.5,0) {};
\vertex (b) at ( 1.5,0) {};
\diagram* {
	(a) --[fermion] (m) --[fermion] (b),
};
\end{feynman}
\end{tikzpicture}
\,\,+\,\,
\begin{tikzpicture}[baseline=-3]
\begin{feynman}[small]
\vertex[draw, circle, minimum size=0.85cm,very thick] (m) at (-0.8, 0) {1};
\vertex[draw, circle, minimum size=0.85cm,very thick] (n) at (0.8, 0) {1};
\vertex (a) at (-2.2,0) {};
\vertex (b) at ( 2.2,0) {};
\diagram* {
	(a) --[fermion] (m);
	(m) --[fermion] (n);
	(n) --[fermion] (b),
};
\end{feynman}
\end{tikzpicture}
\,\,\nonumber\\
&+\,\,\sum_{l\geq 1}
\begin{tikzpicture}[baseline=-3]
\begin{feynman}[small]
\vertex[draw, circle, minimum size=0.85cm,very thick] (m) at (-0.8, 0) {1};
\vertex[draw, circle, minimum size=0.85cm,very thick] (n) at (0.8, 0) {1};
\vertex (a) at (-2.1,0) {};
\vertex (b) at ( 2.1,0) {};
\diagram* {
	(a) --[fermion] (m);
	(m) --[fermion] (n);
	(m) --[half left, out=50, in=130,edge label={\footnotesize $2l+1$}, dashed](n);
	(m) --[out=30, in=150, fermion](n);
	(m) --[out=-30, in=-150, fermion](n);
	(m) --[half left, out=-50, in=-130, dashed](n);
	(n) --[fermion] (b),
};
\end{feynman}
\end{tikzpicture}
\,\,+\,\,
\sum_{l\geq 1} 
\begin{tikzpicture}[baseline=15]
\begin{feynman}
\vertex (m)[draw, circle, minimum size=0.85cm,very thick] at (0, 0) {1};
\vertex (n)[draw, circle, minimum size=0.85cm,very thick] at (0, 1.5) {1};
\vertex (a) at (-1.5, 0) {};
\vertex (b) at ( 1.5, 0) {};
\diagram* {
	(a) -- [fermion] (m) --[fermion](b);
	(m) -- [fermion, out=65, in=-65](n);
	(m) -- [fermion, out=115, in=-115](n);
	(m) -- [out=42, in=-42,edge label'={\footnotesize $2l$},dashed](n);
	(m) -- [out=138, in=-138,dashed](n);
};
\end{feynman}
\end{tikzpicture}
\end{align}
where
\begin{eqnarray} \label{G22Ap}
\begin{tikzpicture}[baseline=-3]
\begin{feynman}[small]
\vertex[draw, circle, minimum size=0.85cm,very thick] (m) at (0, 0) {2};
\vertex (a) at (-1.5,0) {};
\vertex (b) at ( 1.5,0) {};
\diagram* {
	(a) --[fermion] (m) --[fermion] (b),
};
\end{feynman}
\end{tikzpicture}&=& \frac{1}{p^2+M^2} \,\Pi^{(2)}_{2}\, \frac{1}{p^2+M^2}
\\ \label{G22red}
	\begin{tikzpicture}[baseline=-3]
	\begin{feynman}[small]
	\vertex[draw, circle, minimum size=0.85cm,very thick] (m) at (-0.8, 0) {1};
	\vertex[draw, circle, minimum size=0.85cm,very thick] (n) at (0.8, 0) {1};
	\vertex (a) at (-2.2,0) {};
	\vertex (b) at ( 2.2,0) {};
	\diagram* {
		(a) --[fermion] (m);
		(m) --[fermion] (n);
		(n) --[fermion] (b),
	};
	\end{feynman}
	\end{tikzpicture}&=& \frac{1}{p^2+M^2} \,\Pi^{(1)}_{2}\, \frac{1}{p^2+M^2}\, \Pi^{(1)}_{2}\, \frac{1}{p^2+M^2}
\\ \label{G22Cp}
\begin{tikzpicture}[baseline=-3]
\begin{feynman}[small]
\vertex[draw, circle, minimum size=0.85cm,very thick] (m) at (-0.8, 0) {1};
\vertex[draw, circle, minimum size=0.85cm,very thick] (n) at (0.8, 0) {1};
\vertex (a) at (-2.1,0) {};
\vertex (b) at ( 2.1,0) {};
\diagram* {
	(a) --[fermion] (m);
	(m) --[fermion] (n);
	(m) --[half left, out=50, in=130,edge label={\footnotesize $2l+1$}, dashed](n);
	(m) --[out=30, in=150, fermion](n);
	(m) --[out=-30, in=-150, fermion](n);
	(m) --[half left, out=-50, in=-130, dashed](n);
	(n) --[fermion] (b),
};
\end{feynman}
\end{tikzpicture} &=&\frac{1}{p^2+M^2}\,\left[ \Pi^{(1)}_{2+2l}\,\,\Pi^{(1)}_{2+2l}\,\,\frac{I_{2l+1}(p)}{(2l+1)!}\right]\,\frac{1}{p^2+M^2} \quad(l\geq1) \,\,\,\,\,\,\,\,\,\,\,
\\ \label{G22Bp}
\begin{tikzpicture}[baseline=0.60cm]
\begin{feynman}[small]
\vertex (m)[draw, circle, minimum size=0.85cm,very thick] at (0, 0) {1};
\vertex (n)[draw, circle, minimum size=0.85cm,very thick] at (0, 1.5) {1};
\vertex (a) at (-1.5, 0) {};
\vertex (b) at ( 1.5, 0) {};
\diagram* {
	(a) --[fermion] (m) --[fermion] (b);
	(m) -- [fermion, out=65, in=-65](n);
	(m) -- [fermion, out=115, in=-115](n);
	(m) -- [out=42, in=-42,edge label'={\footnotesize $2l$},dashed](n);
	(m) -- [out=138, in=-138,dashed](n);
};
\end{feynman}
\end{tikzpicture} &=&\frac{1}{p^2+M^2}\,\left[\Pi^{(1)}_{2+2l}\,\Pi^{(1)}_{2l}\,\frac{I_{2l}(0)}{(2l)!}\right]\,\frac{1}{p^2+M^2} \quad(l\geq1)
\end{eqnarray}
and $I_r(k)$ indicates the multi-loop integral
\begin{equation}
	I_r(k)\,=\,\int \frac{d^dp_1}{(2\pi)^d}\dots\frac{d^dp_{r-1}}{(2\pi)^d} \frac{1}{p^2_1+M^2}\dots\frac{1}{p^2_{r-1}+M^2}\frac{1}{(k-\sum_{i=1}^{r-1}p_i)^2+M^2}
\end{equation}
where $r\geq2$ is the number of lines between the vertices, and $k$ is the total momentum carried in the loop.

It is worth to note that the second term in the right-hand side of (\ref{G_2^2pdiag}) (i.e. the diagram in (\ref{G22red})) would correspond to the $l=0$ contribution in the sum that appears in the third term. We wrote it separately as it is a reducible diagram, actually the double iteration of the 1PI diagram previously found at $\mathcal{O}(\epsilon)$ (see Eq.(\ref{G2fourier})). 
Finally we observe that in Eqs.\,(\ref{G22Ap})-(\ref{G22Bp}) the amputated diagrams are obtained once the propagators $\frac{1}{p^2+M^2}$ are removed from their extreme left and right sides.

\subsection{Four-points Green's function}
Collecting as before all the $\mathcal{O}(\epsilon^2)$ contributions we get:
\begin{align}
\mathcal{G}^{(2)}_4(x_1,\dots,x_4) &= \,\,
	\begin{tikzpicture}[baseline=-3]
	\begin{feynman} \small
	\vertex[draw, circle, minimum size=0.8cm,very thick] (m) at (0, 0) {2};
	\vertex (a) at (-1.1,1.1)  {$x_1$};
	\vertex (b) at (-1.1,-1.1) {$x_2$};
	\vertex (c) at (1.1,-1.1)  {$x_3$};
	\vertex (d) at (1.1,1.1)   {$x_4$};
	\diagram* {
		(a) -- (m) -- (b);
		(c) -- (m) -- (d);
	};
	\end{feynman}
	\end{tikzpicture}
	\,\,+\,\,\sum_{l\geq 1}^{L_{max}}
	\begin{tikzpicture}[baseline=0]
	\begin{feynman}[inline=(m)] \small
	\vertex (m)[draw, circle, minimum size=0.8cm,very thick] at (0, 0) {1};
	\vertex (n)[draw, circle, minimum size=0.8cm,very thick] at (0, 1.4) {1};
	\vertex (a) at (-1.46, 0.53) {$x_1$};
	\vertex (b) at (-0.89,-1.27) {$x_2$};
	\vertex (c) at (0.89, -1.27) {$x_3$};
	\vertex (d) at ( 1.46, 0.53) {$x_4$};
	\diagram* {
		(a) -- (m) -- (b);
		(c) -- (m) -- (d);
		(m) -- [out=65, in=-65](n);
		(m) -- [out=115, in=-115](n);
		(m) -- [out=44, in=-44,edge label'={\scriptsize $2l$},dashed](n);
		(m) -- [out=136, in=-136,dashed](n);
		};
	\end{feynman}
	\end{tikzpicture}
	\nonumber\\
	&+\sum_{l\geq0}^{L_{max}}\left\{
	\begin{tikzpicture}[baseline=-3]
	\begin{feynman}\small
	\vertex[draw, circle, minimum size=0.8cm,very thick] (m) at (-0.8, 0) {1};
	\vertex[draw, circle, minimum size=0.8cm,very thick] (n) at (0.7, 0) {1};
	\vertex (a) at (-2.1,0) {$x_1$};
	\vertex (b) at ( 1.7,1.1) {$x_2$};
	\vertex (c) at ( 2.1,0) {$x_3$};
	\vertex (d) at ( 1.7,-1.1) {$x_4$};
	\diagram* {
		(a) -- (m);
		(m) -- (n);
		(m) --[out=35, in=145,edge label={\scriptsize $2l+1$}, dashed](n);
		(m) --[out=-35, in=-145, dashed](n);
		(n) -- (b);
		(n) -- (c);
		(n) -- (d);
	};
	\end{feynman}
	\end{tikzpicture}\,\,
	+\,\, 3 \text{ perm.}	
	\right\}
	\nonumber\\
	&+\sum_{l\geq0}^{L_{max}}\left\{
	\begin{tikzpicture}[baseline=-3]
	\begin{feynman}\footnotesize
	\vertex[draw, circle, minimum size=0.8cm, very thick] (m) at (-0.8, 0) {1};
	\vertex[draw, circle, minimum size=0.8cm, very thick] (n) at (0.8, 0) {1};
	\vertex (a) at (-2.0,1.2)  {$x_1$};
	\vertex (b) at (-2.0,-1.2) {$x_2$};
	\vertex (c) at (2.0,-1.2)  {$x_3$};
	\vertex (d) at (2.0,1.2)   {$x_4$};
	\diagram* {
		(a) -- (m) -- (b);
		(c) -- (n) -- (d);
		(m) --[out=25, in=155](n);
		(m) --[out=-25, in=-155](n);
		(m) --[out=46, in=134,edge label={\scriptsize $2l$},dashed](n);
		(m) --[out=-46, in=-134, dashed](n);
	};
	\end{feynman}
	\end{tikzpicture}\,\,
	+\,\, 2 \text{ perm.}	
	\right\}
\end{align}

Going to momentum space, and writing directly the amputated diagrams we have:

\begin{align}\label{G_4^2pdiag}
\widetilde{\mathcal{G}}^{\,(2)}_{4,{\rm amp}}(p_1,\dots,p_4) &= \,\,
\begin{tikzpicture}[baseline=-3]
\begin{feynman} \small
\vertex[draw, circle, minimum size=0.8cm,very thick] (m) at (0, 0) {2};
\vertex (a) at (-0.9,0.9)  ;
\vertex (a1) at (-0.85,0.55);
\vertex (a2) at (-0.55,0.85);
\vertex (b) at (-0.9,-0.9) ;
\vertex (b1) at (-0.85,-0.55);
\vertex (b2) at (-0.55,-0.85);
\vertex (c) at (0.9,-0.9)  ;
\vertex (c1) at (0.85,-0.55);
\vertex (c2) at (0.55,-0.85);
\vertex (d) at (0.9,0.9)   ;
\vertex (d1) at (0.85,0.55);
\vertex (d2) at (0.55,0.85);
\diagram* {
	(a) -- (m) -- (b);
	(c) -- (m) -- (d);
};
\end{feynman}
\draw (a1) to (a2);
\draw (b1) to (b2);
\draw (c1) to (c2);
\draw (d1) to (d2);
\end{tikzpicture}
\,\,+\,\,
\left\{
\begin{tikzpicture}[baseline=-3]
\begin{feynman}\small
\vertex[draw, circle, minimum size=0.8cm,very thick] (m) at (-0.9, 0) {1};
\vertex[draw, circle, minimum size=0.8cm,very thick] (n) at (0.7, 0) {1};
\vertex (a) at (-2.2,0) ;
\vertex (a1) at (-1.94,-0.22);
\vertex (a2) at (-1.94,0.22);
\vertex (b) at ( 1.76,1.06);
\vertex (b1) at (1.392,0.980);
\vertex (b2) at (1.704,0.669);
\vertex (c) at ( 2.2,0) ;
\vertex (c1) at (1.87,-0.22);
\vertex (c2) at (1.87,0.22);
\vertex (d) at ( 1.76,-1.06);
\vertex (d1) at (1.392,-0.980);
\vertex (d2) at (1.704,-0.669);
\diagram* {
	(a) -- (m);
	(m) -- (n);
	(n) -- (b);
	(n) -- (c);
	(n) -- (d);
};
\end{feynman}
\draw (a1) to (a2);
\draw (b1) to (b2);
\draw (c1) to (c2);
\draw (d1) to (d2);
\end{tikzpicture}\,\,
+\,\, 3 \text{ perm.}	
\right\}
\nonumber \\
&+\sum_{l\geq1}^{L_{max}}\left\{
\begin{tikzpicture}[baseline=-3]
\begin{feynman}\small
\vertex[draw, circle, minimum size=0.8cm,very thick] (m) at (-0.9, 0) {1};
\vertex[draw, circle, minimum size=0.8cm,very thick] (n) at (0.7, 0) {1};
\vertex (a) at (-2.2,0) ;
\vertex (a1) at (-1.94,-0.22);
\vertex (a2) at (-1.94,0.22);
\vertex (b) at ( 1.76,1.06);
\vertex (b1) at (1.392,0.980);
\vertex (b2) at (1.704,0.669);
\vertex (c) at ( 2.2,0) ;
\vertex (c1) at (1.87,-0.22);
\vertex (c2) at (1.87,0.22);
\vertex (d) at ( 1.76,-1.06);
\vertex (d1) at (1.392,-0.980);
\vertex (d2) at (1.704,-0.669);
\diagram* {
	(a) -- (m);
	(m) -- (n);
	(m) --[out=35, in=145,edge label={\scriptsize $2l+1$}, dashed](n);
	(m) --[out=-35, in=-145, dashed](n);
	(n) -- (b);
	(n) -- (c);
	(n) -- (d);
};
\end{feynman}
\draw (a1) to (a2);
\draw (b1) to (b2);
\draw (c1) to (c2);
\draw (d1) to (d2);
\end{tikzpicture}\,\,
+\,\, 3 \text{ perm.}	
\right\}
\nonumber\\
\,\,&+\,\,\sum_{l\geq 1}^{L_{max}}\,\,
\begin{tikzpicture}[baseline=0]
\begin{feynman}[inline=(m)] \small
\vertex (m)[draw, circle, minimum size=0.8cm,very thick] at (0, 0) {1};
\vertex (n)[draw, circle, minimum size=0.8cm,very thick] at (0, 1.4) {1};
\vertex (a) at (-1.35, 0.48) ;
\vertex (a1) at (-1.12, 0.17) ;
\vertex (a2) at (-0.98, 0.58);
\vertex (b) at (-0.817,-1.1615) ;
\vertex (b1) at (-0.82,-0.78);
\vertex (b2) at (-0.46,-1.03);
\vertex (c) at (0.817, -1.1615) ;
\vertex (c1) at (0.82,-0.78);
\vertex (c2) at (0.46,-1.03);
\vertex (d) at ( 1.35, 0.48) ;
\vertex (d1) at (1.12, 0.17) ;
\vertex (d2) at (0.98, 0.58);
\diagram* {
	(a) -- (m) -- (b);
	(c) -- (m) -- (d);
	(m) -- [out=65, in=-65](n);
	(m) -- [out=115, in=-115](n);
	(m) -- [out=44, in=-44,edge label'={\scriptsize $2l$},dashed](n);
	(m) -- [out=136, in=-136,dashed](n);
};
\end{feynman}
\draw (a1) to (a2);
\draw (b1) to (b2);
\draw (c1) to (c2);
\draw (d1) to (d2);
\end{tikzpicture}
\,+\,\sum_{l\geq0}^{L_{max}}\left\{
\begin{tikzpicture}[baseline=-3]
\begin{feynman}\footnotesize
\vertex[draw, circle, minimum size=0.8cm, very thick] (m) at (-0.8, 0) {1};
\vertex[draw, circle, minimum size=0.8cm, very thick] (n) at (0.8, 0) {1};
\vertex (a) at (-1.7,0.9)  ;
\vertex (a1) at (-1.65,0.55);
\vertex (a2) at (-1.35,0.85);
\vertex (b) at (-1.7,-0.9) ;
\vertex (b1) at (-1.65,-0.55);
\vertex (b2) at (-1.35,-0.85);
\vertex (c) at (1.7,-0.9)  ;
\vertex (c1) at (1.65,-0.55);
\vertex (c2) at (1.35,-0.85);
\vertex (d) at (1.7,0.9)   ;
\vertex (d1) at (1.65,0.55);
\vertex (d2) at (1.35,0.85);
\diagram* {
	(a) -- (m) -- (b);
	(c) -- (n) -- (d);
	(m) --[out=25, in=155](n);
	(m) --[out=-25, in=-155](n);
	(m) --[out=46, in=134,edge label={\scriptsize $2l$},dashed](n);
	(m) --[out=-46, in=-134, dashed](n);
};
\end{feynman}
\draw (a1) to (a2);
\draw (b1) to (b2);
\draw (c1) to (c2);
\draw (d1) to (d2);
\end{tikzpicture}\,\,
+ \,\, 2 \text{ perm.}	
\right\}
\end{align}
where
\begin{equation}
\label{G24p1}
\begin{tikzpicture}[baseline=-3]
\begin{feynman} \small
\vertex[draw, circle, minimum size=0.8cm,very thick] (m) at (0, 0) {2};
\vertex (a) at (-0.9,0.9)  ;
\vertex (a1) at (-0.85,0.55);
\vertex (a2) at (-0.55,0.85);
\vertex (b) at (-0.9,-0.9) ;
\vertex (b1) at (-0.85,-0.55);
\vertex (b2) at (-0.55,-0.85);
\vertex (c) at (0.9,-0.9)  ;
\vertex (c1) at (0.85,-0.55);
\vertex (c2) at (0.55,-0.85);
\vertex (d) at (0.9,0.9)   ;
\vertex (d1) at (0.85,0.55);
\vertex (d2) at (0.55,0.85);
\diagram* {
	(a) -- (m) -- (b);
	(c) -- (m) -- (d);
};
\end{feynman}
\draw (a1) to (a2);
\draw (b1) to (b2);
\draw (c1) to (c2);
\draw (d1) to (d2);
\end{tikzpicture}
\,\,=\,\,\Pi^{(2)}_4
\end{equation}
\begin{eqnarray} \label{G24pred}
\begin{tikzpicture}[baseline=-3]
\begin{feynman}\small
\vertex[draw, circle, minimum size=0.8cm,very thick] (m) at (-0.9, 0) {1};
\vertex[draw, circle, minimum size=0.8cm,very thick] (n) at (0.7, 0) {1};
\vertex (a) at (-2.2,0) ;
\vertex (a1) at (-1.94,-0.22);
\vertex (a2) at (-1.94,0.22);
\vertex (b) at ( 1.76,1.06);
\vertex (b1) at (1.392,0.980);
\vertex (b2) at (1.704,0.669);
\vertex (c) at ( 2.2,0) ;
\vertex (c1) at (1.87,-0.22);
\vertex (c2) at (1.87,0.22);
\vertex (d) at ( 1.76,-1.06);
\vertex (d1) at (1.392,-0.980);
\vertex (d2) at (1.704,-0.669);
\diagram* {
	(a) -- (m);
	(m) -- (n);
	(n) -- (b);
	(n) -- (c);
	(n) -- (d);
};
\end{feynman}
\draw (a1) to (a2);
\draw (b1) to (b2);
\draw (c1) to (c2);
\draw (d1) to (d2);
\end{tikzpicture}
&=&\Pi^{(1)}_{2}\,\frac{1}{p^2+M^2}\,\Pi^{(1)}_{4}
\\ \label{G24p3}
\begin{tikzpicture}[baseline=-3]
\begin{feynman}\small
\vertex[draw, circle, minimum size=0.8cm,very thick] (m) at (-0.9, 0) {1};
\vertex[draw, circle, minimum size=0.8cm,very thick] (n) at (0.7, 0) {1};
\vertex (a) at (-2.2,0) ;
\vertex (a1) at (-1.94,-0.22);
\vertex (a2) at (-1.94,0.22);
\vertex (b) at ( 1.76,1.06);
\vertex (b1) at (1.392,0.980);
\vertex (b2) at (1.704,0.669);
\vertex (c) at ( 2.2,0) ;
\vertex (c1) at (1.87,-0.22);
\vertex (c2) at (1.87,0.22);
\vertex (d) at ( 1.76,-1.06);
\vertex (d1) at (1.392,-0.980);
\vertex (d2) at (1.704,-0.669);
\diagram* {
	(a) -- (m);
	(m) -- (n);
	(m) --[out=35, in=145,edge label={\scriptsize $2l+1$}, dashed](n);
	(m) --[out=-35, in=-145, dashed](n);
	(n) -- (b);
	(n) -- (c);
	(n) -- (d);
};
\end{feynman}
\draw (a1) to (a2);
\draw (b1) to (b2);
\draw (c1) to (c2);
\draw (d1) to (d2);
\end{tikzpicture}
&=&\Pi^{(1)}_{2+2l}\,\Pi^{(1)}_{4+2l}\,\,I_{2l}(p)
\\ \label{G24p2}
\begin{tikzpicture}[baseline=0]
\begin{feynman}[inline=(m)] \small
\vertex (m)[draw, circle, minimum size=0.8cm,very thick] at (0, 0) {1};
\vertex (n)[draw, circle, minimum size=0.8cm,very thick] at (0, 1.4) {1};
\vertex (a) at (-1.35, 0.48) ;
\vertex (a1) at (-1.12, 0.17) ;
\vertex (a2) at (-0.98, 0.58);
\vertex (b) at (-0.817,-1.1615) ;
\vertex (b1) at (-0.82,-0.78);
\vertex (b2) at (-0.46,-1.03);
\vertex (c) at (0.817, -1.1615) ;
\vertex (c1) at (0.82,-0.78);
\vertex (c2) at (0.46,-1.03);
\vertex (d) at ( 1.35, 0.48) ;
\vertex (d1) at (1.12, 0.17) ;
\vertex (d2) at (0.98, 0.58);
\diagram* {
	(a) -- (m) -- (b);
	(c) -- (m) -- (d);
	(m) -- [out=65, in=-65](n);
	(m) -- [out=115, in=-115](n);
	(m) -- [out=44, in=-44,edge label'={\scriptsize $2l$},dashed](n);
	(m) -- [out=136, in=-136,dashed](n);
};
\end{feynman}
\draw (a1) to (a2);
\draw (b1) to (b2);
\draw (c1) to (c2);
\draw (d1) to (d2);
\end{tikzpicture}&=&\Pi^{(1)}_{4+2l}\,\Pi^{(1)}_{2l} I_{2l}(0)
\\ \label{G24p4}
\begin{tikzpicture}[baseline=-3]
\begin{feynman}\footnotesize
\vertex[draw, circle, minimum size=0.8cm, very thick] (m) at (-0.8, 0) {1};
\vertex[draw, circle, minimum size=0.8cm, very thick] (n) at (0.8, 0) {1};
\vertex (a) at (-1.7,0.9)  ;
\vertex (a1) at (-1.65,0.55);
\vertex (a2) at (-1.35,0.85);
\vertex (b) at (-1.7,-0.9) ;
\vertex (b1) at (-1.65,-0.55);
\vertex (b2) at (-1.35,-0.85);
\vertex (c) at (1.7,-0.9)  ;
\vertex (c1) at (1.65,-0.55);
\vertex (c2) at (1.35,-0.85);
\vertex (d) at (1.7,0.9)   ;
\vertex (d1) at (1.65,0.55);
\vertex (d2) at (1.35,0.85);
\diagram* {
	(a) -- (m) -- (b);
	(c) -- (n) -- (d);
	(m) --[out=25, in=155](n);
	(m) --[out=-25, in=-155](n);
	(m) --[out=46, in=134,edge label={\scriptsize $2l$},dashed](n);
	(m) --[out=-46, in=-134, dashed](n);
};
\end{feynman}
\draw (a1) to (a2);
\draw (b1) to (b2);
\draw (c1) to (c2);
\draw (d1) to (d2);
\end{tikzpicture}&=&\Pi^{(1)}_{2+2l}\,\Pi^{(1)}_{2+2l}\,\,I_{2l}(q)
\end{eqnarray}
where the momentum $p$ carried in the propagator in (\ref{G24pred}) and in the loop integral in (\ref{G24p3}) is one of the external momenta $p_1,\,p_2,\,p_3$ or $p_4$, depending on the specific permutation that we are considering, while the momentum $q$ in (\ref{G24p4}) is $p_1+p_2,\, p_1+p_3$ or $p_1+p_4$, depending on the channel $s$, $t$ or $u$ considered.

As above we observe that the third term in the right-hand side of (\ref{G_4^2pdiag}) (i.e. the diagram in (\ref{G24pred})) would correspond to the $l=0$ contribution in the sum that appears in the fourth term of the same equation. As before, we have not included it in that sum as it is a reducible diagram, where one of the external lines (propagators) is corrected by the $\mathcal{O}(\epsilon)$ bubble (effective vertex) $\Pi^{(1)}_2$ (see Eq.(\ref{Pi_2^1})).

\subsection{Cut-off dependence of the Green's functions in $d=4$}
In order to get a better insight on the cut-off dependence of the radiative corrections to the tree level propagator ${\mathcal{G}^{(0)}_2(x_1,x_2) =\Delta(x_1-x_2)}$, it is worth to consider the familiar $d=4$ case. The four-points Green's function $\mathcal{G}_4$ will be also considered, and the results of this section will be useful for our later purposes.

First of all we note that in the contribution of order $\mathcal{O}(\epsilon^2 g)$, that is in the diagram (\ref{G22Ap}), the dependence on the cut-off is contained in the effective vertex $\Pi^{(2)}_2$ given in Eq.\,(\ref{Pi_2^2}), whose leading divergence, due to the factor $\log(\Delta(0))$ present in $K$\ (defined in (\ref{K})), is:
\begin{equation} \label{G22div1}
\Pi^{(2)}_2 \equiv \begin{tikzpicture}[baseline=-3]
\begin{feynman}
\vertex[draw, circle, minimum size=0.85cm,very thick] (m) at (0, 0) {2};
\vertex (a) at (-1.24421,0) {};
\vertex (b) at (1.24421,0) {};
\vertex (d1) at (0.595294,-0.308824) {};
\vertex (d2) at (0.924706,0.308824) {};
\vertex (d3) at (-0.924706,-0.308824) {};
\vertex (d4) at (-0.595294,0.308824) {};
\diagram* {
	(a) -- (m) -- (b);
};
\draw (d1) to (d2);
\draw (d3) to (d4);
\end{feynman}
\end{tikzpicture} \,=\, -\frac{\epsilon^2}{2}\,g\,\mu^2 \left[K^2 - 1 + \psi'\left(\frac{3}{2}\right)\right]\,\,\,\sim\,\, \left(\log\Lambda\right)^2\,.
\end{equation}

Moving to $\mathcal{O}(\epsilon^2 g^2)$, we note that the reducible diagram in (\ref{G22red}) clearly diverges as $\left(\log\Lambda\right)^2$, as it is the square of the 1PI $\mathcal{O}(\epsilon)$ diagram that, as we know, diverges as $\log\Lambda$
\begin{align} \label{G22div2}
\begin{tikzpicture}[baseline=-3]
\begin{feynman}[small]
\vertex[draw, circle, minimum size=0.85cm,very thick] (m) at (-0.8, 0) {1};
\vertex[draw, circle, minimum size=0.85cm,very thick] (n) at (0.8, 0) {1};
\vertex (a) at (-2.1,0) {};
\vertex (b) at ( 2.1,0) {};
\vertex (d1) at (1.545294,-0.308824) {};
\vertex (d2) at (1.874706,0.308824) {};
\vertex (d3) at (-1.874706,-0.308824) {};
\vertex (d4) at (-1.545294,0.308824) {};
\diagram* {
	(a) -- (m);
	(m) -- (n);
	(n) -- (b),
};
\end{feynman}
\draw (d1) to (d2);
\draw (d3) to (d4);
\end{tikzpicture}\,\,\sim\,\,\left(\log\Lambda\right)^2\,.
\end{align}
Concerning all the other contributions, the dependence on $\Lambda$ is contained in the two effective vertices and in the loop integrals.
For the $\Pi^{(1)}$'s we know the exact cut-off dependence, that is given by the $\Delta(0)$ factors:
\begin{align}
	\Pi^{(1)}_{2}\,&\equiv\,
	\begin{tikzpicture}[baseline=-3]
	\begin{feynman}
	\vertex[draw, circle, minimum size=0.85cm,very thick] (m) at (0, 0) {1};
	\vertex (a) at (-1.24421,0) {};
	\vertex (b) at (1.24421,0) {};
	\vertex (d1) at (0.595294,-0.308824) {};
	\vertex (d2) at (0.924706,0.308824) {};
	\vertex (d3) at (-0.924706,-0.308824) {};
	\vertex (d4) at (-0.595294,0.308824) {};
	\diagram* {
		(a) -- (m) -- (b);
	};
	\draw (d1) to (d2);
	\draw (d3) to (d4);
	\end{feynman}
	\end{tikzpicture} \,=\, -\,\epsilon\,g\,\mu^2\, K \,\,\,\sim\,\, \log(\Lambda)  \label{Pi12div}\\
	\Pi^{(1)}_{n}\,&\equiv\,	\begin{tikzpicture}[baseline=-3]
	\begin{feynman}
	\vertex[draw, circle, minimum size=0.85cm,very thick] (m) at (0, 0) {1};
	\vertex (a) at (-0.9,0.9) {\small$2$};
	\vertex (b) at (-0.9,-0.9) {\small$1$};
	\vertex (c) at (1.24421,0) {\small$n$};
	\vertex (l1) at (0.5,0.35) {};
	\vertex (l2) at (-0.2,0.55) {};
	\vertex (d1) at (0.595294,-0.308824) {};
	\vertex (d2) at (0.924706,0.308824) {};
	\vertex (d3) at (-0.700391,0.226862) {};
	\vertex (d4) at (-0.370979,0.844509) {};
	\vertex (d5) at (-0.700391,-0.226862) {};
	\vertex (d6) at (-0.370979,-0.844509) {};
	\diagram* {
		(a) -- (m) -- (b);
		(m) -- (c);
		(l1) -- [thick,dotted,out=-280, in=-305](l2);
	};
	\draw (d1) to (d2);
	\draw (d3) to (d4);
	\draw (d5) to (d6);
	\end{feynman}
	\end{tikzpicture}\,= \,(-1)^{\frac{n}{2}-1}2^{\frac{n}{2}-1}\,\left(\frac{n}{2}-2\right)!\frac{\epsilon\,g\,\mu^2}{\Delta(0)^{\frac{n}{2}-1}}\,\,\,\sim\,\,\frac{1}{\Lambda^{n-2}} \qquad(n\geq4) \,.\label{Pi1ndiv}
\end{align} 

As for the loop integrals, although we do not know their exact regularized expressions, we can calculate their superficial degree of divergence, thus getting:
\begin{align}
	I_2(k)\,\,&\sim\,\, \log(\Lambda)    \\
	I_r(k)\,\,&\sim\,\, \Lambda^{2(r-2)} \qquad (r>2) \label{Irk}
\end{align}

Collecting the above results, the cut-off dependence of the amputated diagrams corresponding to the second line of (\ref{G_2^2pdiag}), that are the only $\mathcal{O}(\epsilon^2 g^2)$ 1PI diagrams, turns out to be

\begin{eqnarray} \label{G22div3}
\begin{tikzpicture}[baseline=0.60cm]
\begin{feynman}[small]
\vertex (m)[draw, circle, minimum size=0.85cm,very thick] at (0, 0) {1};
\vertex (n)[draw, circle, minimum size=0.85cm,very thick] at (0, 1.5) {1};
\vertex (a) at (-1.5, 0) {};
\vertex (b) at ( 1.5, 0) {};
\vertex (d1) at (0.745294,-0.308824) {};
\vertex (d2) at (1.074706,0.308824) {};
\vertex (d3) at (-1.074706,-0.308824) {};
\vertex (d4) at (-0.745294,0.308824) {};

\diagram* {
	(a) -- (m) -- (b);
	(m) -- [out=65, in=-65](n);
	(m) -- [out=115, in=-115](n);
};
\end{feynman}
\draw (d1) to (d2);
\draw (d3) to (d4);
\end{tikzpicture}&\sim& \frac{\left(\log(\Lambda)\right)^2}{\Lambda^2}
\\ \label{G22div4}
\begin{tikzpicture}[baseline=0.60cm]
\begin{feynman}[small]
\vertex (m)[draw, circle, minimum size=0.85cm,very thick] at (0, 0) {1};
\vertex (n)[draw, circle, minimum size=0.85cm,very thick] at (0, 1.5) {1};
\vertex (a) at (-1.5, 0) {};
\vertex (b) at ( 1.5, 0) {};
\vertex (d1) at (0.745294,-0.308824) {};
\vertex (d2) at (1.074706,0.308824) {};
\vertex (d3) at (-1.074706,-0.308824) {};
\vertex (d4) at (-0.745294,0.308824) {};
\diagram* {
	(a) -- (m) -- (b);
	(m) -- [out=65, in=-65](n);
	(m) -- [out=115, in=-115](n);
	(m) -- [out=42, in=-42,edge label'={\footnotesize $2l$},dashed](n);
	(m) -- [out=138, in=-138,dashed](n);
};
\end{feynman}
\draw (d1) to (d2);
\draw (d3) to (d4);
\end{tikzpicture}&\sim& \frac{1}{\Lambda^2} \qquad \forall l\geq2
\\\label{G22div5}
\begin{tikzpicture}[baseline=-3]
\begin{feynman}[small]
\vertex[draw, circle, minimum size=0.85cm,very thick] (m) at (-0.8, 0) {1};
\vertex[draw, circle, minimum size=0.85cm,very thick] (n) at (0.8, 0) {1};
\vertex (a) at (-2.1,0) {};
\vertex (b) at ( 2.1,0) {};
\vertex (d1) at (1.395294,-0.308824) {};
\vertex (d2) at (1.724706,0.308824) {};
\vertex (d3) at (-1.724706,-0.308824) {};
\vertex (d4) at (-1.395294,0.308824) {};
\diagram* {
	(a) -- (m);
	(m) -- (n);
	(m) --[half left, out=50, in=130,edge label={\footnotesize $2l+1$}, dashed](n);
	(m) --[out=30, in=150](n);
	(m) --[out=-30, in=-150](n);
	(m) --[half left, out=-50, in=-130, dashed](n);
	(n) -- (b),
};
\end{feynman}
\draw (d1) to (d2);
\draw (d3) to (d4);
\end{tikzpicture}&\sim& \frac{1}{\Lambda^2} \qquad \forall l\geq2
\end{eqnarray}

A simple inspection of Eqs.(\ref{G22div3})-(\ref{G22div5}) and (\ref{G22div1}) shows that among the 1PI terms the dominant one is the $\mathcal{O}(g)$ diagram (\ref{G22div1}). The only other $\mathcal{O}(\epsilon^2)$ diagram that is of the same order in $\Lambda$ is the 1P reducible diagram (\ref{G22div2}). We note that the degree of divergence of the $\mathcal{O}(\epsilon^2)$ contribution is higher than the one of $\mathcal{O}(\epsilon)$: $\left(\log(\Lambda)\right)^2$ as compared to $\log(\Lambda)$. 

These observations will be useful when we will later consider resummations of diagrams.

Similar considerations can be made for the $\mathcal{G}_n$'s with higher $n$. In particular we concentrate on the four-point Green's function $\mathcal{G}_4$.
First of all we notice that in the contribution of order $\mathcal{O}(\epsilon^2 g)$, that is the diagram (\ref{G24p1}), the dependence on the cut-off is contained in the effective vertex $\Pi^{(2)}_4$, and we have
\begin{equation} \label{G241div}
\Pi^{(2)}_4 \,\equiv\, \begin{tikzpicture}[baseline=-3]
\begin{feynman} \small
\vertex[draw, circle, minimum size=0.8cm,very thick] (m) at (0, 0) {2};
\vertex (a) at (-0.9,0.9)  ;
\vertex (a1) at (-0.85,0.55);
\vertex (a2) at (-0.55,0.85);
\vertex (b) at (-0.9,-0.9) ;
\vertex (b1) at (-0.85,-0.55);
\vertex (b2) at (-0.55,-0.85);
\vertex (c) at (0.9,-0.9)  ;
\vertex (c1) at (0.85,-0.55);
\vertex (c2) at (0.55,-0.85);
\vertex (d) at (0.9,0.9)   ;
\vertex (d1) at (0.85,0.55);
\vertex (d2) at (0.55,0.85);
\diagram* {
	(a) -- (m) -- (b);
	(c) -- (m) -- (d);
};
\end{feynman}
\draw (a1) to (a2);
\draw (b1) to (b2);
\draw (c1) to (c2);
\draw (d1) to (d2);
\end{tikzpicture}
 = - 2 \epsilon^2 g\mu^2 \frac{K}{\Delta(0)}\,\,\,\sim\,\,\frac{\log(\Lambda)}{\Lambda^2}\,.
\end{equation}  

Moving to the $\mathcal{O}(\epsilon^2 g^2)$, the amputated 1P reducible diagram (\ref{G24pred}) has the cut-off dependence    
\begin{equation}
\begin{tikzpicture}[baseline=-3]
\begin{feynman}\small
\vertex[draw, circle, minimum size=0.8cm,very thick] (m) at (-0.9, 0) {1};
\vertex[draw, circle, minimum size=0.8cm,very thick] (n) at (0.7, 0) {1};
\vertex (a) at (-2.2,0) ;
\vertex (a1) at (-1.94,-0.22);
\vertex (a2) at (-1.94,0.22);
\vertex (b) at ( 1.76,1.06);
\vertex (b1) at (1.392,0.980);
\vertex (b2) at (1.704,0.669);
\vertex (c) at ( 2.2,0) ;
\vertex (c1) at (1.87,-0.22);
\vertex (c2) at (1.87,0.22);
\vertex (d) at ( 1.76,-1.06);
\vertex (d1) at (1.392,-0.980);
\vertex (d2) at (1.704,-0.669);
\diagram* {
	(a) -- (m);
	(m) -- (n);
	(n) -- (b);
	(n) -- (c);
	(n) -- (d);
};
\end{feynman}
\draw (a1) to (a2);
\draw (b1) to (b2);
\draw (c1) to (c2);
\draw (d1) to (d2);
\end{tikzpicture}
\,\,\sim\,\,\frac{\log(\Lambda)}{\Lambda^2}\,,
\end{equation}
being the product of the two subdiagrams $\Pi^{(1)}_2$ and $\Pi^{(1)}_4$ (see Eqs.\,(\ref{Pi12div}) and (\ref{Pi1ndiv})).

Concerning all the other contributions, the dependence on $\Lambda$ is contained in the two effective vertices and in the loop integrals. Making use of Eqs.\,(\ref{Pi12div})-(\ref{Irk}), their cut-off dependence turns out to be

\begin{eqnarray}
\begin{tikzpicture}[baseline=-3]
\begin{feynman}\small
\vertex[draw, circle, minimum size=0.8cm,very thick] (m) at (-0.9, 0) {1};
\vertex[draw, circle, minimum size=0.8cm,very thick] (n) at (0.7, 0) {1};
\vertex (a) at (-2.2,0) ;
\vertex (a1) at (-1.94,-0.22);
\vertex (a2) at (-1.94,0.22);
\vertex (b) at ( 1.76,1.06);
\vertex (b1) at (1.392,0.980);
\vertex (b2) at (1.704,0.669);
\vertex (c) at ( 2.2,0) ;
\vertex (c1) at (1.87,-0.22);
\vertex (c2) at (1.87,0.22);
\vertex (d) at ( 1.76,-1.06);
\vertex (d1) at (1.392,-0.980);
\vertex (d2) at (1.704,-0.669);
\diagram* {
	(a) -- (m);
	(m) -- (n);
	(m) --[out=35, in=145,edge label={\scriptsize $2l+1$}, dashed](n);
	(m) --[out=-35, in=-145, dashed](n);
	(n) -- (b);
	(n) -- (c);
	(n) -- (d);
};
\end{feynman}
\draw (a1) to (a2);
\draw (b1) to (b2);
\draw (c1) to (c2);
\draw (d1) to (d2);
\end{tikzpicture}\,\,&\sim&\,\,\frac{1}{\Lambda^4} \qquad \forall l\geq1
\\
\begin{tikzpicture}[baseline=0]
\begin{feynman}[inline=(m)] \small
\vertex (m)[draw, circle, minimum size=0.8cm,very thick] at (0, 0) {1};
\vertex (n)[draw, circle, minimum size=0.8cm,very thick] at (0, 1.4) {1};
\vertex (a) at (-1.35, 0.48) ;
\vertex (a1) at (-1.12, 0.17) ;
\vertex (a2) at (-0.98, 0.58);
\vertex (b) at (-0.817,-1.1615) ;
\vertex (b1) at (-0.82,-0.78);
\vertex (b2) at (-0.46,-1.03);
\vertex (c) at (0.817, -1.1615) ;
\vertex (c1) at (0.82,-0.78);
\vertex (c2) at (0.46,-1.03);
\vertex (d) at ( 1.35, 0.48) ;
\vertex (d1) at (1.12, 0.17) ;
\vertex (d2) at (0.98, 0.58);
\diagram* {
	(a) -- (m) -- (b);
	(c) -- (m) -- (d);
	(m) -- [out=65, in=-65](n);
	(m) -- [out=115, in=-115](n);
};
\end{feynman}
\draw (a1) to (a2);
\draw (b1) to (b2);
\draw (c1) to (c2);
\draw (d1) to (d2);
\end{tikzpicture}
\,\,&\sim&\,\,\frac{\left(\log(\Lambda)\right)^2}{\Lambda^4}
\\
\begin{tikzpicture}[baseline=0]
\begin{feynman}[inline=(m)] \small
\vertex (m)[draw, circle, minimum size=0.8cm,very thick] at (0, 0) {1};
\vertex (n)[draw, circle, minimum size=0.8cm,very thick] at (0, 1.4) {1};
\vertex (a) at (-1.35, 0.48) ;
\vertex (a1) at (-1.12, 0.17) ;
\vertex (a2) at (-0.98, 0.58);
\vertex (b) at (-0.817,-1.1615) ;
\vertex (b1) at (-0.82,-0.78);
\vertex (b2) at (-0.46,-1.03);
\vertex (c) at (0.817, -1.1615) ;
\vertex (c1) at (0.82,-0.78);
\vertex (c2) at (0.46,-1.03);
\vertex (d) at ( 1.35, 0.48) ;
\vertex (d1) at (1.12, 0.17) ;
\vertex (d2) at (0.98, 0.58);
\diagram* {
	(a) -- (m) -- (b);
	(c) -- (m) -- (d);
	(m) -- [out=65, in=-65](n);
	(m) -- [out=115, in=-115](n);
	(m) -- [out=44, in=-44,edge label'={\scriptsize $2l$},dashed](n);
	(m) -- [out=136, in=-136,dashed](n);
};
\end{feynman}
\draw (a1) to (a2);
\draw (b1) to (b2);
\draw (c1) to (c2);
\draw (d1) to (d2);
\end{tikzpicture}
\,\,&\sim&\,\,\frac{1}{\Lambda^4} \qquad \forall l\geq2
\\
\begin{tikzpicture}[baseline=-3]
\begin{feynman}\footnotesize
\vertex[draw, circle, minimum size=0.8cm, very thick] (m) at (-0.8, 0) {1};
\vertex[draw, circle, minimum size=0.8cm, very thick] (n) at (0.8, 0) {1};
\vertex (a) at (-1.7,0.9)  ;
\vertex (a1) at (-1.65,0.55);
\vertex (a2) at (-1.35,0.85);
\vertex (b) at (-1.7,-0.9) ;
\vertex (b1) at (-1.65,-0.55);
\vertex (b2) at (-1.35,-0.85);
\vertex (c) at (1.7,-0.9)  ;
\vertex (c1) at (1.65,-0.55);
\vertex (c2) at (1.35,-0.85);
\vertex (d) at (1.7,0.9)   ;
\vertex (d1) at (1.65,0.55);
\vertex (d2) at (1.35,0.85);
\diagram* {
	(a) -- (m) -- (b);
	(c) -- (n) -- (d);
	(m) --[out=25, in=155](n);
	(m) --[out=-25, in=-155](n);
};
\end{feynman}
\draw (a1) to (a2);
\draw (b1) to (b2);
\draw (c1) to (c2);
\draw (d1) to (d2);
\end{tikzpicture}\,\,&\sim&\,\, \frac{\log(\Lambda)}{\Lambda^4}
\\
\begin{tikzpicture}[baseline=-3]
\begin{feynman}\footnotesize
\vertex[draw, circle, minimum size=0.8cm, very thick] (m) at (-0.8, 0) {1};
\vertex[draw, circle, minimum size=0.8cm, very thick] (n) at (0.8, 0) {1};
\vertex (a) at (-1.7,0.9)  ;
\vertex (a1) at (-1.65,0.55);
\vertex (a2) at (-1.35,0.85);
\vertex (b) at (-1.7,-0.9) ;
\vertex (b1) at (-1.65,-0.55);
\vertex (b2) at (-1.35,-0.85);
\vertex (c) at (1.7,-0.9)  ;
\vertex (c1) at (1.65,-0.55);
\vertex (c2) at (1.35,-0.85);
\vertex (d) at (1.7,0.9)   ;
\vertex (d1) at (1.65,0.55);
\vertex (d2) at (1.35,0.85);
\diagram* {
	(a) -- (m) -- (b);
	(c) -- (n) -- (d);
	(m) --[out=25, in=155](n);
	(m) --[out=-25, in=-155](n);
	(m) --[out=46, in=134,edge label={\scriptsize $2l$},dashed](n);
	(m) --[out=-46, in=-134, dashed](n);
};
\end{feynman}
\draw (a1) to (a2);
\draw (b1) to (b2);
\draw (c1) to (c2);
\draw (d1) to (d2);
\end{tikzpicture}
\,\,&\sim &\,\, \frac{1}{\Lambda^4} \qquad \forall l\geq2
\end{eqnarray}

Similarly to what we have seen for the two-points function, leaving aside the reducible diagram, the leading term is the $\mathcal{O}(g)$ diagram (\ref{G241div}). As compared to the $\mathcal{O}(\epsilon)$ result (see Eq.\,(\ref{Gnzero}) for $n=4$), the cut-off dependence of this $\mathcal{O}(\epsilon^2)$ contribution to the radiative correction to $\mathcal{G}_4$ is enhanced (less vanishing) with respect to the lower order result: $\frac{\log \Lambda}{\Lambda^2}$ as compared to $\frac{1}{\Lambda^2}$.

As stressed above for $\mathcal{G}_2$, these observations will be crucial for our later developments, where we will be interested in resummations of selected classes of diagrams at each order in $\epsilon$.
To this end we need to gain some knowledge on the higher orders contributions to the $\mathcal{G}_n$. The next Section is devoted to derive some general results in this direction.

\section{Higher orders}
The next step of the expansion in $\epsilon$ consists in considering the $\mathcal{O}(\epsilon^3)$ contribution. Starting from this order, however, the procedure outlined in the previous sections presents increasing difficulties, and at present it is not clear to us how to overcome these problems for the most general case. Nevertheless, a restricted but important subclass of diagrams can still be treated by resorting to the same techniques previously introduced.

Let us shed some light on this point by considering the different cases that we can face when we move to the generic $\mathcal{O}(\epsilon^k)$. In principle one should start the calculation from the Feynman rules derived in Section 1 and perform all the steps of the replica trick procedure described in that Section. According to those rules, the $\mathcal{O}(\epsilon^k)$ contributions to the $\mathcal{G}_n$'s are obtained by taking all the combinations of vertices $\lambda_{i}$ such that the total degree in $\epsilon$ of the resulting diagrams is $\epsilon^k$. Three cases have to be distinguished:
\begin{enumerate}
	\item Single-vertex diagram. This diagram is obtained when we take the vertex $\lambda_{k}$.
	 It is practically the same as those appearing in Eqs.\,(\ref{Gn1}) and (\ref{Gn2A}), with the difference that the elementary vertex is $\lambda_{k}$ rather than $\lambda_{1}$ or $\lambda_{2}$, and that we have the k-derivative with respect to $N$ rather than the first or second derivative. As a consequence, this diagram is easily given in terms of the $\mathcal{O}(\epsilon^k)$ $n$-legs effective vertex:
		\begin{align} \label{Pi_n^kdef}
		\Pi^{(k)}_{n}\,=\,
		\begin{tikzpicture}[baseline=-3]
		\begin{feynman}
		\vertex[draw, circle, minimum size=0.85cm,very thick] (m) at (0, 0) {k};
		\vertex (a) at (-0.9,0.9) {\small$2$};
		\vertex (b) at (-0.9,-0.9) {\small$1$};
		\vertex (c) at (1.24421,0) {\small$n$};
		\vertex (l1) at (0.5,0.35) {};
		\vertex (l2) at (-0.2,0.55) {};
		\vertex (d1) at (0.595294,-0.308824) {};
		\vertex (d2) at (0.924706,0.308824) {};
		\vertex (d3) at (-0.700391,0.226862) {};
		\vertex (d4) at (-0.370979,0.844509) {};
		\vertex (d5) at (-0.700391,-0.226862) {};
		\vertex (d6) at (-0.370979,-0.844509) {};
		\diagram* {
			(a) -- (m) -- (b);
			(m) -- (c);
			(l1) -- [thick,dotted,out=-280, in=-305](l2);
		};
		\draw (d1) to (d2);
		\draw (d3) to (d4);
		\draw (d5) to (d6);
		\end{feynman}
		\end{tikzpicture} \,&\equiv\,
		\lim_{N \rightarrow 1} \frac{d^k}{dN^k} \left\{-\lambda_{k}(N)\left[\Delta(0)\right]^{N-\frac{n}{2}} C_n(N)\right\}\nonumber\\
		&=-\frac{\epsilon^k}{k!} g\,\mu^2 \left[\frac{2}{\Delta(0)}\right]^{\frac{n}{2}-1}\lim_{N \to 1} \frac{d^k}{dN^k} f_n(N)\,.
		\end{align}
	Diagrams of this kind are handled with the same techniques described before.
	\item 2-vertices diagrams.
	These diagrams are obtained by taking a vertex $\lambda_{i}$ and a vertex $\lambda_{k-i}$ (with $1\leq i \leq k-1$). 
	They are then very similar to those appearing in Eqs.\,(\ref{Gn2Be}) and (\ref{Gn2Bo}), with the difference that the elementary vertices are now $\lambda_{i}$ and $\lambda_{k-i}$ (rather than twice $\lambda_{1}$), and that the derivatives with respect to $N$ and $M$ are  of degree $i$ and $k-i$ (rather than first order derivatives in both cases). In this case we perform the analytic extensions of the appropriate functions similarly to what we have done for the $\mathcal{O}(\epsilon^2\,g^2)$. Naturally, we are assuming again that the series defining our Green's functions are truncated at a given cut-off $L_{max}$, and under this assumption these diagrams are given in terms of the $\mathcal{O}(\epsilon^i)$ and $\mathcal{O}(\epsilon^{k-i})$ effective vertices. Referring again to the (similar) Feynman diagrams in x-space given in Eqs.\,(\ref{Gn2Bediag}) and (\ref{Gn2Bodiag}), these contributions are written as
	\begin{equation} \label{twovertex}
	\begin{tikzpicture}[baseline=(m)]
	\begin{feynman}[inline=(m)]
	\vertex[draw, circle, minimum size=1cm,very thick] (m) at (-1, 0) {\footnotesize $i$};
	\vertex[draw, circle, minimum size=1cm,very thick] (n) at (1, 0) {\footnotesize $k\!-\!i$};
	\vertex (a) at (-2.3,1.3) {$x_{1}$};
	\vertex (b) at (-2.3,-1.3) {$x_{2j}$};
	\vertex (c) at (2.3,1.3) {$x_{2j+1}$};
	\vertex (d) at (2.3,-1.3) {$x_{n}$};
	\vertex (l1) at (1.8,0.7) {};
	\vertex (l2) at (1.8,-0.7) {};
	\vertex (p1) at (-1.8,0.7) {};
	\vertex (p2) at (-1.8,-0.7) {};
	\diagram* {
		(a) -- (m) -- (b);
		(m) -- [half left, out=17, in=163](n);
		(m) -- [half left, out=-17, in=-163](n);
		(m) -- [half left, out=36, in=144,dashed](n);
		(m) -- [half left, out=-36, in=-144,edge label'={\small $2l$},dashed](n);
		(c) -- (n) -- (d);	
		(l1) -- [thick,dotted,out=-60, in=60](l2);
		(p2) -- [thick,dotted,out=120, in=-120](p1);
	};
	\end{feynman}
	\end{tikzpicture}
\qquad \qquad
	\begin{tikzpicture}[baseline=(m)]
	\begin{feynman}[inline=(m)]
	\vertex[large,blob,very thick,pattern color=white] (m) at (-1, 0) {\footnotesize $i$};
	\vertex[large,blob,very thick,pattern color=white] (n) at (1, 0) {\footnotesize $k\!-\!i$};	
	\vertex (a) at (-2.3,1.3) {$x_{1}$};
	\vertex (b) at (-2.3,-1.3) {$x_{2j+1}$};
	\vertex (c) at (2.3,1.3) {$x_{2j+2}$};
	\vertex (d) at (2.3,-1.3) {$x_{n}$};
	\vertex (l1) at (1.8,0.7) {};
	\vertex (l2) at (1.8,-0.7) {};
	\vertex (p1) at (-1.8,0.7) {};
	\vertex (p2) at (-1.8,-0.7) {};
	\diagram* {
		(a) -- (m) -- (b);
		(m) -- [half left,out=30, in=150, dashed](n);
		(m) -- [half left,out=-30, in=-150, edge label'={\footnotesize $2l+1$}, dashed](n);
		(m) -- (n);
		(c) -- (n) -- (d);	
		(l1) -- [thick,dotted,out=-60, in=60](l2);
		(p2) -- [thick,dotted,out=120, in=-120](p1);
	};
	\end{feynman}
	\end{tikzpicture}
	\end{equation}
	\item 3 or more vertices. These diagrams are more complicated. The only point we can clearly show is that the combinatorial coefficients are factorized as before. However, the analytic extensions of the series are much more involved, and we have not been able to perform this step in all its generality.
	Actually, while for the two-vertex diagrams the sum over the number of links between the two vertices almost straightforwardly gives rise to the hypergeometric function ${}_2{F}_{1}$, in these latter cases we have $m$  vertices, with $m>2$. As a consequence, there are $\binom{m}{2}$ sums over the number of links among these vertices, and it is much more difficult to gain control on the analytic extensions of the corresponding series, mainly in connection with their convergence properties. More precisely we have checked that, when it is possible to define these analytic extensions, they should be again given in terms of hypergeometric functions, even though of a higher number of variables (multiple hypergeometric functions).
	Clearly in these cases the problem with the convergence of the series is more severe than before, and the possibility of defining the analytic extension of these series again resorts on the assumption that they can be consistently truncated. If this is doable, the derivatives and the limits with respect to the variables $N_i$ can be done term by term, and the corresponding contributions to the Green's functions are given again in terms of the effective vertices $\Pi^{(i)}_n$.
\end{enumerate}

Once all the contributions to the Green's functions are written in terms of the effective vertices $\Pi$'s, it is possible to analyse their dependence on the momentum cut-off $\Lambda$ similarly to what we have previously done for the first and second order contributions.

Moreover we note that, when the systematic renormalization program of the theory at each order in $\epsilon$ will be undertaken, the possibility of writing all the contributions in terms of effective vertices would make it possible to apply to the expansion in $\epsilon$ the results of the BPHZ theorem concerning the subleading divergences. This is however beyond the scopes of the present work and is left for future investigations.

As usual we are interested in the amputated 1PI diagrams, with total momentum conservation already factored out. Moreover, in the  diagrams it is always understood that at each effective vertex an even number of legs is attached, even though for notational simplicity we do not indicate it explicitly.

The leading behaviour of the 1-vertex diagram contributing to the $\mathcal{G}_n$ at $\mathcal{O}(\epsilon^k)$ is:
\begin{align}\label{1vert}
\begin{tikzpicture}[baseline=-3]
\begin{feynman}
\vertex[draw, circle, minimum size=0.8cm,very thick] (m) at (0, 0) {$k$};
\vertex (a) at (-0.9,0.9) {\small$2$};
\vertex (b) at (-0.9,-0.9) {\small$1$};
\vertex (c) at (1.24421,0) {\small$n$};
\vertex (l1) at (0.5,0.35) {};
\vertex (l2) at (-0.2,0.55) {};
\vertex (d1) at (0.595294,-0.308824) {};
\vertex (d2) at (0.924706,0.308824) {};
\vertex (d3) at (-0.700391,0.226862) {};
\vertex (d4) at (-0.370979,0.844509) {};
\vertex (d5) at (-0.700391,-0.226862) {};
\vertex (d6) at (-0.370979,-0.844509) {};
\diagram* {
	(a) -- (m) -- (b);
	(m) -- (c);
	(l1) -- [thick,dotted,out=-280, in=-305](l2);
};
\draw (d1) to (d2);
\draw (d3) to (d4);
\draw (d5) to (d6);
\end{feynman}
\end{tikzpicture} \,&=\,\,\,\Pi^{(k)}_{n}\quad\sim\,\,\, \frac{\log^{k-1}\Lambda}{\Lambda^{n-2}}
\end{align}
where we adopt the convention that for $n=2$, $\Lambda^0$ has to be read as $\log\Lambda$ in the numerator. This convention will be also used for all the diagrams considered below.

From (\ref{1vert}) we see that for $n>2$ even if we increase the order $k$, at any finite order we get a vanishing contribution to $\mathcal{G}_n$, as no power of $\log\Lambda$ can underdo the suppression due to the inverse power of $\Lambda$.

Concerning the diagrams with two effective vertices, we have 
\begin{align} \label{2vert}
\begin{tikzpicture}[baseline=-3]
\begin{feynman}[small]
\vertex[draw, circle, minimum size=0.9cm,very thick] (m) at (-0.8,0) {\footnotesize $i$};
\vertex[draw, circle, minimum size=0.9cm,very thick] (n) at (0.8, 0) {\scriptsize $k\!-\!i$};	
\vertex (a) at (-2,1.2) {\footnotesize $x_{1}$};
\vertex (b) at (-2,-1.2) {\footnotesize $x_{a}$};
\vertex (c) at (2,1.2) {\footnotesize $x_{a+1}$};
\vertex (d) at (2,-1.2) {\footnotesize $x_{n}$};
\vertex (l1) at (1.6,0.65) {};
\vertex (l2) at (1.6,-0.65) {};
\vertex (p1) at (-1.6,0.65) {};
\vertex (p2) at (-1.6,-0.65) {};
\diagram* {
	(a) -- (m) -- (b);
	(m) -- [half left, out=17, in=163](n);
	(m) -- [half left, out=-17, in=-163](n);
	(m) -- [half left, out=36, in=144,dashed](n);
	(m) -- [half left, out=-36, in=-144,edge label'={\footnotesize $r$},dashed](n);
	(c) -- (n) -- (d);	
	(l1) -- [thick,dotted,out=-60, in=60](l2);
	(p2) -- [thick,dotted,out=120, in=-120](p1);
};
\end{feynman}
\end{tikzpicture} =\,\,\, \Pi^{(i)}_{a+r}\,\Pi^{(k-i)}_{n-a+r} \frac{I_{r}(p)}{r!} \quad \sim\,\,\, 
\frac{\left(\log \Lambda\right)^{i-1}}{\Lambda^{a+r-2}} \frac{\left(\log \Lambda\right)^{k-i-1}}{\Lambda^{n-a+r-2}} \Lambda^{2(r-2)} 
\end{align}
where $r\geq 2$ as we are considering only 1PI diagrams. 
The powers of $\Lambda$ in \ref{2vert} are collected in three different pieces because they come from three different contributions and to each of them the convention on $\Lambda^0$ introduced above has to be applied.

Moving now to the case of three effective vertices, the diagrams to be considered are:
\begin{align}\label{3vert}
\begin{tikzpicture}[baseline=-25]
\begin{feynman}[small]
\vertex[draw, circle, minimum size=0.9cm,very thick] (m) at (-1,0) {\footnotesize $i_1$};
\vertex[draw, circle, minimum size=0.9cm,very thick] (n) at (1, 0) {\footnotesize $i_2$};
\vertex[draw, circle, minimum size=0.9cm,very thick] (o) at (0,-1.73) {\footnotesize $i_3$};	
\vertex (a) at (-2.2,0) {\footnotesize $x_{1}$};
\vertex (b) at (-1.6,1.04) {\footnotesize $x_{a}$};
\vertex (c) at (2.2,0) {\footnotesize $y_{1}$};
\vertex (d) at (1.6,1.04) {\footnotesize $y_{b}$};
\vertex (e) at (-0.6,-2.77) {\footnotesize $z_{1}$};
\vertex (f) at (0.6,-2.77) {\footnotesize $z_{c}$};
\vertex (l1) at (-1.896,0.078) {};
\vertex (l2) at (-1.516, 0.737) {};
\vertex (l3) at (1.896,0.078) {};
\vertex (l4) at (1.516, 0.737) {};
\vertex (l5) at (-0.38,-2.546) {};
\vertex (l6) at (0.38,-2.546) {};
\diagram* {
	(a) -- (m) -- (b);
	(c) -- (n) -- (d);
	(e) -- (o) -- (f);
	(m) -- (n) -- (o) -- (m);
	(m)	-- [out=20, in=160, dashed,edge label={\footnotesize $r$}](n);
	(m)	-- [out=-20, in=-160, dashed](n);
	(n)	-- [out=-100, in=40, dashed,edge label={\footnotesize $s$}](o);
	(n)	-- [out=-140, in=80, dashed](o);
	(o)	-- [out=100, in=-40, dashed](m);
	(o)	-- [out=140, in=-80, dashed,edge label={\footnotesize $t$}](m);
	(l1) -- [thick,dotted,out=80, in=-140](l2);
	(l3) -- [thick,dotted,out=100, in=-40](l4);
	(l5) -- [thick,dotted,out=-20, in=-160](l6);
};
\end{feynman}
\end{tikzpicture}
\end{align}
where $a+b+c=n$ and $i_1+i_2+i_3=k$. Moreover, as we are taking into account only 1PI diagrams, we must have either $r=0$ and $s,t\geq2$ (and similar permutations) or $r,s,t,\geq1$. 

The particular case where $r\!=\!s\!=\!t\!=\!1$ (see Eq.\,(\ref{3vertA}) below) has a convergent triangle loop, so that the leading behaviour in terms of the cut-off is:
\begin{align}\label{3vertA}
\begin{tikzpicture}[baseline=-25]
\begin{feynman}[small]
\vertex[draw, circle, minimum size=0.9cm,very thick] (m) at (-1,0) {\footnotesize $i_1$};
\vertex[draw, circle, minimum size=0.9cm,very thick] (n) at (1, 0) {\footnotesize $i_2$};
\vertex[draw, circle, minimum size=0.9cm,very thick] (o) at (0,-1.73) {\footnotesize $i_3$};	
\vertex (a) at (-2.2,0) {\footnotesize $x_{1}$};
\vertex (b) at (-1.6,1.04) {\footnotesize $x_{a}$};
\vertex (c) at (2.2,0) {\footnotesize $y_{1}$};
\vertex (d) at (1.6,1.04) {\footnotesize $y_{b}$};
\vertex (e) at (-0.6,-2.77) {\footnotesize $z_{1}$};
\vertex (f) at (0.6,-2.77) {\footnotesize $z_{c}$};
\vertex (l1) at (-1.896,0.078) {};
\vertex (l2) at (-1.516, 0.737) {};
\vertex (l3) at (1.896,0.078) {};
\vertex (l4) at (1.516, 0.737) {};
\vertex (l5) at (-0.38,-2.546) {};
\vertex (l6) at (0.38,-2.546) {};
\diagram* {
	(a) -- (m) -- (b);
	(c) -- (n) -- (d);
	(e) -- (o) -- (f);
	(m) -- (n) -- (o) -- (m);
	(l1) -- [thick,dotted,out=80, in=-140](l2);
	(l3) -- [thick,dotted,out=100, in=-40](l4);
	(l5) -- [thick,dotted,out=-20, in=-160](l6);
};
\end{feynman}
\end{tikzpicture}
\quad \sim \quad \frac{\left(\log\Lambda\right)^{i_1-1}}{\Lambda^{a}}\, 
\frac{\left(\log\Lambda\right)^{i_2-1}}{\Lambda^{b}}\,
\frac{\left(\log\Lambda\right)^{i_3-1}}{\Lambda^{c}}
\end{align}

In all the other cases, the superficial degree of divergence of the loop integrals is non-negative, more specifically $D=4(r+s+t-2)-2(r+s+t)=2(r+s+t-4)\geq0$. The (superficial) cut-off dependence of the diagram is then (for $r+s+t\geq4$):
\begin{align}\label{3vertB}
\begin{tikzpicture}[baseline=-25]
\begin{feynman}[small]
\vertex[draw, circle, minimum size=0.9cm,very thick] (m) at (-1,0) {\footnotesize $i_1$};
\vertex[draw, circle, minimum size=0.9cm,very thick] (n) at (1, 0) {\footnotesize $i_2$};
\vertex[draw, circle, minimum size=0.9cm,very thick] (o) at (0,-1.73) {\footnotesize $i_3$};	
\vertex (a) at (-2.2,0) {\footnotesize $x_{1}$};
\vertex (b) at (-1.6,1.04) {\footnotesize $x_{a}$};
\vertex (c) at (2.2,0) {\footnotesize $y_{1}$};
\vertex (d) at (1.6,1.04) {\footnotesize $y_{b}$};
\vertex (e) at (-0.6,-2.77) {\footnotesize $z_{1}$};
\vertex (f) at (0.6,-2.77) {\footnotesize $z_{c}$};
\vertex (l1) at (-1.896,0.078) {};
\vertex (l2) at (-1.516, 0.737) {};
\vertex (l3) at (1.896,0.078) {};
\vertex (l4) at (1.516, 0.737) {};
\vertex (l5) at (-0.38,-2.546) {};
\vertex (l6) at (0.38,-2.546) {};
\diagram* {
	(a) -- (m) -- (b);
	(c) -- (n) -- (d);
	(e) -- (o) -- (f);
	(m) -- (n) -- (o) -- (m);
	(m)	-- [out=20, in=160, dashed,edge label={\footnotesize $r$}](n);
	(m)	-- [out=-20, in=-160, dashed](n);
	(n)	-- [out=-100, in=40, dashed,edge label={\footnotesize $s$}](o);
	(n)	-- [out=-140, in=80, dashed](o);
	(o)	-- [out=100, in=-40, dashed](m);
	(o)	-- [out=140, in=-80, dashed,edge label={\footnotesize $t$}](m);
	(l1) -- [thick,dotted,out=80, in=-140](l2);
	(l3) -- [thick,dotted,out=100, in=-40](l4);
	(l5) -- [thick,dotted,out=-20, in=-160](l6);
};
\end{feynman}
\end{tikzpicture}
\quad \sim \quad 
\frac{\left(\log\Lambda\right)^{i_1-1}}{\Lambda^{a+r+t-2}}\, 
\frac{\left(\log\Lambda\right)^{i_2-1}}{\Lambda^{b+r+s-2}}\,
\frac{\left(\log\Lambda\right)^{i_3-1}}{\Lambda^{c+s+t-2}}\,
\Lambda^{2(r+s+t-4)}
\end{align}

To summarize, leaving aside the subleading dependence on $\log\Lambda$, the leading behaviour with the cut-off of the 1PI diagrams is given by:
\begin{itemize}
\item 1-vertex: $\,\sim\, \Lambda^{2-n}$
\item 2-vertices: $\,\sim\, \Lambda^{-n}$
\item 3-vertices: $\,\sim\, \Lambda^{-n}$ (Eq.(\ref{3vertA})) \quad or\quad $\Lambda^{-2-n}$ (Eq.(\ref{3vertB}))
\end{itemize}

We can now extend the previous analysis to a generic number of effective vertices $V$. Indicating as usual with $N_I$ the number of internal lines, and referring to a generic dimension $d$, the superficial degree of divergence $D$ coming from the loop integrals is:
\begin{equation} \label{Loopscontr}
	D=d(N_I-V+1)-2N_I\,,
\end{equation}
where the integral is (superficially) convergent when $D<0$.

As for the contribution of the effective vertices, we have $V$ vertices with a total number of legs $n+2N_I$.
Therefore, referring to (\ref{1vert}) with generic $d$, that gives 
\begin{equation}\label{1vertd}
	\Pi^{(k)}_{n}\sim \frac{\left(\log\Lambda\right)^{k-1}}{\Lambda^{(\frac{d-2}{2})(n-2)}} \sim \Lambda^{-(\frac{d-2}{2})(n-2)}
\end{equation}
we have
\begin{equation} \label{Picontr}
\prod_{i=1}^{V} \Pi^{k_i}_{a_i} \sim \prod_{i=1}^{V} \frac{\left(\log\Lambda\right)^{k_i-1}}{\Lambda^{(\frac{d-2}{2})(a_i-2)}} \sim \frac{1}{\Lambda^{(\frac{d-2}{2})(n+2N_I-2V)}} = \frac{1}{\Lambda^{\left[(\frac{d-2}{2})n +(d-2)N_I-(d-2)V)\right]}}\,,
\end{equation}
where in the last members of (\ref{1vertd}) and (\ref{Picontr}) the subleading logarithms are neglected.

We have to distinguish the two cases $D\geq0$ and $D<0$.
In the first case, the cut-off dependence of the diagram with $n$ external lines, $V$ vertices and $N_I$ internal lines, comes from both the effective vertices (\ref{Picontr}) and the loop integrals (\ref{Loopscontr}). In $d$ dimensions it is:
\begin{equation}
\qquad \Lambda^{-[\frac{d-2}{2}n+2V-d]}\,\qquad (\,\,\overset{d=4}{\rightarrow} \,\,\Lambda^{4-n-2V}\,)
\end{equation}
and it is clear that the dominant (less suppressed) diagram is the one with $V=1$, i.e. (\ref{1vertd}).

In the second case, when $D<0$, the cut-off dependence is given uniquely by (\ref{Picontr}). For any fixed value of $V$, the condition $D<0$ is verified when $V\leq N_I < \frac{d}{d-2}(V-1)$, where the lower limit comes from considering only 1PI diagrams.
For $V=2$, the condition $D<0$ is never verified.
For $V>2$, among the diagrams with $D<0$, the leading one (i.e. the one less suppressed by powers of $\Lambda$) is the diagram with the minimal possible value of $N_I$, that is $N_I=V$. It goes as:
\begin{equation}\label{caseD<0}
\Lambda^{-(\frac{d-2}{2})n} \qquad (\,\,\overset{d=4}{\rightarrow} \,\, \Lambda^{-n}\,)
\end{equation}
Comparing (\ref{caseD<0}) with (\ref{1vertd}), again we see that the dominant contribution comes from (\ref{1vertd}), that is from the diagram with only one effective vertex.

In the present section, extending the analysis previously applied to  the $\mathcal{O}(\epsilon^2)$, we have been able to identify the leading contributions to the $\mathcal{G}_n$ at any order in $\epsilon$. However, to our dissatisfaction, we have seen that at any finite order in $\epsilon$ the radiative corrections give a too strong suppression factor with the physical cut-off $\Lambda$ so that, disregarding possible unconventional renormalization prescriptions (flows) on which we have arleady commented, it seems that all the physical amplitude vanish, thus resulting in a non-interacting theory.

In our opinion this is due to the fact that truncating the expansion to a finite order in $\epsilon$we get a too mild impact of the interaction term on the Green's functions. However we have seen that, increasing the power of $\epsilon$, the suppression factor that causes the vanishing of the $\mathcal{G}_n$ becomes milder and milder, thus rising the hope that an infinite resummation of different orders in $\epsilon$ could rescue the theory.

With this in mind, in the next section we will focus our attention on the leading order (in $\Lambda$) contributions at each order in $\epsilon$ and attempt to a resummation of these diagrams.
The way for this analysis has been paved in this section.

\section{Resumming the leading diagrams}
We now proceed to the resummation of the leading contributions to the Green's functions $\mathcal{G}_n$, and note that we have to treat separately (as it was for the previous sections) the $n=2$ and $n>2$ cases.

\subsection{Two-points Green's function $\mathcal{G}_2$}
In the previous sections we have seen that the $\mathcal{O}(\epsilon)$ contribution to $\widetilde{\mathcal{G}}_2$ is:
\begin{equation} \label{2geps}
\begin{tikzpicture}[baseline=-3]
\begin{feynman}[small]
\vertex[draw, circle, minimum size=0.85cm,very thick] (m) at (0, 0) {1};
\vertex (a) at (-1.5,0) {};
\vertex (b) at ( 1.5,0) {};
\diagram* {
	(a) --[fermion] (m) --[fermion] (b),
};
\end{feynman}
\end{tikzpicture}
\end{equation}
that diverges as $\log \Delta(0)$. At order $\epsilon^2$ there are two leading diagrams,
\begin{equation} \label{2geps2}
\begin{tikzpicture}[baseline=-3]
\begin{feynman}[small]
\vertex[draw, circle, minimum size=0.85cm,very thick] (m) at (0, 0) {2};
\vertex (a) at (-1.5,0) {};
\vertex (b) at ( 1.5,0) {};
\diagram* {
	(a) --[fermion] (m) --[fermion] (b),
};
\end{feynman}
\end{tikzpicture}
\qquad
\begin{tikzpicture}[baseline=-3]
\begin{feynman}[small]
\vertex[draw, circle, minimum size=0.85cm,very thick] (m) at (-0.8, 0) {1};
\vertex[draw, circle, minimum size=0.85cm,very thick] (n) at (0.8, 0) {1};
\vertex (a) at (-2.2,0) {};
\vertex (b) at ( 2.2,0) {};
\diagram* {
	(a) --[fermion] (m);
	(m) --[fermion] (n);
	(n) --[fermion] (b),
};
\end{feynman}
\end{tikzpicture}
\end{equation}
both diverging as $\log^2 \Delta(0)$, while for the order $\epsilon^3$ the leading divergences are contained in the three diagrams:
\begin{equation} \label{2geps3}
\begin{tikzpicture}[baseline=-3]
\begin{feynman}[small]
\vertex[draw, circle, minimum size=0.85cm,very thick] (m) at (0, 0) {3};
\vertex (a) at (-1.3,0) {};
\vertex (b) at ( 1.3,0) {};
\diagram* {
	(a) --[fermion] (m) --[fermion] (b),
};
\end{feynman}
\end{tikzpicture}
\quad\,
\begin{tikzpicture}[baseline=-3]
\begin{feynman}[small]
\vertex[draw, circle, minimum size=0.85cm,very thick] (m) at (-0.75, 0) {2};
\vertex[draw, circle, minimum size=0.85cm,very thick] (n) at (0.75, 0) {1};
\vertex (a) at (-2.1,0) {};
\vertex (b) at ( 2.1,0) {};
\diagram* {
	(a) --[fermion] (m);
	(m) --[fermion] (n);
	(n) --[fermion] (b),
};
\end{feynman}
\end{tikzpicture}
\quad\,
\begin{tikzpicture}[baseline=-3]
\begin{feynman}[small]
\vertex[draw, circle, minimum size=0.85cm,very thick] (m) at (-1.5, 0) {1};
\vertex[draw, circle, minimum size=0.85cm,very thick] (n) at (0, 0) {1};
\vertex[draw, circle, minimum size=0.85cm,very thick] (o) at (1.5, 0) {1};
\vertex (a) at (-2.8,0) {};
\vertex (b) at ( 2.8,0) {};
\diagram* {
	(a) --[fermion] (m);
	(m) --[fermion] (n);
	(n) --[fermion] (o);
	(o) --[fermion] (b),
};
\end{feynman}
\end{tikzpicture}
\end{equation}
all of them diverging as $\log^3\Delta(0)$. Similarly for the higher orders.

Our goal is to proceed to the resummation of the infinite ``triangle" of diagrams whose first instances are given in Eqs.\,(\ref{2geps})-(\ref{2geps3}). This can be done by resumming first the 1PI diagrams in the left side of the triangle, and then considering the geometric series arising from their iteration, that means resumming all the diagrams in the triangle.

Considering then the amputated 1PI bubbles
$\begin{tikzpicture}[baseline=-3]
\begin{feynman}
\vertex[draw, circle, minimum size=0.85cm,very thick] (m) at (0, 0) {k};
\vertex (a) at (-1.24421,0) {};
\vertex (b) at (1.24421,0) {};
\vertex (d1) at (0.595294,-0.308824) {};
\vertex (d2) at (0.924706,0.308824) {};
\vertex (d3) at (-0.924706,-0.308824) {};
\vertex (d4) at (-0.595294,0.308824) {};
\diagram* {
	(a) -- (m) -- (b);
};
\draw (d1) to (d2);
\draw (d3) to (d4);
\end{feynman}
\end{tikzpicture}$
, i.e. the effective vertices $\Pi_2^{(k)}$, we obtain the resummed two-legs vertex function (i.e. proper self-energy) that we name $\Gamma_2$:
\begin{align} \label{G2res}
\Gamma_2\,\equiv\,\sum_{k=1}^{\infty}\,\Bigg \{
\begin{tikzpicture}[baseline=-3]
\begin{feynman}
\vertex[draw, circle, minimum size=0.85cm,very thick] (m) at (0, 0) {k};
\vertex (a) at (-1.24421,0) {};
\vertex (b) at (1.24421,0) {};
\vertex (d1) at (0.595294,-0.308824) {};
\vertex (d2) at (0.924706,0.308824) {};
\vertex (d3) at (-0.924706,-0.308824) {};
\vertex (d4) at (-0.595294,0.308824) {};
\diagram* {
	(a) -- (m) -- (b);
};
\draw (d1) to (d2);
\draw (d3) to (d4);
\end{feynman}
\end{tikzpicture}
\Bigg \}
&=\sum_{k=1}^{\infty}\, \Pi_2^{(k)} \,=\, \sum_{k=1}^{\infty} \lim_{N \rightarrow 1} \frac{d^k}{dN^k} \left\{-\lambda_{k}(N)\left[\Delta(0)\right]^{N-1} C_2(N)\right\}\nonumber\\
&=- g\,\mu^2  \sum_{k=1}^{\infty} \frac{\epsilon^k}{k!} \lim_{N \to 1} \frac{d^k}{dN^k} f_2(N) \,=\,- g\,\mu^2 \left[f_2(1+\epsilon) - f_2(1)\right]
\nonumber\\
&=-g\,\mu^2 \left[(\epsilon+1) \frac{\Gamma(\epsilon+\frac{3}{2})}{\Gamma(\frac{3}{2})}\left[2\mu^{2-d}\Delta(0)\right]^{\epsilon}- 1\right]\,,
\end{align} 
where the resummation in the second line of (\ref{G2res}) is possible due to the analyticity on the positive real axis of the function $f_2(x)$ in (\ref{f2N}), and the final result is given in the last member. 

It is worth to stress that the above result is obtained for generic real positive values of $\epsilon$. At the same time, what is physically interesting at the end of the calculation is to consider integer values of $\epsilon$ that correspond to relevant interacting theories as $\phi^4$ ($\epsilon=1$), $\phi^6$ ($\epsilon=2$), $\dots$.
In these latter cases, being $\frac{\Gamma(\epsilon+\frac{3}{2})}{\Gamma(\frac{3}{2})}=2^{-\epsilon} (2\epsilon+1)!!$, Eq.(\ref{G2res}) becomes:
\begin{align}\label{G2res2}
\Gamma_2\,\overset{\epsilon \in \mathbb{Z_+}}{=}\, g \mu^2 -(\epsilon +1)(2\epsilon+1)!!\,g \mu^{2+(2-d)\epsilon}\, \Delta(0)^{\epsilon}\,.
\end{align}

Few comments are in order. Let us start by recalling that the lagrangian defining the theory (see Eq.(\ref{eq:L})) is:
\begin{align} \label{Lagrangian}
\mathcal{L} &= \frac{1}{2}(\nabla\phi)^2+\frac{1}{2}m^2\phi^2+\frac{1}{2}g\mu^{2+(2-d)\epsilon}\phi^{2+2\epsilon} \nonumber \\
&= \frac{1}{2}(\nabla\phi)^2+\frac{1}{2}\left(m^2+g\mu^2\right)\phi^2+\left[\frac{1}{2}g\mu^{2+(2-d)\epsilon}\phi^{2+2\epsilon} - \frac{1}{2}g\mu^2\phi^2 \right]\,,
\end{align}
where the second line is written according to the splitting in free and interaction Lagrangians as given in Eqs.(\ref{L0})-(\ref{lagrnew}). 
Considering then as $\mathcal{L}_0$ the first two terms of the second line in Eq.(\ref{Lagrangian}), and as $\mathcal{L}_{int}$ the terms in the square brackets, we find the central result of this subsection, namely that Eq.(\ref{G2res2}) is nothing but the proper self-energy obtained within the framework of the weak coupling expansion. 
Indeed, at the first perturbative order in $g$ we obtain the two diagrams:
\begin{eqnarray} \label{G2weakgcounter}
\begin{tikzpicture}[baseline=-3]
\begin{feynman}
\vertex[crossed dot, minimum size=0.3cm] (m) at (0, 0) {};
\vertex (a) at (-1.24421,0) {};
\vertex (b) at (1.24421,0) {};
\vertex (d1) at (0.595294,-0.308824) {};
\vertex (d2) at (0.924706,0.308824) {};
\vertex (d3) at (-0.924706,-0.308824) {};
\vertex (d4) at (-0.595294,0.308824) {};
\diagram* {
	(a) -- (m) -- (b);
};
\draw (d1) to (d2);
\draw (d3) to (d4);
\end{feynman}
\end{tikzpicture} &=& g\,\mu^2
\\ \label{G2weakgtadpole}
\begin{tikzpicture}[baseline=5]
\begin{feynman}[small]
\vertex (b) [dot] at (0, 0) {};
\vertex (a) at (-1.2, 0) {};
\vertex (c) at ( 1.2, 0) {};
\vertex (l1) at (-0.3,0.8) {};
\vertex (l2) at (0.3,0.8) {};
\vertex (d1) at (0.595294,-0.308824) {};
\vertex (d2) at (0.924706,0.308824) {};
\vertex (d3) at (-0.924706,-0.308824) {};
\vertex (d4) at (-0.595294,0.308824) {};
\diagram* {
	(a) -- (b) -- [out=30, in=80, loop, min distance=1.3cm,edge label'={\footnotesize $\epsilon$}] b -- [out=100, in=150, loop, min distance=1.3cm,edge label'={\footnotesize $1$}] b -- (c);
	(l1) -- [thick,dotted,out=30, in=150](l2);
};
\end{feynman}
\draw (d1) to (d2);
\draw (d3) to (d4);
\end{tikzpicture} &=& -\frac{1}{2}g\mu^{2+(2-d)\epsilon} (2\epsilon+2)\,(2\epsilon+1)!!\,\Delta(0)^{\epsilon} 
\end{eqnarray}
whose sum gives the proper self energy in (\ref{G2res2}).
Note that the diagram in (\ref{G2weakgcounter}) comes from the term $- \frac{1}{2}g\mu^2\phi^2$ in $\mathcal{L}_{int}$ and that the mass term in the loop integral $\Delta(0)$ is $M^2=m^2+g\mu^2$. We would like to stress that would we have used the splitting in $\mathcal{L}_{0}$ and $\mathcal{L}_{int}$ according to the first line of (\ref{Lagrangian}), in (\ref{G2res2}) the term $g\mu^2$ would be absent, and the propagator $\Delta(0)$ would contain the mass $m^2$. This is simply related to the freedom of moving $\phi^2$ terms from the free to the interaction lagrangian.

We are now ready to proceed to the resummation of the whole ``triangle" of diagrams in Eqs.\,(\ref{2geps})-(\ref{2geps3}). Iterating the proper self-energy insertion (\ref{G2res}), we obtain for the full propagator the approximation (for integer values of $\epsilon$ (\ref{G2res}) coincides with (\ref{G2res2})):
\begin{align}
\begin{tikzpicture}[baseline=-3]
\begin{feynman}[small]
\vertex[blob, minimum size=0.85cm] (m) at (0, 0) {};
\vertex (a) at (-1.4,0) {};
\vertex (b) at ( 1.4,0) {};
\diagram* {
	(a) --[fermion] (m) --[fermion] (b),
};
\end{feynman}
\end{tikzpicture} 
&= \frac{1}{p^2+M^2} \Bigg\{1 + \left[g \mu^2 -2^{\epsilon}\,(\epsilon +1)\,\frac{\Gamma(\epsilon+\frac{3}{2})}{\Gamma(\frac{3}{2})}\,g \mu^{2+(2-d)\epsilon}\, \Delta(0)^{\epsilon}\right] \frac{1}{p^2+M^2} \nonumber\\
&+\left(\left[g \mu^2 -2^{\epsilon}\,(\epsilon +1)\frac{\Gamma(\epsilon+\frac{3}{2})}{\Gamma(\frac{3}{2})}\,g \mu^{2+(2-d)\epsilon}\, \Delta(0)^{\epsilon}\right]\frac{1}{p^2+M^2}\right)^2 +\, \cdots\Bigg\} \nonumber\\
&= \frac{1}{p^2+M^2-\left[g \mu^2 -2^{\epsilon}\,(\epsilon +1)\frac{\Gamma(\epsilon+\frac{3}{2})}{\Gamma(\frac{3}{2})}\,g \mu^{2+(2-d)\epsilon}\, \Delta(0)^{\epsilon}\right]}
\end{align}
from which the renormalized mass turns out to be:
\begin{align} \label{mRg}
m_R^2&=M^2-g\mu^2+2^{\epsilon}\,(\epsilon +1)\frac{\Gamma(\epsilon+\frac{3}{2})}{\Gamma(\frac{3}{2})}\,g \mu^{2+(2-d)\epsilon}\, \Delta(0)^{\epsilon} \nonumber\\
&=m^2+2^{\epsilon}\,(\epsilon +1)\frac{\Gamma(\epsilon+\frac{3}{2})}{\Gamma(\frac{3}{2})}\,g \mu^{2+(2-d)\epsilon}\, \Delta(0)^{\epsilon}
\end{align}

Eq.(\ref{mRg}) is one of the central results of the present work and we now explain this point making a certain number of comments.

First of all we stress that the above resummation, containing iterated insertions of (\ref{G2weakgcounter}), provides the term $-g\mu^2$ that, combined with $M^2$, restores $m^2$ as the tree level mass term.

We note however that the mass in the loop integral $\Delta(0)$ is still $M^2=m^2+g\mu^2$. As we will see in the next section, the reason is that in order to get even in $\Delta(0)$ the same shift $M^2\to m^2$ we have to go further in the systematic approximation of the proper self-energy. More specifically, we will see that the resummation of subleading diagrams will provide, among others, iterated insertions of (\ref{G2weakgcounter}) inside the loops, thus progressively shifting the mass $M^2$ towards $m^2$ also in the radiative corrections.

We are now ready to appreciate the importance of (\ref{mRg}). 
To this end, we compare the result for $m^2_R$ at $\mathcal{O}(\epsilon)$ (see Eq.(\ref{mReps}) in Section 2) with the result obtained after resummation in (\ref{mRg}).
In the first case, the radiative correction to the mass is proportional to $\epsilon\,\log(\Delta(0))$, which for $d\geq2$ implies a logarithmic divergence $\log(\Lambda)$ that does not depend on $\epsilon$, while after resummation the radiative correction goes as $\Lambda^{\frac{d-2}{2}\epsilon}$.
 
A better insight on this result is gained if we consider integer values $\epsilon=1,2,\dots$ that correspond to interacting theories $\phi^4,\,\phi^6,\dots$. 
For the sake of definiteness let us specify to the case $\epsilon=1$ and $d=4$, i.e. to the 4-dimensional $\phi^4$ theory. In this case the renormalized mass turns out to be:
\begin{align} \label{mRphi4}
m_R^2=m^2+\,6\,g \, \Delta(0)\,,
\end{align}
where the second term is due to the tadpole diagram in (\ref{G2weakgtadpole}) with only one loop ($\epsilon=1$). Using then Eq.(\ref{delta0reg}) for $\Delta(0)$, we get
\begin{equation} \label{mRphi4reg}
m_R^2=m^2 + \frac{3\,g}{8 \pi^2}\left(\Lambda^2-M^2\log{\frac{\Lambda^2}{M^2}}\right)
\end{equation}
that coincides with Eq.(\ref{mR1ord}) obtained in the framework of the weak-coupling expansion, with the only difference that in the logarithmic divergence $m^2$ is replaced by $M^2$. The explanation for this difference is given above (it is cured when higher order approximations are considered).

In a sense Eq.\,(\ref{mRphi4reg}) is as a deceptive result. In fact at $\mathcal{O}(\epsilon)$ we found that the radiative correction to the mass goes as $\log(\Lambda)$ (Eq.(\ref{mR1epsreg})) and this gave rise to the hope that within the $\epsilon$-expansion the hierarchy problem could be enormously alleviated. However Eq.\,(\ref{mRphi4reg}) shows that the resummation of the leading diagrams at any order in $\epsilon$ restores the weak-coupling result, i.e. a quadratically divergent correction.

We may ask ourselves, how is it possible that such a resummation switches a logarithmic to a quadratic divergence. The answer is contained in Eq.\,(\ref{G2res}), whose leading behaviour is given by (note that here $\epsilon=1$ and $\Delta(0)\sim\Lambda^2$): 
\begin{align} \label{G2res30}
&\sum_{k}\,\Bigg \{
\begin{tikzpicture}[baseline=-3]
\begin{feynman}
\vertex[draw, circle, minimum size=0.85cm,very thick] (m) at (0, 0) {k};
\vertex (a) at (-1.24421,0) {};
\vertex (b) at (1.24421,0) {};
\vertex (d1) at (0.595294,-0.308824) {};
\vertex (d2) at (0.924706,0.308824) {};
\vertex (d3) at (-0.924706,-0.308824) {};
\vertex (d4) at (-0.595294,0.308824) {};
\diagram* {
	(a) -- (m) -- (b);
};
\draw (d1) to (d2);
\draw (d3) to (d4);
\end{feynman}
\end{tikzpicture}
\Bigg \}
\sim \sum_{k}\,\frac{\epsilon^k}{k!}\log[\Delta(0)]^{k} \sim [\Delta(0)]^{\epsilon}
\end{align}

This result sounds like a warning and apparently confirms the suspicion that we have expressed above, namely that when we consider the expansion of the lagrangian (\ref{Lagrangian}) in powers of $\epsilon$ and limit ourselves to a finite order in $\epsilon$, this could result in grasping a too poor truncation of the physical interaction. In the following we will further investigate this crucial point.

\subsection{Green's functions $\mathcal{G}_n$ for $n>2$}
As we already noted in Section 4, the diagrams with one effective vertex have a leading behaviour with $\Lambda$, but there are also reducible diagrams that have the same superficial cut-off dependence. The contribution of the latters, however,  is already taken into account when dealing with Green's functions with a smaller number of points $n$. For this reason, we focus on the resummation of the effective vertices only, that are 1PI diagrams. Amputating the external propagators and factoring out the $\delta$-momentum conservation, we get the resummed $n$-legs vertex function that we name $\Gamma_n$:
\begin{align} \label{Gnres}
\Gamma_n=\sum_{k=1}^{\infty}\,\left\{
\begin{tikzpicture}[baseline=-3]
\begin{feynman}
\vertex[draw, circle, minimum size=0.85cm,very thick] (m) at (0, 0) {k};
\vertex (a) at (-0.9,0.9) {\footnotesize$2$};
\vertex (b) at (-0.9,-0.9) {\footnotesize$1$};
\vertex (c) at (1.24421,0) {\footnotesize$n$};
\vertex (l1) at (0.5,0.35) {};
\vertex (l2) at (-0.2,0.55) {};
\vertex (d1) at (0.595294,-0.308824) {};
\vertex (d2) at (0.924706,0.308824) {};
\vertex (d3) at (-0.700391,0.226862) {};
\vertex (d4) at (-0.370979,0.844509) {};
\vertex (d5) at (-0.700391,-0.226862) {};
\vertex (d6) at (-0.370979,-0.844509) {};
\diagram* {
	(a) -- (m) -- (b);
	(m) -- (c);
	(l1) -- [thick,dotted,out=-280, in=-305](l2);
};
\draw (d1) to (d2);
\draw (d3) to (d4);
\draw (d5) to (d6);
\end{feynman}
\end{tikzpicture} 
\right\}
&=\sum_{k=1}^{\infty}\,\Pi_n^{(k)} =\sum_{k=1}^{\infty} \lim_{N \rightarrow 1} \frac{d^k}{dN^k} \left\{-\lambda_{k}(N)\left[\Delta(0)\right]^{N-1} C_n(N)\right\}\nonumber\\
&=- g\,\mu^2 \left[\frac{2}{\Delta(0)}\right]^{\frac{n}{2}-1}  \sum_{k=1}^{\infty} \frac{\epsilon^k}{k!} \lim_{N \to 1} \frac{d^k}{dN^k} f_n(N) \nonumber\\ 
&=- g\,\mu^2 \left[\frac{2}{\Delta(0)}\right]^{\frac{n}{2}-1} \left[f_n(1+\epsilon) - f_n(1)\right]
\end{align} 
where, again, the last step can be done due to the analyticity of the function $f_n(x)$ on the positive real axis. Referring to Eq.(\ref{fnNcont}), we see that $f_n(1)=0\,\, \forall n>2$ (due to the vanishing of the falling factorial) so that we finally have:
\begin{align} \label{Gnres2}
\Gamma_n=\sum_{k=1}^{\infty} \Pi_n^{(k)}
\, &= \,
-g\,\mu^2\,\left[\frac{2}{\Delta(0)}\right]^{\frac{n}{2}-1}\left\{(\epsilon+1)_{\frac{n}{2}} \frac{\Gamma(\epsilon+\frac{3}{2})}{\Gamma(\frac{3}{2})}\left[2\mu^{2-d}\Delta(0)\right]^{\epsilon} \right\} \,.
\end{align}
As for the case of $\mathcal{G}_2$, this result is obtained for generic positive real values of $\epsilon$. Moving to the physically relevant cases, i.e. to integer values of $\epsilon$, Eq.(\ref{Gnres2}) becomes:
\begin{equation} \label{Gnres3}
	\Gamma_n\,\overset{\epsilon \in \mathbb{Z_+}}{=}\,-g\,\mu^{2+(2-d)\epsilon}\, 2^{\frac{n}{2}-1}(\epsilon+1)_{\frac{n}{2}} (2\epsilon+1)!! \Delta(0)^{\epsilon+1-\frac{n}{2}}
\end{equation}

This is the main result of the present subsection. As can be immediately checked, Eq.(\ref{Gnres3}) is nothing but the $\mathcal{O}(g)$ contribution to the $n$-points Green's functions within the framework of the weak-coupling expansion:
\begin{equation} \label{Gnweakg}
\begin{tikzpicture}[baseline=-3]
\begin{feynman}[small]
\vertex (b) [dot] at (0, 0) {};
\vertex (a) at (-1.3, 0) {\footnotesize $1$};
\vertex (c) at ( 1.3, 0) {\footnotesize $n$};
\vertex (d) at (-1,-0.83) {\footnotesize $2$};
\vertex (e) at (1.1,-0.91) {\footnotesize $n-1$};
\vertex (l1) at (-0.3,0.8) {};
\vertex (l4) at (0.35,-0.35) {};
\vertex (l3) at (-0.35,-0.35) {};
\vertex (l2) at (0.3,0.8) {};
\vertex (d1) at (0.8,-0.31) {};
\vertex (d2) at (0.8,0.31) {};
\vertex (d3) at (-0.8,-0.31) {};
\vertex (d4) at (-0.8,0.31) {};
\vertex (d5) at (-0.839,-0.241) {};
\vertex (d6) at (-0.392,-0.780) {};
\vertex (d7) at (0.839,-0.241) {};
\vertex (d8) at (0.392,-0.780) {};
\diagram* {
	(a) -- (b) -- [out=30, in=80, loop, min distance=1.3cm,edge label'={\footnotesize $\epsilon$}] b -- [out=100, in=150, loop, min distance=1.3cm,edge label'={\footnotesize $1$}] b -- (c);
	(d)--(b)--(e);
	(l1) -- [thick,dotted,out=30, in=150](l2);
	(l3) -- [thick,dotted,out=-35, in=-145](l4);
};
\end{feynman}
\draw (d1) to (d2);
\draw (d3) to (d4);
\draw (d5) to (d6);
\draw (d7) to (d8);
\end{tikzpicture} = -\frac{1}{2}g\mu^{2+(2-d)\epsilon} (2\epsilon+2)_{n} \,(2\epsilon+1-n)!!\,\Delta(0)^{\epsilon+1-\frac{n}{2}} \,.
\end{equation}

From the lagrangian (\ref{Lagrangian}) we see that the mass term in the loops is $M^2=m^2+g\mu^2$. In passing we note that considering also loops with iterated insertions of the crossed diagram (\ref{G2weakgcounter}) we would obtain the shift $M^2\to m^2$ (see comments below Eq.\,(\ref{mRg})).

Referring to (\ref{Gnres}) we stress that the resummation (\ref{Gnres2}) gives only one term as $f_n(1)=0$, and this reflects the fact that the term $-\frac{1}{2}g\mu^2\phi^2$ does not give connected diagrams for $n>2$, that in other words means that in this case there is no diagram corresponding to the crossed diagram of (\ref{G2weakgcounter}).

A striking result is that with the resummation (\ref{Gnres2}) the $\mathcal{G}_n$ with $n>2\epsilon+2$ vanish (due to the vanishing of the falling factorial). This is in agreement with the well-known fact that, at first order in the weak-coupling expansion, for $n>2\epsilon+2$ we cannot draw connected diagrams as those in (\ref{Gnweakg}). 

Last but (certainly) not least we note that, similarly to what has already been seen for the two-points Green's function, having the resummation restored the weak-coupling result, the behaviour of the $\mathcal{G}_n$ with respect to their dependence from the momentum cut-off $\Lambda$ is drastically changed. Indeed Eq.\,(\ref{Gnres2}) shows that the resummation of the leading diagrams generates powers of $\Delta(0)$ in the numerator so to compensate the suppressing powers of $\Delta(0)$ in the denominator found at any finite order in $\epsilon$:
\begin{align} \label{G2res3}
\sum_{k=1}^{\infty}\,\left\{
\begin{tikzpicture}[baseline=-3]
\begin{feynman}
\vertex[draw, circle, minimum size=0.85cm,very thick] (m) at (0, 0) {k};
\vertex (a) at (-0.9,0.9) {\footnotesize$2$};
\vertex (b) at (-0.9,-0.9) {\footnotesize$1$};
\vertex (c) at (1.24421,0) {\footnotesize$n$};
\vertex (l1) at (0.5,0.35) {};
\vertex (l2) at (-0.2,0.55) {};
\vertex (d1) at (0.595294,-0.308824) {};
\vertex (d2) at (0.924706,0.308824) {};
\vertex (d3) at (-0.700391,0.226862) {};
\vertex (d4) at (-0.370979,0.844509) {};
\vertex (d5) at (-0.700391,-0.226862) {};
\vertex (d6) at (-0.370979,-0.844509) {};
\diagram* {
	(a) -- (m) -- (b);
	(m) -- (c);
	(l1) -- [thick,dotted,out=-280, in=-305](l2);
};
\draw (d1) to (d2);
\draw (d3) to (d4);
\draw (d5) to (d6);
\end{feynman}
\end{tikzpicture} 
\right\}\sim
\frac{1}{[\Delta(0)]^{\frac{n}{2}-1}} \sum_{k=1}\,\frac{\epsilon^k}{k!}\log[\Delta(0)]^{k-1} \sim [\Delta(0)]^{\epsilon-\frac{n}{2}+1}\,.
\end{align}

Few comments are in order. Eq.\,(\ref{G2res3}) shows that suitable resummations of diagrams allow to recover the interacting character of the theory. In particular, having chosen to resum the specific class of diagrams that have a leading behaviour with $\Lambda$, we recovered the results of the weak-coupling expansion. More precisely, as a by-product of our analysis, we have established the bridge between the expansion in the non-linearity parameter $\epsilon$ and the weak-coupling expansion at $\mathcal{O}(g)$. 

As already hinted, it seems to us that from these and previous results we can infer that, before proceeding to a systematic program for the renormalization of the theory, it is necessary to consider appropriate resummations of diagrams (not necessarily the one considered in the present section). In fact, as we have already seen in a specific example at the end of section 2, the attempt of renormalizing the theory order-by-order in the $\epsilon$-expansion seems to give rise to a ``too weird" behaviour of the parameters that define the theory. Just to mention an example and sticking on the $d=4$ case, we remind that, to get a finite $\mathcal{G}_4$ at order $\epsilon$, we had to give to the coupling constant $g$ a dependence on $\Lambda$ that goes as $g\sim\Lambda^2$, that in turn gives for the radiative correction $\delta m^2$ to the mass the cut-off dependence $\delta m^2 \sim \Lambda^2 \log\Lambda$ (see comments below Eq.\,(\ref{Gnzero}))

On the contrary, the resummation performed in this section ``cures" the problem with the vanishing of the cross sections, restoring the divergence structure of the weak-coupling expansion, actually the weak-coupling results tout-court.

Motivated by these results, in the following we push further the analysis on the link between the two expansions. To this end, in the next section we will perform higher order resummations, considering in particular diagrams with two effective vertices $\Pi_i$.

\section{Resummation of two-vertex diagrams}
In the previous section we have considered the resummation of diagrams with one effective vertex. We want to move now a step further in the approximation of the Green's functions, by considering the resummation of diagrams with two effective vertices. As explained in Section 4, the latters contain next-to-leading contributions, and even though there are other diagrams with the same subleading behaviour, these are the only $\mathcal{O}(g^2)$ ones. Therefore, for the purposes of our comparison between the expansion in $\epsilon$ and the weak-coupling expansion, these are the only diagrams that we have to take into account.

First of all we have to remind that, when dealing with the analytic continuation of contributions coming from diagrams with two auxiliary vertices, the $l$-series that sums over the links between the two vertices brings an ultraviolet divergence in the loop integral for all the Green's function with $n>4$. In Sections 3 and 4 we regularized these expressions with the help of a finite upper limit $L_{max}$. So doing, all these $l$-series contributions are easily written as a sum of diagrams with two $\Pi$ effective vertices. 

At a generic order $\epsilon^k$ (with $k\geq2$), the total contribution to the $n$-points Green's function of the diagrams with two vertices, that we indicate with $\mathcal{G}^{(k,g^2)}_{n}$, is given by the sum of all the diagrams of the kind given in Eq.\,(\ref{twovertex}), where we distinguish the two classes of ``even" and ``odd" diagrams with respect to the parity of the number of links between the two vertices. Working on the two classes separately, we have (with obvious notations) $\mathcal{G}^{(k,g^2)}_{n}=\mathcal{G}^{(k,g^2)}_{n,\,E}+\mathcal{G}^{(k,g^2)}_{n,\,O}$ where:
\begin{align} \label{Gnkg2e}
\mathcal{G}^{(k,g^2)}_{n,\,E} &= \frac{1}{2} \,\sum_{\alpha=1}^{k-1} \,\sum_{j=0}^{\frac{n}{2}}\, \sum_{l=1}^{L_{max}} 
\left\{\begin{tikzpicture}[baseline=(m)]
\begin{feynman}[inline=(m)]
\vertex[draw, circle, minimum size=1cm,very thick] (m) at (-1, 0) {\footnotesize $\alpha$};
\vertex[label={center:{\footnotesize $u$}}] (m0) at (-1,-0.7) {};
\vertex[draw, circle, minimum size=1cm,very thick] (n) at (1, 0) {\scriptsize $k\!-\!\alpha$};
\vertex[label={center:{\footnotesize $w$}}] (n0) at (1,-0.7) {};	
\vertex (a) at (-2.3,1.3) {$x_{1}$};
\vertex (b) at (-2.3,-1.3) {$x_{2j}$};
\vertex (c) at (2.3,1.3) {$x_{2j+1}$};
\vertex (d) at (2.3,-1.3) {$x_{n}$};
\vertex (l1) at (1.8,0.7) {};
\vertex (l2) at (1.8,-0.7) {};
\vertex (p1) at (-1.8,0.7) {};
\vertex (p2) at (-1.8,-0.7) {};
\diagram* {
	(a) -- (m) -- (b);
	(m) -- [half left, out=17, in=163](n);
	(m) -- [half left, out=-17, in=-163](n);
	(m) -- [half left, out=36, in=144,dashed](n);
	(m) -- [half left, out=-36, in=-144,edge label'={\small $2l$},dashed](n);
	(c) -- (n) -- (d);	
	(l1) -- [thick,dotted,out=-60, in=60](l2);
	(p2) -- [thick,dotted,out=120, in=-120](p1);
};
\end{feynman}
\end{tikzpicture}
+\,\,\binom{n}{2j}-1 \,\, \text{perm.}\right\} \nonumber \\
&= \frac{1}{2} \sum_{\alpha=1}^{k-1} \sum_{j=0}^{n/2} \sum_{l=1}^{L_{max}} \,\,\,\Pi^{(\alpha)}_{2j+2l} \,\,\, \Pi^{(k-\alpha)}_{n-2j+2l}\,\,\frac{1}{(2l)!}\nonumber\\
&\times \left[\int d^du \, d^dw\, \prod_{i=1}^{2j}\Delta(x_i-u) \prod_{h=2j+1}^n \!\Delta(x_h-w) \,\, \Delta(u-w)^{2l}\,\, + \, \binom{n}{2j}-1 \, {\rm perm.}\right]
\end{align}

\begin{align} \label{Gnkg2o}
\mathcal{G}^{(k,g^2)}_{n,\,O} &=\frac{1}{2} \,\sum_{\alpha=1}^{k-1}\,\sum_{j=0}^{\frac{n}{2}-1}\,\sum_{l=0}^{L_{max}} \left\{\begin{tikzpicture}[baseline=(m)]
\begin{feynman}[inline=(m)]
\vertex[draw, circle, minimum size=1cm,very thick] (m) at (-1, 0) {\footnotesize $\alpha$};
\vertex[label={center:{\footnotesize $u$}}] (m0) at (-1,-0.7) {};
\vertex[draw, circle, minimum size=1cm,very thick] (n) at (1, 0) {\scriptsize $k\!-\!\alpha$};
\vertex[label={center:{\footnotesize $w$}}] (n0) at (1,-0.7) {};	
\vertex (a) at (-2.3,1.3) {$x_{1}$};
\vertex (b) at (-2.3,-1.3) {$x_{2j+1}$};
\vertex (c) at (2.3,1.3) {$x_{2j+2}$};
\vertex (d) at (2.3,-1.3) {$x_{n}$};
\vertex (l1) at (1.8,0.7) {};
\vertex (l2) at (1.8,-0.7) {};
\vertex (p1) at (-1.8,0.7) {};
\vertex (p2) at (-1.8,-0.7) {};
\diagram* {
	(a) -- (m) -- (b);
	(m) -- [half left,out=30, in=150, dashed](n);
	(m) -- [half left,out=-30, in=-150, edge label'={\footnotesize $2l+1$}, dashed](n);
	(m) -- (n);
	(c) -- (n) -- (d);	
	(l1) -- [thick,dotted,out=-60, in=60](l2);
	(p2) -- [thick,dotted,out=120, in=-120](p1);
};
\end{feynman}
\end{tikzpicture}
+\,\,\binom{n}{2j\!+\!1}-1 \,\, \text{perm.}\right\} \nonumber\\
&=\frac{1}{2} \sum_{\alpha=1}^{k-1} \sum_{j=0}^{\frac{n}{2}-1} \sum_{l=0}^{L_{max}} \,\,\,\Pi^{(\alpha)}_{2j+2l+2} \,\,\, \Pi^{(k-\alpha)}_{n-2j+2l}\,\,\frac{1}{(2l+1)!}\times\nonumber\\
&\times \left[\int d^du \, d^dw\, \prod_{i=1}^{2j+1}\Delta(x_i-u) \prod_{h=2j+2}^n \!\Delta(x_h-w) \,\, \Delta(u-w)^{2l+1}\,\, + \,\, \binom{n}{2j\!+\!1}-1 \,\, {\rm perm.}\right]\,.
\end{align}

Our goal is to evaluate the correction to the Green's functions given by the sum of these contributions coming from all orders in $\epsilon$, i.e. to calculate the sum of the series $\mathcal{G}^{(g^2)}_{n} = \sum_{k=2}^{\infty} \mathcal{G}^{(k,g^2)}_{n}$. The even and odd terms can be resummed separately, as long as these series converges.

Starting from the even case we have:
\begin{align} \label{Gng2e-0}
\mathcal{G}^{(g^2)}_{n,\,E} &= \frac{1}{2} \sum_{k=2}^{\infty}\,\sum_{\alpha=1}^{k-1} \sum_{j=0}^{n/2} \sum_{l=1}^{L_{max}} \,\Pi^{(\alpha)}_{2j+2l} \, \Pi^{(k-\alpha)}_{n-2j+2l}\,\frac{1}{(2l)!} \left[I_{2l}(x_1{,\scriptstyle \dots},x_{2j};x_{2j+1}{,\scriptstyle \dots},x_n) \,+\, \binom{n}{2j}-1 \, {\rm perm.}\right]
\end{align}
where we indicated with $I_{2l}(x_1,\dots,x_{2j};x_{2j+1},\dots,x_n)$ the integral of the propagators written in the last line of Eq.\,(\ref{Gnkg2e}).
Due to the presence of the cut-off $L_{max}$, the general term of the double series in $k$ and $\alpha$ is a sum of a finite number of terms (i.e. the sum over $j$ and $l$) so that, as long as the double series converges for each of these terms, it can be splitted as follows:
\begin{align} \label{Gng2e-1}
\mathcal{G}^{(g^2)}_{n,\,E} &= \frac{1}{2} \sum_{j=0}^{n/2} \sum_{l=1}^{L_{max}} \left\{\sum_{k=2}^{\infty}\,\sum_{\alpha=1}^{k-1}\, \Pi^{(\alpha)}_{2j+2l} \, \Pi^{(k-\alpha)}_{n-2j+2l}\right\}\frac{1}{(2l)!} \left[I_{2l}(x_1{,\scriptstyle \dots},x_{2j};x_{2j+1}{,\scriptstyle \dots},x_n) \,\,+\,\, \binom{n}{2j}-1 \, {\rm perm.}\right]\,.
\end{align}
At this point we need to evaluate the double series
\begin{align}
\sum_{k=2}^{\infty}\,\sum_{\alpha=1}^{k-1}\,\, \Pi^{(\alpha)}_{2j+2l} \,\, \Pi^{(k-\alpha)}_{n-2j+2l}\,.
\end{align}
It is easy to see that this double series is nothing but the Cauchy product:
\begin{align} \label{splittingseries}
\sum_{k=2}^{\infty}\,\sum_{\alpha=1}^{k-1}\,\, \Pi^{(\alpha)}_{2j+2l} \,\, \Pi^{(k-\alpha)}_{n-2j+2l} \,=\, \left(\sum_{\alpha=1}^{\infty}\Pi^{(\alpha)}_{2j+2l}\right) \cdot \left(\sum_{\beta=1}^{\infty}\Pi^{(\beta)}_{n-2j+2l}\right)
\end{align}
where $\beta=k-\alpha$. Therefore, as long as the two series are convergent and at least one of them is absolutely convergent, we can safely calculate the sum in the l.h.s. of Eq.(\ref{splittingseries}) as the product of the two series of $\Pi$'s. We already evaluated these series in the previous section, getting the resummed vertex functions that we denoted with $\Gamma$'s (see Eqs.\,(\ref{G2res}) and (\ref{Gnres})-(\ref{Gnres2})).

We note that, as these series are nothing but the Taylor expansions around the point $x=1$ of functions that are analytic in the the complex half-plane $Re(x)>-\frac{1}{2}$ (due to the presence of the factor $\Gamma(x+\frac{1}{2})$), they have radius of convergence $\frac{3}{2}$, so that the validity of Eq.\,(\ref{splittingseries}) is guaranteed for $\epsilon<\frac{3}{2}$ (as in this case the absolute convergence is guaranteed). The possibility of extending this result to larger values of $\epsilon$ is an interesting question that we have not pursued. 

Eq.\,(\ref{splittingseries}) is crucial for this section. In fact, due to the splitting of the double series, the resummation of the subleading class of diagrams with two effective vertices is reduced to the resummations of the $\Pi$'s at each vertex of the diagrams. This in turn means that the result will be expressible in terms of diagrams composed of resummed $\Gamma$ vertices, and (as we will see in a moment) this is ultimately the reason why the resummation of this class of diagrams will yield to the $\mathcal{O}(g^2)$ weak-coupling result.

Thanks to Eq.(\ref{splittingseries}) we can write the final result for the resummation of all the ``even" diagrams with two effective vertices as:
\begin{align} \label{Gng2e-2}
\mathcal{G}^{(g^2)}_{n,\,E} =& \frac{1}{2} \sum_{j=0}^{n/2} \sum_{l=1}^{L_{max}} \left(\sum_{\alpha=1}^{\infty}\Pi^{(\alpha)}_{2j+2l}\right) \left(\sum_{\beta=1}^{\infty}\Pi^{(\beta)}_{n-2j+2l}\right)  \frac{1}{(2l)!} \nonumber\\
\times& \left[I_{2l}(x_1,\dots,x_{2j};x_{2j+1},\dots,x_n) \,\,+\,\, \binom{n}{2j}-1 \, {\rm perm.}\right]\nonumber\\
=&\frac{1}{2} \sum_{j=0}^{n/2} \sum_{l=1}^{L_{max}} \frac{\Gamma_{2j+2l}\,\,\Gamma_{n-2j+2l}}{(2l)!} \left[I_{2l}(x_1{,\scriptstyle \dots},x_{2j};x_{2j+1}{,\scriptstyle \dots},x_n) \,\,+\,\, \binom{n}{2j}-1 \, {\rm perm.}\right]
\end{align}

Following the same steps, an analogous result is obtained for the odd contribution:
\begin{align} \label{Gng2o-2}
\mathcal{G}^{(g^2)}_{n,\,O} =& \frac{1}{2} \sum_{j=0}^{\frac{n}{2}-1} \sum_{l=0}^{L_{max}} \left(\sum_{\alpha=1}^{\infty}\Pi^{(\alpha)}_{2j+2l+2}\right) \left(\sum_{\beta=1}^{\infty}\Pi^{(\beta)}_{n-2j+2l}\right)  \frac{1}{(2l+1)!} \nonumber\\
\times& \left[I_{2l+1}(x_1,\dots,x_{2j+1};x_{2j+2},\dots,x_n) \,\,+\,\, \binom{n}{2j+1}-1 \, {\rm perm.}\right]\nonumber\\
=& \frac{1}{2} \sum_{j=0}^{\frac{n}{2}-1} \sum_{l=0}^{L_{max}} \frac{\Gamma_{2j+2l+2}\,\, \Gamma_{n-2j+2l}}{(2l+1)!}\left[I_{2l+1}(x_1{,\scriptstyle \dots},x_{2j+1};x_{2j+2}{,\scriptstyle \dots},x_n) \,\,+\,\, \binom{n}{2j+1}-1 \, {\rm perm.}\right]
\end{align}

The final result is clearly obtained once we replace in both (\ref{Gng2e-2}) and (\ref{Gng2o-2}) the resummed vertices $\Gamma$ derived in the previous section, that for the reader's convenience we report here for the physically relevant cases of integer values of $\epsilon$\,:
\begin{equation} \label{Piresum}
\Gamma_n=\sum_{k=1}^{\infty}\Pi_n^{(k)} \,\overset{\epsilon \in \mathbb{Z_+}}{=}\, 
\begin{cases*}
g \mu^2 -(\epsilon +1)(2\epsilon+1)!!\,g \mu^{2+(2-d)\epsilon}\, \Delta(0)^{\epsilon} & for $n=2$ \\
-g\,\mu^{2+(2-d)\epsilon}\, 2^{\frac{n}{2}-1}(\epsilon+1)_{\frac{n}{2}} (2\epsilon+1)!! \Delta(0)^{\epsilon+1-\frac{n}{2}} & for $n\geq4$
\end{cases*}
\end{equation}

It is worth to note that for a fixed positive integer value of $\epsilon$ the falling factorial $(\epsilon+1)_{\frac{n}{2}}$ vanishes for $n>2\epsilon+2$. Therefore, once the resummation is performed then, the number of links between the vertices is limited by this restriction, so that the cut-off $L_{max}$ does not play any role for integer values of $\epsilon$. As we will see in moment, such a result is crucial to have the coincidence between the sum of (\ref{Gng2e-2}) and (\ref{Gng2o-2}) and the $\mathcal{O}(g^2)$ weak-coupling result, as in the weak coupling expansion the number of legs at each vertex is naturally limited by the specific interaction term. 

To better appreciate the results in (\ref{Gng2e-2}) and (\ref{Gng2o-2}), we now focus on the two- and four-points Green's functions, considering integer values of $\epsilon$.

Starting from the two-points function and replacing (\ref{Piresum}) in (\ref{Gng2e-2}) and (\ref{Gng2o-2}), we get:
\begin{align} \label{G2g2e}
\mathcal{G}^{(g^2)}_{2,\,E} &= \sum_{l=1}^{L_{max}}   \,\frac{\Gamma_{2l} \Gamma_{2+2l}}{(2l)!}\,\, I_{2l}(x_1,x_2;)\nonumber \\
&= \left[g \mu^2 -(\epsilon +1)(2\epsilon+1)!!\,g \mu^{2+(2-d)\epsilon}\, \Delta(0)^{\epsilon}\right]
\left[-g \mu^{2+(2-d)\epsilon}\, 2(\epsilon+1) \epsilon (2\epsilon+1)!! \Delta(0)^{\epsilon-1}\right] \nonumber\\
&\qquad \times \frac{1}{2} \int d^du \, d^dw\, \Delta(x_1-u)\Delta(x_2-u)\Delta(u-w)^2 \nonumber\\
&+ \sum_{l=2}^{\epsilon} \left[-g \mu^{2+(2-d)\epsilon}\, 2^{l-1}(\epsilon+1)_{l} (2\epsilon+1)!! \Delta(0)^{\epsilon+1-l}\right]\left[-g \mu^{2+(2-d)\epsilon}\, 2^{l}(\epsilon+1)_{l+1} (2\epsilon+1)!! \Delta(0)^{\epsilon-l}\right]\nonumber\\
&\qquad \times\frac{1}{(2l)!} \int d^du\,d^dw\, \Delta(x_1-u)\Delta(x_2-u)\Delta(u-w)^{2l}
\end{align}
and
\begin{align} \label{G2g2o}
\mathcal{G}^{(g^2)}_{2,\,O} &= \sum_{l=0}^{L_{max}}  \frac{\left(\Gamma_{2+2l}\right)^2}{(2l+1)!}\,\, I_{2l+1}(x_1;x_2)\nonumber \\
&= \left[g \mu^2 -(\epsilon +1)(2\epsilon+1)!!\,g \mu^{2+(2-d)\epsilon}\, \Delta(0)^{\epsilon}\right]^2 \int d^du \, d^dw\, \Delta(x_1-u)\Delta(u-w)\Delta(x_2-w)\nonumber\\ 
&+ \sum_{l=1}^{\epsilon}\left[-g \mu^{2+(2-d)\epsilon}\, 2^{l}(\epsilon+1)_{l+1} (2\epsilon+1)!! \Delta(0)^{\epsilon-l}\right]^2 \,\frac{1}{(2l+1)!}\nonumber \\
&\quad\times\int d^du \, d^dw\, \Delta(x_1-u)\Delta(u-w)^{2l+1} \Delta(x_2-w)\,.
\end{align}
It is easy to verify that the sum of (\ref{G2g2e}) and (\ref{G2g2o}) is nothing but the result that we would have obtained at the second order of the weak-coupling expansion for the two-points Green's function of the $\phi^{2+2\epsilon}$ theory given by the Lagrangian (\ref{Lagrangian}).

For the sake of concreteness, let us show this coincidence for the particular case $\epsilon=1$, i.e. for the $\phi^4$ theory. In this case, starting from the splitting in free and interaction terms of the Lagrangian given in Eq.\,(\ref{Lagrangian}), the $\mathcal{O}(g^2)$ correction to the two-points Green's function is given by the following diagrams:

\begin{eqnarray} \label{G2weakg2-diag1}
\begin{tikzpicture}[baseline=-3]
\begin{feynman}
\vertex[crossed dot, minimum size=0.3cm] (w) at (0, 1) {};
\vertex[dot] (u) at (0,0) {};
\vertex (a) at (-1.24421,0) {$x_1$};
\vertex (b) at (1.24421,0) {$x_2$};
\vertex (d1) at (0.595294,-0.308824) {};
\vertex (d2) at (0.924706,0.308824) {};
\vertex (d3) at (-0.924706,-0.308824) {};
\vertex (d4) at (-0.595294,0.308824) {};
\diagram* {
	(a) -- (u) -- (b);
	(u) -- [out=50, in=-50] (w) -- [out=-130, in=130] (u);
};
\end{feynman}
\end{tikzpicture} &=& -6\, g^2\mu^{6-d} \int d^du\,d^dw\, \Delta(x_1-u)\Delta(x_2-u)\Delta(u-w)^{2}
\\ \label{G2weakg2-diag2}
\begin{tikzpicture}[baseline=-3]
\begin{feynman}
\vertex[dot] (w) at (0, 0.9) {};
\vertex[dot] (u) at (0,0) {};
\vertex (a) at (-1.24421,0) {$x_1$};
\vertex (b) at (1.24421,0) {$x_2$};
\vertex (d1) at (0.595294,-0.308824) {};
\vertex (d2) at (0.924706,0.308824) {};
\vertex (d3) at (-0.924706,-0.308824) {};
\vertex (d4) at (-0.595294,0.308824) {};
\diagram* {
	(a) -- (u) -- (b);
	(u) -- [out=40, in=-40] (w) -- [out=-140, in=140] (u);
	(w) -- [out=50, in=130, loop, min distance=1.3cm] w;
};
\end{feynman}
\end{tikzpicture} &=& 36 \left(g\mu^{4-d}\right)^2 \Delta(0) \int d^du\,d^dw\, \Delta(x_1-u)\Delta(x_2-u)\Delta(u-w)^{2}\nonumber\\
\\ \label{G2weakg2-diag3}
\begin{tikzpicture}[baseline=-3]
\begin{feynman}
\vertex[crossed dot, minimum size=0.3cm] (u) at (-0.5, 0) {};
\vertex[crossed dot, minimum size=0.3cm] (w) at (0.5, 0) {};
\vertex (a) at (-1.64421,0) {$x_1$};
\vertex (b) at (1.64421,0) {$x_2$};
\vertex (d1) at (0.595294,-0.308824) {};
\vertex (d2) at (0.924706,0.308824) {};
\vertex (d3) at (-0.924706,-0.308824) {};
\vertex (d4) at (-0.595294,0.308824) {};
\diagram* {
	(a) -- (u) -- (w) -- (b);
};
\end{feynman}
\end{tikzpicture} &=& \left(g\mu^2\right)^2 \int d^du\,d^dw\, \Delta(x_1-u)\Delta(u-w)\Delta(x_2-w)
\\ \label{G2weakg2-diag4}
\begin{tikzpicture}[baseline=-3]
\begin{feynman}
\vertex[crossed dot, minimum size=0.3cm] (u) at (-0.5, 0) {};
\vertex[dot] (w) at (0.5, 0) {};
\vertex (a) at (-1.64421,0) {$x_1$};
\vertex (b) at (1.64421,0) {$x_2$};
\vertex (d1) at (0.595294,-0.308824) {};
\vertex (d2) at (0.924706,0.308824) {};
\vertex (d3) at (-0.924706,-0.308824) {};
\vertex (d4) at (-0.595294,0.308824) {};
\diagram* {
	(a) -- (u) -- (w) -- (b);
	(w) -- [out=50, in=130, loop, min distance=1.5cm] w;
};
\end{feynman}
\end{tikzpicture} &=& -12 \, g^2\mu^{6-d} \Delta(0) \int d^du\,d^dw\, \Delta(x_1-u)\Delta(u-w)\Delta(x_2-w) \nonumber \\
\\ \label{G2weakg2-diag5}
\begin{tikzpicture}[baseline=-3]
\begin{feynman}
\vertex[dot] (u) at (-0.5, 0) {};
\vertex[dot] (w) at (0.5, 0) {};
\vertex (a) at (-1.64421,0) {$x_1$};
\vertex (b) at (1.64421,0) {$x_2$};
\vertex (d1) at (0.595294,-0.308824) {};
\vertex (d2) at (0.924706,0.308824) {};
\vertex (d3) at (-0.924706,-0.308824) {};
\vertex (d4) at (-0.595294,0.308824) {};
\diagram* {
	(a) -- (u) -- (w) -- (b);
	(u) -- [out=50, in=130, loop, min distance=1.5cm] u;
	(w) -- [out=50, in=130, loop, min distance=1.5cm] w;
};
\end{feynman}
\end{tikzpicture} &=& 36 \left(g\mu^{4-d}\right)^2 \Delta(0)^2 \int d^du\,d^dw\, \Delta(x_1-u)\Delta(u-w)\Delta(x_2-w) \nonumber \\
\\ \label{G2weakg2-diag6}
\begin{tikzpicture}[baseline=-3]
\begin{feynman}
\vertex[dot] (u) at (-0.5, 0) {};
\vertex[dot] (w) at (0.5, 0) {};
\vertex (a) at (-1.64421,0) {$x_1$};
\vertex (b) at (1.64421,0) {$x_2$};
\vertex (d1) at (0.595294,-0.308824) {};
\vertex (d2) at (0.924706,0.308824) {};
\vertex (d3) at (-0.924706,-0.308824) {};
\vertex (d4) at (-0.595294,0.308824) {};
\diagram* {
	(a) -- (u) -- (w) -- (b);
	(u) -- [out=90, in=90, min distance=0.6cm] (w);
	(u) -- [out=-90, in=-90, min distance=0.6cm] (w);
};
\end{feynman}
\end{tikzpicture} &=& 24 \left(g\mu^{4-d}\right)^2 \int d^du\,d^dw\, \Delta(x_1-u)\Delta(u-w)^3\Delta(x_2-w)
\end{eqnarray}
where the first two diagrams coincide with the two terms corresponding to $l=1$ in Eq.\,(\ref{G2g2e}) (which is the only possible value of $l$ for $\epsilon=1$), while the three following diagrams coincide with the three possible terms corresponding to $l=0$ in Eq.\,(\ref{G2g2o}), and the last one corresponds to the $l=1$ odd term.
Let us finally note that the above diagrams are the usual second order Feynman diagrams for $\mathcal{G}_2$ of a $\phi^4$ theory, with the addition of diagrams that contain the insertion of two-legs vertices due to the splitting of the mass term.
Moreover, the third, fourth and fifth diagrams are reducible (and in fact they are already considered in the geometrical series of the previous section related to the radiative correction to the mass) while the first, second and sixth diagrams are 1PI, and genuinely provide the $\mathcal{O}(g^2)$ contribution to the proper self-energy.

Moving to the 4-points Green's function, the resummation gives:

\begin{align} \label{G4g2e}
&\mathcal{G}^{(g^2)}_{4,\,E} 
= \sum_{l=1}^{L_{max}} \frac{\Gamma_{2l}\Gamma_{4+2l}}{(2l)!} I_{2l}(x_1,\dots,x_{4};)+ \sum_{l=1}^{L_{max}} \frac{\left( \Gamma_{2+2l}\right)^2}{(2l)!}\left[I_{2l}(x_1,x_2;x_3,x_4) \,\,+\,\, 2 \,{\rm perm.}\right]\nonumber\\
&= \left[g \mu^2 -(\epsilon +1)(2\epsilon+1)!!\,g \mu^{2+(2-d)\epsilon}\, \Delta(0)^{\epsilon}\right] \left[-g\,\mu^{2+(2-d)\epsilon}\, 4(\epsilon+1)\epsilon(\epsilon-1) (2\epsilon+1)!! \Delta(0)^{\epsilon-2}\right] \nonumber\\ & \qquad \times \int d^du \, d^dw\, \Delta(x_1-u)\dots\Delta(x_4-u)\Delta(u-w)^2 \nonumber\\
&+ \sum_{l=2}^{\epsilon-1} \left[-g\,\mu^{2+(2-d)\epsilon}\, 2^{l-1}(\epsilon+1)_{l} (2\epsilon+1)!! \Delta(0)^{\epsilon+1-l}\right]  \left[-g\,\mu^{2+(2-d)\epsilon}\, 2^{l+1}(\epsilon+1)_{l+2} (2\epsilon+1)!! \Delta(0)^{\epsilon-1-l}\right] \nonumber \\ &\qquad \times \frac{1}{(2l)!} \int d^du \, d^dw\, \Delta(x_1-u)\dots\Delta(x_4-u)\Delta(u-w)^{2l}
\nonumber\\
&+ \sum_{l=1}^{\epsilon} \left[-g\,\mu^{2+(2-d)\epsilon}\, 2^{l}(\epsilon+1)_{l+1} (2\epsilon+1)!! \Delta(0)^{\epsilon-l}\right]^2  \nonumber \\
&\qquad \times \frac{1}{(2l)!} \left[ \int d^du \, d^dw\, \Delta(x_1-u)\Delta(x_2-u) \Delta(u-w)^{2l} \Delta(x_3-w)\Delta(x_4-w)\,\,+\,\, 2 \,{\rm perm.}\right]
\end{align}
\begin{align} \label{G4g2o}
&\mathcal{G}^{(g^2)}_{4,\,O} 
= \sum_{l=0}^{L_{max}}  \frac{\Gamma_{2+2l}\Gamma_{4+2l}}{(2l+1)!} \left[I_{2l+1}(x_1;x_2,x_3,x_4)\,\,+\,\, 3 \,{\rm perm.}\right] \nonumber\\
&= \left[g \mu^2 -(\epsilon +1)(2\epsilon+1)!!\,g \mu^{2+(2-d)\epsilon}\, \Delta(0)^{\epsilon}\right] \left[-g\,\mu^{2+(2-d)\epsilon}\, 2 (\epsilon+1)\epsilon (2\epsilon+1)!! \Delta(0)^{\epsilon-1}\right] \nonumber\\
&\qquad \times \left[\int d^du \, d^dw\, \Delta(x_1-u)\Delta(u-w)\Delta(x_2-w)\dots\Delta(x_4-w)\,\,+\,\, 3\, {\rm perm.}\right] \nonumber\\
&+\sum_{l=1}^{\epsilon-1} \left[-g\,\mu^{2+(2-d)\epsilon}\, 2^{l}(\epsilon+1)_{l+1} (2\epsilon+1)!! \Delta(0)^{\epsilon-l}\right] \left[-g\,\mu^{2+(2-d)\epsilon}\, 2^{l+1}(\epsilon+1)_{l+2} (2\epsilon+1)!! \Delta(0)^{\epsilon-l-1}\right]  \nonumber\\
&\qquad \times \frac{1}{(2l+1)!}\left[\int d^du \, d^dw\, \Delta(x_1-u)\Delta(u-w)^{2l+1}\Delta(x_2-w)\dots\Delta(x_4-w)\,\,+\,\, 3\, {\rm perm.}\right]
\end{align}
Again, it is easy to check that the sum of (\ref{G4g2e}) and (\ref{G4g2o}) coincides with the $\mathcal{O}(g^2)$ correction to the 4-points Green's function in the weak-coupling expansion of the $\phi^{2+2\epsilon}$ theory described by the Lagrangian (\ref{Lagrangian}). As before, specifying to the case $\epsilon=1$,  we obtain:
\begin{align}
&\begin{tikzpicture}[baseline=-3]
\begin{feynman}
\vertex[dot] (m) at (-0.8, 0) {};
\vertex[dot] (n) at (0.8, 0) {};	
\vertex (a) at (-1.9,1.1) {$x_{1}$};
\vertex (b) at (-1.9,-1.1) {$x_{2}$};
\vertex (c) at (1.9,1.1) {$x_{3}$};
\vertex (d) at (1.9,-1.1) {$x_{4}$};
\diagram* {
	(a) -- (m) -- (b);
	(m) -- [half left, out=45, in=135](n);
	(m) -- [half left, out=-45, in=-135](n);
	(c) -- (n) -- (d);	
};
\end{feynman}
\end{tikzpicture}  \nonumber\\
&= 72 \left(g\mu^{4-d}\right)^2 \int d^du \, d^dw\, \Delta(x_1-u)\Delta(x_2-u)\Delta(u-w)^{2}  \Delta(x_3-w)\Delta(x_4-w)
\end{align}
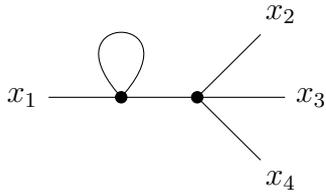
\begin{align} \label{G4tadpole}
&\begin{tikzpicture}[baseline=-3]
\begin{feynman}
\vertex[dot] (m) at (-0.4, 0) {};
\vertex[dot] (n) at (0.6, 0) {};
\vertex (a) at (-1.7,0) {$x_1$};
\vertex (b) at ( 1.7,1.1) {$x_2$};
\vertex (c) at ( 2.1,0) {$x_3$};
\vertex (d) at ( 1.7,-1.1) {$x_4$};
\diagram* {
	(a) -- (m);
	(m) -- [out=50, in=130, loop, min distance=1.4cm] m;
	(m) -- (n);
	(n) -- (b);
	(n) -- (c);
	(n) -- (d);
};
\end{feynman}
\end{tikzpicture} \nonumber \\
&= 72 (g\mu^{4-d})^2 \Delta(0) \int d^du \, d^dw\, \Delta(x_1-u) \Delta(u-w) \Delta(x_2-w)\Delta(x_3-w)\Delta(x_4-w)
\\ \label{G4insertion}
&
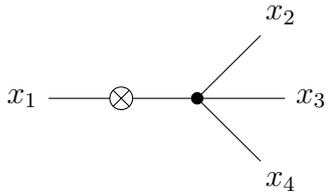
\begin{tikzpicture}[baseline=-3]
\begin{feynman}
\vertex[crossed dot, minimum size=0.3cm] (m) at (-0.4, 0) {};
\vertex[dot] (n) at (0.6, 0) {};
\vertex (a) at (-1.7,0) {$x_1$};
\vertex (b) at ( 1.7,1.1) {$x_2$};
\vertex (c) at ( 2.1,0) {$x_3$};
\vertex (d) at ( 1.7,-1.1) {$x_4$};
\diagram* {
	(a) -- (m);
	(m) -- (n);
	(n) -- (b);
	(n) -- (c);
	(n) -- (d);
};
\end{feynman}
\end{tikzpicture} \nonumber\\
&= 12\, g^2\mu^{6-d} \int d^du \, d^dw\, \Delta(x_1-u) \Delta(u-w)
\Delta(x_2-w)\Delta(x_3-w)\Delta(x_4-w)
\end{align}
where the first diagram corresponds to the $l=1$ term of the sum in the last line of Eq.(\ref{G4g2e}) (that is the only non-vanishing ``even" term for $\epsilon$=1), while the second and third diagrams correspond to the two terms in the first line of Eq.(\ref{G4g2o}) (for $\epsilon=1$ all the other odd terms vanish).
Similarly to the previous case, the diagrams (\ref{G4tadpole}) and (\ref{G4insertion}) are one particle reducible and are related to the external propagator correction, while the first diagram is the genuine $\mathcal{O}(g^2)$ correction to $\mathcal{G}_4$. 

Although in the above lines we have explicitly considered the results of the resummations in the cases of $\mathcal{G}_2$ and $\mathcal{G}_4$ and shown that they coincide with those of the $\mathcal{O}(g^2)$ weak-coupling expansion for  the $\phi^4$ theory (i.e. $\epsilon=1$), these results are of a more general validity: this coincidence holds for all the Green's functions and for all the integer values of $\epsilon$ for which (\ref{splittingseries}) is fulfilled. This result is actually contained in the general Eqs.\,(\ref{Gng2e-2}) and (\ref{Gng2o-2}), where the terms appearing in the weak coupling expansion, including the combinatorial factors, although differently organized, are all present in the $\Gamma$'s resummed vertices and in the integrals contained in the square brackets.

Pushing the analysis of Section 5, in the present section we showed how the bridge between the expansion in the non-linearity parameter $\epsilon$ and the weak coupling expansion is realized at $\mathcal{O}(g^2)$: it is obtained by resumming the diagrams with two effective vertices.

In particular we have seen that, as compared to the $\mathcal{O}(g)$ case, the peculiarity of the $\epsilon$-expansion brings some technical difficulties that are in part related to the necessity of introducing the cut-off $L_{max}$ when considering generic values of $\epsilon$. Interestingly, when we move to integer values of $\epsilon$, the cut-off $L_{max}$ does not play any longer a role, and we easily obtain the weak coupling results.
Moreover, a delicate point of this analysis concerns the convergence and/or absolute convergence of the series involved in the calculation, more precisely the Cauchy product of (\ref{splittingseries}). In particular, a word of caution has to be said for values of $\epsilon>3/2$, where the splitting  for the double series is not guaranteed (see comments below Eq.\,(\ref{splittingseries})), while for $\epsilon<3/2$ the results are well-established. In this respect, it is worth to remind that the interacting $\phi^4$ theory is obtained for $\epsilon=1$.

Before moving to the conclusions, we would like to observe that, following the same line of reasoning of this section, we expect that the bridge between the two expansions actually extends also to higher orders in $g$. In particular, sticking on the assumption that the contributions to the $\mathcal{G}_n$ coming from diagrams with three or more vertices can still be regularized with the help of $L_{max}$ cut-offs, the resummation of diagrams with a generic number of vertices $m$ should yield to the results of the weak-coupling expansion at order $g^m$.

\section{Summary, conclusions and outlooks}
Renormalization and Renormalization Group techniques are crucial tools for extracting meaningful result from interacting quantum field theories. In a recent paper \cite{Bender:2018pbv} it was suggested that a previously introduced formal expansion in a parameter that measures the ``distance" between the free and the interacting theory \cite{Bender:1987dn}, i.e. the parameter $\epsilon$ introduced in Section 2, could be the key for a systematic renormalization program of non-hermitian PT-symmetric theories. 

Motivated by this suggestion, in the present paper we took a step forward in the analysis of such an expansion. However, in order to avoid the complications related to the peculiar structure of non-hermitian theories, we have performed the analysis for ordinary interacting scalar theories, leaving the investigation of non-hermitian PT-symmetric theories is left for future work\,\cite{future}.

In Section 2 we started by reviewing the $\mathcal{O}(\epsilon)$ results \cite{Bender:1987dn,Bender:1988rq}, and then moved to the analysis of the renormalization properties of the theory. In particular, considering the two-points Green's function, we noted that, as compared to the weak-coupling expansion at $\mathcal{O}(g)$, the radiative correction to the mass at $\mathcal{O}(\epsilon)$ is significantly milder. Implementing the regularization with a physical cut-off $\Lambda$, and considering $d>2$ dimensions, we showed that the $\mathcal{O}(\epsilon)$ correction goes as $\log(\Lambda)$, irrespectively of the power of the interacting term (i.e. irrespectively of $\epsilon$), while for instance for a $\phi^4$ theory in $d=4$ dimensions the $\mathcal{O}(g)$ radiative correction to the mass goes as $\Lambda^2$. The comparison between these two results seemed to suggest that within the expansion in $\epsilon$ the hierarchy problem could be enormously alleviated.
However we have shown (see Section 5) that the resummation of the leading diagrams of each order in $\epsilon$ restores the weak-coupling result, that is a radiative correction that goes as $\Lambda^2$, thus presenting (at least at this order of approximation) the same unnaturalness problem already encountered within the weak-coupling expansion. 

As for the higher order Green's functions ($n\geq 4$), considering again the case $d\geq2$, we have seen that at $\mathcal{O}(\epsilon)$ all the $\mathcal{G}_n$ (and as a consequence all the scattering amplitudes) behave as inverse powers (increasing with $n$) of the cut-off, i.e. vanish with $\Lambda\to \infty$, thus implying that at this order in $\epsilon$ the theory is non-interacting.
However, similarly to the case of $\mathcal{G}_2$, in Section 5 we showed that the resummation of the leading diagrams from all orders in $\epsilon$ restores the weak-coupling results, with the known dependence on $\Lambda$ of the $\mathcal{G}_n$'s.

The systematic analysis of the order $\epsilon^2$ was performed in Section 3. We showed that this calculation presents several delicate aspects mainly connected with the analytic extension of results related to diagrams that contain two vertices.
In fact, while in the $\mathcal{O}(\epsilon)$ calculation the functions that needed to be extended were factored out of space-time integrals, for the $\mathcal{O}(\epsilon^2)$ case we have to consider analytic extensions of functions that are still included in space-time integrals. 
More specifically, the analytic extension of sums over links between two vertices gives rise to hypergeometric functions that could bring ultraviolet divergences for the Green's functions.
Clearly this is a delicate problem that deserves further investigations, beyond the scope of the present work. In our analysis, we took care of these divergences by introducing a numerical cut-off $L_{max}$ as upper limit of the power series, so rendering all the expressions finite and expressible as a sum of diagrams with two effective vertices.

However we have shown (Section 6) that a very interesting result is obtained when the resummation of these two-vertex diagrams for integer value of $\epsilon$ is considered. In this case, due to the vanishing of the resummed $\Gamma$ vertices, the series related to the number $l$ of internal lines are automatically truncated starting from a certain value of $l$, thus rendering the presence of $L_{max}$ harmless.

We then studied the dependence on the cut-off $\Lambda$ of the Green's functions at $\mathcal{O}(\epsilon^2)$ (Section 3) and then extended this analysis to higher orders in $\epsilon$ (Section 4). In this latter case diagrams with an increasing number of vertices appear, and the problem (already encountered at order $\epsilon^2$) of the divergences that arise from the analytic extensions of the series becomes more severe. We again assumed that a suitable regularization can be made by truncating the series with the help of numerical cut-offs $L_{max}$. So doing, we were able to express all these contributions in terms of sums of diagrams built with effective vertices $\Pi$'s (see Eq.\,(\ref{Pi_n^kdef})). It was then possible to study the dependence of the Green's functions on the physical cut-off $\Lambda$. 

The results of Sections 3 and 4 together showed that when considering higher and higher orders in $\epsilon$ the two-points Green's function $\mathcal{G}_2$ receives contributions that go as higher and higher powers of $\log(\Lambda)$, and similarly the other Green's functions $\mathcal{G}_n$ ($n\geq4$) become less and less suppressed in terms of $\Lambda$, although for every finite order in $\epsilon$ they still vanish in the $\Lambda\to \infty$ limit (see Eq.\,(\ref{1vert})).

In our opinion the vanishing of the higher order Green's functions at any finite order in $\epsilon$ can cast doubts on the  possibility of realizing the renormalization of the theory within the framework of this expansion. This is certainly a very delicate issue that deserves further investigation.

An attempt to implement a renormalization program in this context has been done in \cite{Bender:1988ux,Bender:1988ig}, where a $\phi^{2+2\epsilon}$ theory in $d=3,4$ dimensions is considered. 
However, while the approach followed in these papers is apparently successful at $\mathcal{O}(\epsilon)$, it seems to lack of systematicity when moving to higher orders in $\epsilon$.
More specifically, the authors begin by considering the $\mathcal{O}(\epsilon)$ contribution and, in order to keep the 4-points Green's function finite when $\Lambda\to\infty$, they introduce an unusual multiplicative renormalization\footnote{In \cite{Bender:1988ux,Bender:1988ig} an x-space regularization, with $a\sim\frac{1}{\Lambda}$, is used.}.
When subsequently the order $\epsilon^2$ contributions are added, the authors consider together the $\mathcal{O}(\epsilon)$ and $\mathcal{O}(\epsilon^2)$ terms, looking for a renormalization that can make $\mathcal{G}_4$ finite (non-vanishing). However the different orders in $\epsilon$ have not the same ultraviolet behavior, and it turns out that in order to realize this program it is necessary to make finite the terms of the highest order, i.e. the $\mathcal{O}(\epsilon^2)$ terms. This in turn results in the vanishing of the lower order contributions, i.e. the $\mathcal{O}(\epsilon)$ terms, when $\Lambda\to\infty$. We have checked that this behavior persists also when higher orders in $\epsilon$ are considered, as it turns out that the highest order contributions are always the dominant ones in terms of $\Lambda$.
We believe that, in order to overcame this mismatch of terms, an order-by-order renormalization procedure needs to be implemented, and in our opinion this is a necessary ingredient for an expansion in a ``small" parameter. Such a program has not yet been undertaken and is left for future work. 
Naturally this is not the last word on the possibility of finding a systematic way to renormalize the theory within the framework of the expansion in $\epsilon$, and alternative interpretations and/or approaches could be considered.

In this work we focused on the resummation of classes of diagrams in the expansion in $\epsilon$ that could provide better approximations to the Green's functions. Having found that the diagrams with a single effective vertex $\Pi$ are leading with respect to their dependence on the physical cut-off $\Lambda$, we first considered the resummation of these diagrams. 
So doing we got for the Green's functions the weak-coupling results at order $g$.
This is one of the central results of the present paper. The resummation that we have performed provides a connection between the expansion in the non-linearity parameter $\epsilon$ and the weak-coupling expansion. 

Motivated by this finding, we tried to put forward a selection criterion for the diagrams related to the expansion in the parameter $\epsilon$ in order to check whether this connection could be extended to higher orders in $g$. With this in mind, as a second step we considered the resummation of all those diagrams with two effective vertices $\Pi$. Interestingly it turned out that these diagrams are next-to-leading with respect to the previous ones\footnote{There are other diagrams with the same dependence on the physical cut-off $\Lambda$, but they are not $\mathcal{O}(g^2)$. In other words, the diagrams that are at the same time next-to-leading with respect to those with only one effective vertex $\Pi$ and order $g^2$ are only those with two effective vertices $\Pi$.}. 

Compared to the resummation that leads to the $\mathcal{O}(g)$ result, in this case we faced the additional difficulty of having to resum double series, and for that we needed to assure the convergence and absolute convergence of each of the involved series. In particular, we found that when $\epsilon<3/2$ these requirements are fulfilled and the resummation led to the known order $g^2$ weak-coupling results. This is of great interest, as for instance the physically relevant $\phi^4$ theory lies within the convergence radius $\epsilon=3/2$. 

Finally, we also hinted the result for the higher orders in $g$, indicating which class of diagrams should be resummed in order to get the weak-coupling results. 

In our opinion the results summarized above suggest that the ``mild" behaviour with the physical ultraviolet cut-off $\Lambda$ that the Green's functions present at any order in $\epsilon$, that would point towards a non-interacting theory at any order in $\epsilon$, is actually due to a ``too poor" truncation of the interaction term in the lagrangian (\ref{eq:Lexp}), that is intrinsically rooted in this peculiar expansion. Our results indicate that a possible way to find suitable approximations of the theory within the context of this expansion, and in particular to dig out the interacting character of the theory, could be found in realizing appropriate resummations of diagrams from \textit{all orders} in $\epsilon$. In this work we have found the resummation that reproduces the weak coupling results.

\end{document}